\documentclass[12pt]{article}

\usepackage{cite}
\usepackage{amsmath}
\usepackage{amssymb}
\usepackage{bm}
\usepackage{mathtools}
\usepackage{graphicx}
\numberwithin{equation}{section}

\def\Res{{\rm Res}}
\def\H{{\widehat H}}
\def\Q{{\widehat Q}}
\def\P{{\widehat P}}
\def\A{{\cal A}}
\def\E{{\cal E}}
\DeclareMathOperator{\Ai}{Ai}
\DeclareMathOperator{\Det}{Det}
\DeclareMathOperator{\Tr}{Tr}
\DeclareMathOperator{\Area}{Area}

\begin{document}

\renewcommand{\thefootnote}{\fnsymbol{footnote}}

\begin{titlepage}
\begin{flushright}
{\footnotesize OCU-PHYS 520, NITEP 73}
\end{flushright}
\bigskip
\begin{center}
{\Large\bf Spectral Theories and Topological Strings\\[6pt]
on del Pezzo Geometries
}\\
\bigskip\bigskip
{\large
Sanefumi Moriyama\footnote{moriyama@sci.osaka-cu.ac.jp}}\\
\bigskip
{\it
${}^*$Department of Physics, Graduate School of Science, Osaka City University,\\
${}^*$Nambu Yoichiro Institute of Theoretical and Experimental Physics (NITEP),\\
${}^*$Osaka City University Advanced Mathematical Institute (OCAMI),\\
3-3-138 Sugimoto, Sumiyoshi, Osaka 558-8585, Japan\\
}
\end{center}

\begin{abstract}
Motivated by understanding M2-branes, we propose to reformulate partition functions of M2-branes by quantum curves.
Especially, we focus on the backgrounds of del Pezzo geometries, which enjoy Weyl group symmetries of exceptional algebras.
We construct quantum curves explicitly and turn to the analysis of classical phase space areas and quantum mirror maps.
We find that the group structure helps in clarifying previous subtleties, such as the shift of the chemical potential in the area and the identification of the overall factor of the spectral operator in the mirror map.
We list the multiplicities characterizing the quantum mirror maps and find that the decoupling relation known for the BPS indices works for the mirror maps.
As a result, with the group structure we can present explicitly the statement for the correspondence between spectral theories and topological strings on del Pezzo geometries.

\centering\includegraphics[scale=0.35,angle=-90]{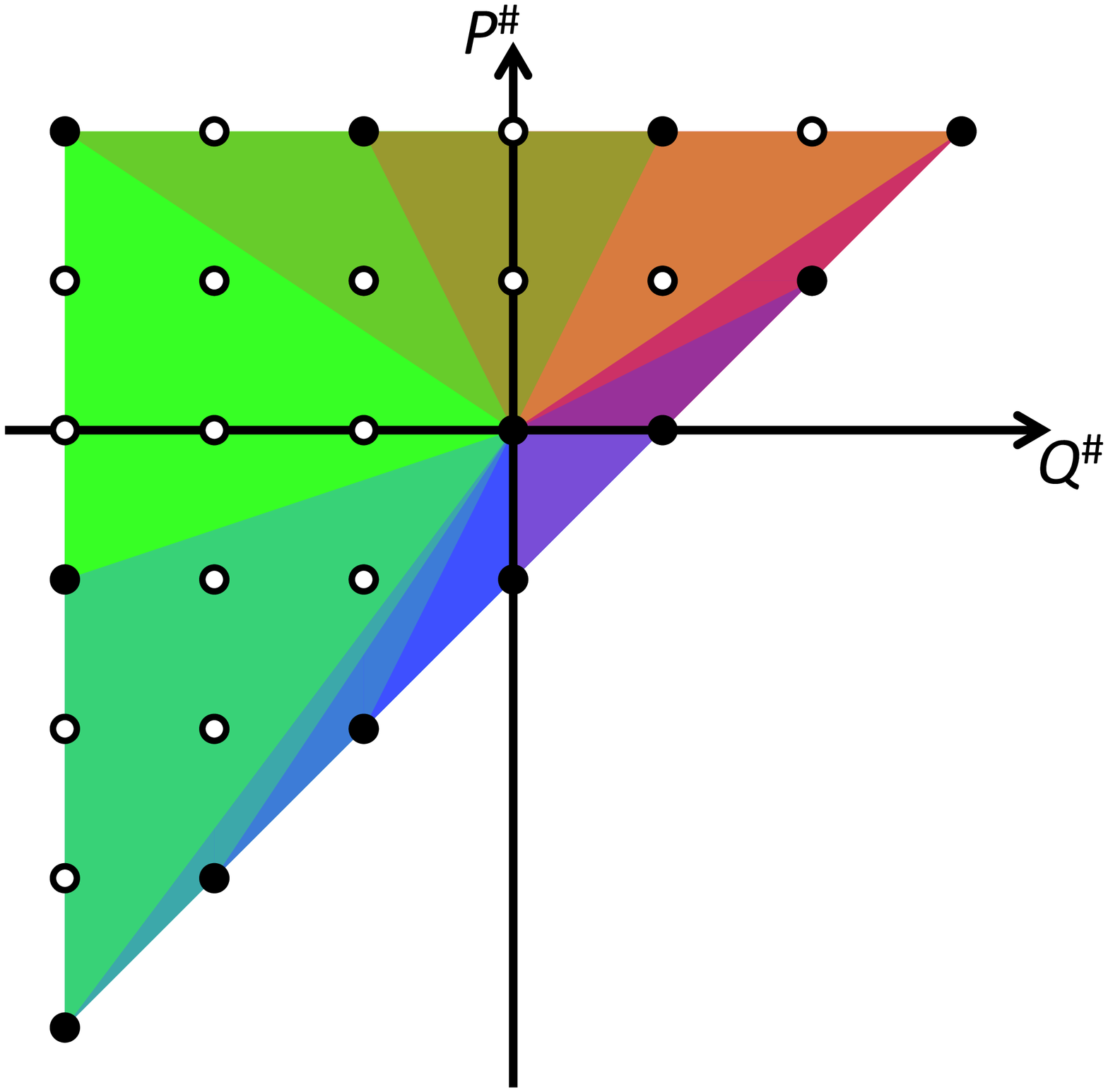}
\end{abstract}
\end{titlepage}

\setcounter{footnote}{0}

\section{Introduction and motivation}

In this paper, we formulate explicitly the correspondence between spectral theories and topological strings on del Pezzo geometries using their group-theoretical structure.
In the introduction, we first explain why this is interesting from the viewpoint of the M2-brane physics.

Recently much progress is made in understanding M-theory, especially a system of multiple M2-branes.
Classically it was known from the gravity analysis \cite{KT} that the degrees of freedom of the worldvolume theory on $N$ coincident M2-branes behave as $N^{\frac{3}{2}}$ in the large $N$ limit.
Following previous ideas on supersymmetrizing the Chern-Simons theory \cite{S,GW}, it was proposed in \cite{ABJM} that the worldvolume of $N$ coincident M2-branes on ${\mathbb C}^4/{\mathbb Z}_k$ is described by the ${\cal N}=6$ supersymmetric Chern-Simons theory with gauge group $\text{U}(N)_k\times\text{U}(N)_{-k}$ and two pairs of bifundamental matters where the subscripts $(k,-k)$ denote the Chern-Simons levels.
The proposal is made based on a brane configuration in the IIB string theory.
The brane configuration consists of $N$ D3-branes on a compact circle $S^1$ with an NS5-brane and a $(1,k)$5-brane located at different positions on $S^1$, perpendicular to the D3-branes and tilted relatively by an angle determined by $k$.
After taking T-duality and lifting to M-theory, we recover the multiple M2-brane system. 
It was then desirable to reproduce the $N^{\frac{3}{2}}$ behavior of the degrees of freedom and figure out the corrections from the worldvolume theory

Soon after, the partition function $Z_k(N)$ on $S^3$ was computed.
By applying the localization technique for supersymmetric theories \cite{P}, the partition function, originally defined by the infinite-dimensional path integral, reduces to a finite-dimensional multiple integral, called the ABJM matrix model \cite{KWY}.
The 't Hooft large $N$ expansion of the matrix model was studied and the degrees of freedom $N^{\frac{3}{2}}$ were reproduced \cite{DMP1} from the free energy $\log Z_k(N)\simeq C^{-\frac{1}{2}}N^{\frac{3}{2}}$ with $C=2/(\pi^2k)$.
Besides, the perturbative corrections were studied \cite{DMP2} and it was found that all of the perturbative corrections are summed up to the Airy function $Z_k(N)\simeq\Ai(C^{-\frac{1}{3}}N)$ \cite{FHM}.

The resulting Airy function was suggestive.
Since its integral representation by a simple cubic monomial
\begin{align}
\Ai(N)=\int\frac{d\mu}{2\pi i}e^{\frac{1}{3}\mu^3-N\mu},
\end{align}
is reminiscent of the inverse transformation from the grand canonical ensemble
\begin{align}
Z_k(N)=\int\frac{d\mu}{2\pi i}e^{J_k(\mu)-N\mu},
\end{align}
the result immediately suggests us to move our analysis to the reduced grand potential $J_k(\mu)\simeq C\mu^3/3$ \cite{MP}.
Here we define the grand partition function $\Xi_k(z)$ and the reduced grand potential $J_k(\mu)$ by
\begin{align}
\Xi_k(z)=\sum_{N=0}^\infty z^NZ_k(N),\quad
\Xi_k(e^\mu)=\sum_{n=-\infty}^\infty e^{J_k(\mu+2\pi in)}.
\label{reducedgp}
\end{align}
Usually, the grand potential $\widetilde J_k(\mu)$ is defined from the grand partition function $\Xi_k(e^\mu)$ by $\Xi_k(e^\mu)=e^{\widetilde J_k(\mu)}$, which enjoys a trivial $2\pi i$ shift symmetry in $\mu$.
Unlike the usual one, the reduced grand potential is defined so that the trivial $2\pi i$ shift symmetry is taken care of from the beginning.
To avoid the ambiguity in choosing the reduced grand potential out of infinitely many candidates with $\mu$ shifted by $2\pi i$, we require that the quadratic term of $\mu$ in the reduced grand potential $J_k(\mu)$ is absent.
We need, however, to keep in mind to question whether such a candidate always exits or not.

There are two interesting aspects for the ABJM matrix model originating from \cite{MPtop}.
On one hand the grand partition function can be rewritten into the spectral determinant \cite{MP}
\begin{align}
\Xi_k(z)=\Det(1+z\H^{-1}),
\label{fredholmdet}
\end{align}
where the spectral operator $\H$ is given by
\begin{align}
\H=\widehat{\cal Q}\widehat{\cal P},\quad
\widehat{\cal Q}=\Q^{\frac{1}{2}}+\Q^{-\frac{1}{2}},\quad
\widehat{\cal P}=\P^{\frac{1}{2}}+\P^{-\frac{1}{2}},
\label{Habjm}
\end{align}
(with $\widehat{\cal Q}$ and $\widehat{\cal P}$ corresponding respectively to the $(1,k)$5-brane and the NS5-brane in the brane configuration) in terms of the exponentiated canonical operators
\begin{align}
(\Q,\P)=(e^{\widehat q},e^{\widehat p}),\quad[\widehat q,\widehat p]=i\hbar,
\label{QPexp}
\end{align}
under the identification $\hbar=2\pi k$.
After a similarity transformation and a rescaling transformation, the spectral operator $\H$ \eqref{Habjm} becomes a quantum analogue of the defining equation for the algebraic curve ${\mathbb P}^1\times{\mathbb P}^1$.
On the other hand, using the expression of the spectral determinant \eqref{fredholmdet}, we can perform a full non-perturbative analysis for the matrix model.
By combining various results from the 't Hooft expansion \cite{DMP1}, the WKB expansion \cite{MP,CM} and the analysis of exact values \cite{HMO1,PY}, we are able to detect worldsheet instantons \cite{DMP1}, membrane instantons \cite{DMP2,MP,CM,HMO2} and their bound states \cite{HMO3} for the non-perturbative effects.
Finally, it was found that, by choosing the K\"ahler parameters suitably, the full expression is given by the free energy of topological strings on local ${\mathbb P}^1\times{\mathbb P}^1$ \cite{HMMO}.
It is interesting to stress that in \cite{MP} the computation of the spectral determinant \eqref{fredholmdet} in the perturbation theory reduces directly to the evaluation of the phase space area for the corresponding Fermi gas system.

The exploration of the non-perturbative effects is especially interesting in the following aspects.
Although the coefficients of all the instanton effects are divergent at infinitely many values of $k$, after summing up all of them, the divergences cancel among themselves \cite{HMO2}.
This cancellation mechanism serves as a clue to clarify the whole structure of the instanton effects.
Among others, it was found that, by redefining the chemical potential $\mu$ into an effective one $\mu_\text{eff}$ \cite{HMO3}, all of the bound states are taken care of by the worldsheet instantons.
Thus the non-perturbative effects consist effectively only of the worldsheet instantons $J^\text{WS}_k(\mu_\text{eff})$ and the membrane instantons $J^\text{MB}_k(\mu_\text{eff})$.
This implies that the cancellation of the divergences from all of the instanton effects can be simplified as the cancellation purely between the worldsheet instantons and the membrane instantons.
Finally, all of these simplifications lead to the description of the non-perturbative effects by the free energy of topological strings \cite{HMMO}.

Fortunately, the structure is captured clearly from a geometrical viewpoint \cite{MiMo,ACDKV}.
For algebraic curves, there are two sets of cycles known as A-cycles and B-cycles.
The A-periods obtained from integration along the A-cycles give the redefinitions (known as mirror maps), while the B-periods combined with the mirror maps give the derivatives of the free energy.
After taking care of the mirror maps, the free energy is fully characterized by a set of non-negative integers known as BPS indices \cite{HKP}.
In contrast, the explicit expression for the mirror maps is missing.
Since the mirror maps serve half of the roles in understanding the description by topological strings, it is crucial to see whether and how the mirror maps also reduce to a set of non-negative integers.

Interestingly, by removing the role of the matrix model, a correspondence between spectral theories (ST) and topological strings (TS) was proposed \cite{GHM1}, which roughly states that the spectral determinant \eqref{fredholmdet} is given by the free energy of topological strings.
See \cite{KaMa,GHM1,WZH} for another related proposal on the spectrum.
Here, on the ST side, the spectral operator $\H$ in \eqref{fredholmdet} is the quantum analogue of the algebraic curve described by the exponentiated canonical operators $(\Q,\P)$ \eqref{QPexp}, while, on the TS side, the curve serves as the geometrical background.
The algebraic curves include those known as del Pezzo geometries \cite{HKRS,GKMR}, which are classified by Weyl group symmetries of exceptional algebras.
Since the defining equations of curves appear as the quantum spectral operators in the ST/TS correspondence, this notion is often referred to as quantum curves \cite{MiMo}.
Compared with the original discussions with matrix models, the relation to M2-branes may not be as clear.
Nevertheless, one advantage of the ST/TS correspondence on the del Pezzo geometries is the beautiful group structure in behind.
It is then desirable to take the group structure seriously and see how this helps in studying the correspondence.
In fact in \cite{MNY,KMN} the BPS indices for each combination of degrees and spins assemble into representations of the exceptional algebras.
It is then natural to ask, if the mirror maps are described by non-negative integers, whether these integers also assemble into representations.

To clarify the relation to M2-branes, spectral operators for some del Pezzo geometries were related to matrix models or brane configurations.
For example, partition functions of the ${\cal N}=4$ supersymmetric Chern-Simons theories with gauge group $\text{U}(N)_k\times\text{U}(N)_0\times\text{U}(N)_{-k}\times\text{U}(N)_0$ and $\text{U}(N)_k\times\text{U}(N)_{-k}\times\text{U}(N)_k\times\text{U}(N)_{-k}$ were associated to spectral operators $\H=\widehat{\cal Q}^2\widehat{\cal P}^2$ \cite{MN1,MN3} and $\H=\widehat{\cal Q}\widehat{\cal P}\widehat{\cal Q}\widehat{\cal P}$ \cite{HM} respectively.
These spectral operators fall into the $D_5$ quantum curves, which is consistent with the result that they are described by the free energy of topological strings on the $D_5$ local del Pezzo geometry \cite{MN3}.

It may seem at the first sight that the ST/TS correspondence is only relevant to M2-branes at certain parameters.
To fully relate to M2-branes, rank deformations for the matrix models were considered \cite{MM,HO,MNN}.
It was known in \cite{HLLLP2,ABJ} that, for the ABJM matrix model, we can introduce a relative rank difference for the gauge group as in $\text{U}(N_1)_k\times\text{U}(N_2)_{-k}$ preserving the same supersymmetry ${\cal N}=6$, where the rank difference is interpreted as the number of fractional branes in the brane configuration.
It was found that with rank deformations, the results are still given by the free energy of topological strings with different values of the K\"ahler parameters.
This implies that the rank deformations should also be taken care of by the coefficients of the quantum curve \cite{KMZ,closed,KM,K}.
These coefficients are often parameterized by asymptotic values of the curve known as point configurations \cite{Sakai,KNY}.
It is then interesting to ask how the space of brane configurations ${\cal C}_\text{B}$ is embedded into the space of point configurations ${\cal C}_\text{P}$.
This question was investigated in \cite{KM} and finally it was found that, for the cases of $\H=\Q^2\P^2$ and $\H=\Q\P\Q\P$ connected by rank deformations via the Hanany-Witten transitions \cite{HW}, the three-dimensional space ${\cal C}_\text{B}$ with three relative ranks is embedded into the five-dimensional space ${\cal C}_\text{P}$ for the $D_5$ curve.
Note that, in embedding ${\cal C}_\text{B}$ into ${\cal C}_\text{P}$, the parameters of the curve are identified as phase functions of the relative ranks \cite{KMN,KM}.
Due to this reason the phase space area can generally be imaginary, which would contradict with our perturbative assumption for the reduced grand potential $J_k(\mu)$.

Considering that the spectral determinants of all quantum curves share the cubic behavior $J_k(\mu)\simeq C\mu^3/3$ leading to the degrees of freedom $N^{\frac{3}{2}}$ and the Airy function, it is natural to regard ``the space of quantum curves'' ${\cal C}_\text{P}$ (instead of the subspace realized by rank deformations ${\cal C}_\text{B}$) as ``the moduli space of the backgrounds for M2-branes''.
Besides, as a bonus, instead of the original space of more physical brane configurations ${\cal C}_\text{B}$, the space of point configurations ${\cal C}_\text{P}$ enjoys the full-fledged Weyl group symmetry.
Thus, we are naturally led to the picture that {\it the partition functions of M2-branes should be reformulated by the ST/TS correspondence}.
Stated differently, the geometrical backgrounds for M2-branes may not be simple fractional branes, but can be more obscure objects.

From this viewpoint, in order to fully investigate the M2-branes by extending the background constructed from fractional branes into a more general geometrical background, it is crucial to study the ST/TS correspondence for all of the del Pezzo geometries.
Since the del Pezzo geometries are governed strongly by the Weyl group symmetries, we are especially interested in the quantum curves of higher ranks such as $E_6$, $E_7$ and $E_8$.
To state the ST/TS correspondence for these curves, there are several important questions remaining to be clarified.
On the ST side, as far as we know, even the corresponding quantum spectral operators $\H$ are not all written down explicitly.

On the TS side, we first note that currently even the perturbative coefficients are not given in an expression fully respecting the group structure.
Due to this reason, it was difficult to proceed to state the full conjecture for the ST/TS correspondence explicitly.
For the non-perturbative effects, as we mention earlier, after redefining the chemical potential using the mirror map, the results are given by the BPS indices.
Compared with the BPS indices, however, the progress in the mirror map is only made gradually.
It is especially unclear (1) whether the mirror maps are characterized by non-negative integers, (2) if yes, whether these integers are interpreted as multiplicities of representations, (3) how these representations are related to those for the BPS indices and (4) how the multiplicities for different curves relate to one another.
The mirror map for the $D_5$ quantum curve was studied in \cite{FMS} and it was found that the mirror map is given by multiplicities of the $D_5$ representations as in the case of the BPS indices.
It was also observed that the representations appearing in the mirror map are the same as those appearing in the BPS indices (except for the trivial case of degree one).
Note however that, in the analysis of \cite{FMS}, the overall factor $\alpha$ of the spectral operator $\H$ has to be identified with the combination of parameters of the curve which transforms identically as $\alpha$.
This apparently needs clarifications.
In this paper, we extend the analysis to the curves of higher ranks to further investigate the mirror maps.

In the introduction we have raised three questions:
\begin{itemize}
\item
The reduced grand potential in \eqref{reducedgp} is defined by requiring that the quadratic term of $\mu$ is absent.
It is unclear whether such a candidate always exists or not.
\item
Since the coefficients of the curve are identified as phase functions of the relative ranks, the phase space area can generally be imaginary.
\item
In expressing the mirror map in terms of group characters, we need to identify the overall factor $\alpha$ of the spectral operator $\H$ as the combination of parameters of the curve transforming identically as $\alpha$.
\end{itemize}
We shall see in section \ref{weyl} that all of these questions can be understood from the group-theoretical viewpoint.

This paper is organized as follows.
In the next section, we first summarize our main results of the paper.
Namely, we present the ST/TS correspondence on the del Pezzo geometries explicitly and explain the new insights we try to add to this correspondence from the group-theoretical viewpoint.
To accomplish the proposal, since the remaining ingredients mainly concern quantizations of the curves, most of the current paper is devoted to the analysis of the quantum curves.
After shortly recapitulating the $D_5$ quantum curve in section \ref{d5}, we proceed to the analysis of the quantum curves of $E_6$, $E_7$ and $E_8$ (for identifying the quantum curves, studying the Weyl groups and analyzing the mirror maps) in the subsequent three sections.
With all of the results we make an observation of the decoupling relation in section \ref{decoupling}.
Finally we conclude by listing some interesting future directions in section \ref{conclusion}.
Two appendices are devoted respectively to the classical analysis of the phase space areas for the corresponding Fermi gas systems and a summary of the BPS indices.

\section{Summaries: ST/TS correspondence}\label{summary}

As explained in the introduction, the study of the partition functions of M2-branes can be reformulated as the correspondence between spectral theories and topological strings.
Following the analysis of the ABJM model \cite{HMMO}, by removing the role of matrix models, it was proposed in \cite{GHM1} that spectral theories and topological strings correspond to each other.
Especially, it was proposed that, after removing $\Xi_k(z)$ from \eqref{reducedgp} and \eqref{fredholmdet}, the spectral determinant of a spectral operator $\H$ associated to a geometrical background is given by the reduced grand potential $J_k(\mu)$,
\begin{align}
\Det(1+z\H^{-1})=\sum_{n=-\infty}^\infty e^{J_k(\mu+2\pi in)},\quad (z=e^\mu),
\label{sttseq}
\end{align}
which is described by the free energy of topological strings on the corresponding background.

In this paper we study the case when the backgrounds are the del Pezzo geometries by fully utilizing their Weyl group structures.
We stress that, although much progress was made so far after the proposal, we believe that we still have some new insights to add to this correspondence from the group-theoretical viewpoint.
Especially, we can present explicitly the statement of the conjecture for the correspondence between spectral theories and topological strings on the del Pezzo geometries.
We shall summarize our main results in the following respectively from the aspects of spectral theories, topological strings and Weyl groups.

\subsection{Spectral theories}

Let us start from the left-hand side of \eqref{sttseq} for spectral theories.
Although the quantum curves (or the spectral operators $\H$) of $D_5$ and $E_7$ were given in \cite{KMN}, to the best of our knowledge, the quantum curves of $E_6$ and $E_8$ have not been written down explicitly yet.
Especially, it is notoriously difficult for the $E_8$ quantum curves.
Here we are able to provide an explicit expression for the quantum curves of $E_6$ and $E_8$.
The spectral operator $\H$ for the $E_6$ curve is given in \eqref{e6triP} or \eqref{e6rectP}, that for the $E_7$ curve in \eqref{e7tricurve} or \eqref{e7rectcurve} and that for the $E_8$ curve in \eqref{e8trieq} or \eqref{e8recteq}.
Note that, although the classical limit of this subject plays an important role in understanding the BPS state counting for the E-string theory \cite{KMV,MNVW,ES} and the Weierstrass normal forms of the curves were given previously, here we pursue toric-like expressions which are more suitable for our purpose of quantizations.
The main difficulties in the construction and the resolutions are as follows.

Firstly, it was known that the del Pezzo geometry of the largest rank which can be described by the toric diagram is the $E_6$ curve and the curves of $E_7$ and $E_8$ are not toric.
Due to this reason, it was difficult to write down all the curves even classically.
Recently this situation was improved and nice references for physicists can be found in \cite{BBT,KY}.
Namely, by allowing degeneracies of the curves, we can describe the curves of $E_7$ and $E_8$ as well.

Secondly, even though the classical curve enjoys the Weyl group symmetry of the exceptional algebra, it is not clear how to lift the algebraic curve and the Weyl group action quantum-mechanically.
For the curves of $D_5$ and $E_7$, the Weyl group actions were studied in \cite{KMN} and it was found that the factor $q=e^{i\hbar}$ needs to be incorporated for the Weyl actions to be symmetries.
However, simply from \cite{KMN} it is difficult to understand the general rule to incorporate the $q$ factor.
In this paper we follow the prescription of the Weyl ordering to incorporate the $q$ factor systematically.

Thirdly, even with the standard Weyl order, it is still unclear how to take care of the degenerate curves $E_7$ and $E_8$.
For clarification, we first rewrite the expression for the $E_6$ curve without degeneracies into that with degeneracies using the similarity transformation in order to learn how to handle the degeneracies quantum-mechanically.
After the rewriting, we find that we can incorporate the $q$ factor for the degenerate curve by replacing the integral coefficients coming from the degeneracies into the so-called $q$-integers.
Stated alternatively, instead of the multiple roots appearing in the defining equation for the classical degenerate curve, we encounter roots with step-by-step increasing/decreasing powers of $q$ for the quantum curve.
This was partially observed in \cite{KMN}, though we can present the analysis much more systematically.

Fourthly, technically it is difficult to treat the large number of variables for the curves of higher ranks especially when they are degenerate.
Here we introduce a coloring for the toric diagrams (see figure \ref{e6tri} and figure \ref{e6rect} for the $E_6$ curve, figure \ref{e7tri} and figure \ref{e7rect} for the $E_7$ curve and figure \ref{e8tri} and figure \ref{e8rect} for the $E_8$ curve), which can visualize the process of fixing the coefficients.
Namely the color of an inner dot (located in the bulk) of the toric diagram indicates the independent asymptotic value responsible for determining the coefficient of the dot.
With this clarification, we are able to write down all of the quantum curves explicitly.

Starting by recapitulating the $D_5$ curve in section \ref{d5}, in the subsequent sections (section \ref{e6qc}, section \ref{e7qc} and section \ref{e8qc}) we follow these strategies and subsequently generalize our study to the curves of $E_6$, $E_7$ and $E_8$ to find out the quantum curves explicitly.
For each curve we observe some lessons helpful for studying the curves of higher ranks.
For example, in the study of the $E_6$ curve we learn how to handle degeneracies in curves and how to apply these techniques to the curves of higher ranks.
Also, in the study of the curves of $E_6$ and $E_7$, we learn the process of determining the coefficients of the degenerate curves by coloring the toric diagrams, which is applicable to the $E_8$ curve.

\subsection{Topological strings}

Now let us turn to the right-hand side of \eqref{sttseq} for topological strings.
We present the explicit expressions of the reduced grand potential $J_k(\mu)$ on the del Pezzo geometries, by conjecturing the perturbative coefficients, redefining the chemical potential with the mirror map and presenting the non-perturbative effects.
The expressions in this subsection are conjectured by consulting various previous computations and should be proved or disproved in the future studies.
We stress that it is only after we clarify the group structure that we can present the exact statement of the conjecture for the correspondence explicitly.

\subsubsection{Perturbations}\label{perturbations}

We first concentrate on the perturbative analysis.
It was proposed that, out of infinitely many candidates, the reduced grand potential $J_k(\mu)$ is defined to be that without quadratic terms of $\mu$.
It is difficult, however, to argue its existence in general.
After fixing the expression for the curves, we can compute the perturbative expression explicitly at least in the classical limit.

Since the reduced grand potential $J_k(\mu)$ coincides with the usual grand potential $\widetilde J_k(\mu)$ perturbatively if the quadratic term of $\mu$ is missing, let us restrict ourselves to $\widetilde J_k(\mu)$ for now.
As in \cite{MP}, the derivative of the grand potential $\widetilde J_k(\mu)$ with respect to the chemical potential $\mu=\log z$ reduces to
\begin{align}
\frac{\partial\widetilde J_k(\mu)}{\partial\mu}=\frac{\partial}{\partial\mu}\log\Det(1+z\H^{-1})=\Tr\frac{1}{1+\H/z}=\frac{\Area(|H|<z)}{2\pi\hbar},
\label{derivative}
\end{align}
since $(1+H/z)^{-1}$ is essentially $1$ or $0$ classically for the canonical variables $(q,p)$ satisfying $|H|\ll z$ or $|H|\gg z$ respectively.
Using \eqref{derivative}, we can compute the grand potential $\widetilde J_k(\mu)$ from the phase space area surrounded by the curve at least classically and perturbatively.
In the canonical variables $(q,p)$ instead of their exponentiations $(Q,P)$, the computation of the phase space area in the logarithmic scale amounts to considering the so-called tropical geometry.
The computations are provided in appendix \ref{area} and the results are given in \eqref{d5area}, \eqref{e6area}, \eqref{e7area} and \eqref{e8area} for the curves of $D_5$, $E_6$, $E_7$ and $E_8$ respectively.

As we have stressed in the introduction, since the parameters $f_i=e^{i\widetilde f_i}$, $g_i=e^{i\widetilde g_i}$, $h_i=e^{i\widetilde h_i}$ of the spectral operator $\H$ are identified as phase functions of the rank differences \cite{KM}, in general the phase space area can be complex.
Indeed, the linear terms of the chemical potential $\mu$ are pure imaginary and linear also in the rank differences $(\widetilde f_i,\widetilde g_i,\widetilde h_i)$, which would lead to the grand potential $\widetilde J_k(\mu)$ with imaginary quadratic terms of $\mu$.
It is only after we redefine the chemical potential $\mu$ by shifts as in \eqref{d5muredef}, \eqref{e6muredef}, \eqref{e7muredef} and \eqref{e8muredef} respectively for the curves of $D_5$ $E_6$, $E_7$ and $E_8$ (see \cite{AHK,BPTY,MPTY}) that the classical area becomes real, free of linear terms and given by group invariants,
\begin{align}
\frac{\partial\widetilde J_k(\mu)}{\partial\mu}\bigg|^\text{cl}=C^\text{cl}\mu^2+B^\text{cl},
\end{align}
where the coefficients are given by
\begin{align}
C^\text{cl}=\frac{9-n}{4\pi\hbar},\quad
B^\text{cl}=\frac{\Omega(E_n)}{4\pi\hbar},
\label{classicalCB}
\end{align}
for the $E_n$ curve with $\Omega(E_n)$ being the quadratic Casimir element of the exceptional algebra $E_n$.
Then, after the integration we obtain the general perturbative expression of the reduced grand potential
\begin{align}
\widetilde J_k^\text{p}(\mu)=\frac{C}{3}\mu^3+B\mu+A,
\end{align}
without quadratic terms of $\mu$.
We will return to the interpretation of the above shift of the chemical potential $\mu$ in section \ref{weyl}.

In this paper we conjecture that, in full quantum corrections, the perturbative coefficients $C$ and $B$ consist of a classical part which is determined from the above classical analysis of the phase space area and an invariant part which is common for the geometry and does not depend on parameters of the curve.
More explicitly, we conjecture that the coefficients are given by
\begin{align}
C=\frac{9-n}{8\pi^2k},\quad
B=\frac{\Omega(E_n)}{8\pi^2k}+\frac{(3-n)}{12k}-\frac{(3+n)k}{24},
\label{CB}
\end{align}
for the $E_n$ curve (by substituting $\hbar=2\pi k$).
The general expression for $A$ is not known yet.

This conjecture matches well with many computations obtained so far.
For the ${\mathbb P}^2$ curve (or $E_0$) without parameter deformations, the coefficients $C=9/(8\pi^2k)$ and $B=1/(4k)-k/8$ were obtained from the WKB analysis (see (4.6) in \cite{GHM1}).
For the ABJM case (or $E_1$) with the spectral operator $\H=(\Q^{\frac{1}{2}}+\Q^{-\frac{1}{2}})(\P^{\frac{1}{2}}+\P^{-\frac{1}{2}})$ subject to relative rank deformations by $M$, the coefficients $C=2/(\pi^2k)$ and $B=(M-k/2)^2/(2k)+1/(3k)-k/12$ were found (see (1.3) in \cite{MM}).
If we redefine $\widehat q'=(\widehat q-\widehat p)/2$ and $\widehat p'=(\widehat q+\widehat p)/2$ as the usual notation for ${\mathbb P}^1\times{\mathbb P}^1$, the Planck constant is rescaled by $k'=k/2$, which reproduces correctly $C=1/(\pi^2k')$ and $B=\Omega/(8\pi^2k')+1/(6k')-k'/6$.
For the $D_5$ curve (or $E_5$), the coefficients $C=1/(2\pi^2k)$ and $B=\Omega/(8\pi^2k)-1/(6k)-k/3$ were already proposed in \cite{KM} based on the analysis in \cite{MN1,MN3,MNN}.

Note that in \cite{GKMR} other examples of the del Pezzo geometries were examined.
It was found that, for some examples, the quadratic terms of $\mu$ are present, which may contradict with our conjecture \eqref{CB} at the first sight.\footnote{We are grateful to Marcos Marino and Futoshi Yagi for valuable discussions.}
Our proposal for the resolution is to introduce overall factors for the defining equations of the curves from the beginning to remove these quadratic terms.
We will come back to this proposal in section \ref{weyl}.
At this stage we simply shift the chemical potentials suitably as in \eqref{d5muredef}, \eqref{e6muredef}, \eqref{e7muredef} and \eqref{e8muredef} to remove the quadratic terms.
Then, (4.71) in \cite{GKMR} for the $F_1$ curve (or $\widetilde E_1$) gives $C=1/(\pi^2k)$ and $B=\Omega/(8\pi^2k)+1/(6k)-k/6$, while (4.86) in \cite{GKMR} for the $B_2$ curve (or $E_2$) gives $C=7/(8\pi^2k)$ and $B=\Omega/(8\pi^2k)+1/(12k)-5k/12$.
All of these results match correctly with our conjecture \eqref{CB}.
Note that, since overall factors of the quadratic Casimir elements depend on the definition, the match of the first term $\Omega/(8\pi^2k)$ in $B$ is not very meaningful for these curves of lower ranks.

\subsubsection{Mirror map}

For the non-perturbative part, most of the general structures were already proposed previously.
After redefining the chemical potential with the mirror map, the reduced grand potential $J_k(\mu)$ is expressed by the BPS indices known in \cite{HMMO,HKP,MNY,KM}.
After obtaining the quantum curves of higher ranks, here we can proceed to study the mirror maps to fill in the missing part of the conjectural correspondence.
In the following sections, we check the proposed structures of the mirror maps for these curves, find the concrete expressions (section \ref{e6mirror}, section \ref{e7mirror} and section \ref{e8mirror}) and observe a relation among them (section \ref{decoupling}).

For the non-perturbative effects, first we need to redefine the chemical potential $\mu$ into the effective one $\mu_\text{eff}$ by the mirror map.
The mirror map we need here for the quantum curves (known as the quantum mirror map) is obtained by evaluating the quantum A-period, which is defined as the residue of the symplectic form constructed from the wave function for the corresponding Schr\"odinger equation \cite{ACDKV}.
The inverse transformation of the quantum mirror map is given by \cite{HMMO,FMS}
\begin{align}
\mu=\mu_\text{eff}+\sum_{\ell=1}^\infty(-1)^\ell\E_\ell e^{-\ell\mu_\text{eff}}
+\log\Bigl(1-Ee^{-\mu_\text{eff}-\sum_{\ell=1}^\infty(-1)^\ell\E_\ell e^{-\ell\mu_\text{eff}}}\Bigr),
\label{Eell}
\end{align}
where $E$ is the constant term of the spectral operator $\H$ while $\E_\ell$ is a function given in terms of the factor $q=e^{i\hbar}=e^{2\pi ik}$ and parameters of the curve including the overall factor $\alpha$.
In \eqref{Eell} (and \eqref{wsmb} below) we display the relation for the case when the exponent $r$ introduced in \cite{GHM1,GKMR} is trivial, $r=1$.\footnote{We are grateful to Yasuyuki Hatsuda and Hiroaki Kanno for valuable discussions.}
For non-trivial cases, we need to rescale all of the chemical potentials by $r$.
To express the mirror map by group characters, we need to construct characters and identify the arguments of characters with parameters of the curve.
Besides, remarkably it was found \cite{FMS} that we need to identify the overall factor $\alpha$ of the spectral operator $\H$ as the combination of parameters of the curve which transforms identically as $\alpha$ (to which we will return in section \ref{weyl}).

The quantum mirror map for the $D_5$ curve was investigated in \cite{FMS} by evaluating the quantum A-period explicitly and it was found that the results of $\E_\ell$ are summarized by group characters if the overall factor $\alpha$ of the spectral operator $\H$ is identified as in \eqref{d5alpha}.
For the $D_5$ curve a multi-covering structure for $\E_\ell$
\begin{align}
\E_\ell=\sum_{n|\ell}\frac{(-1)^{n+1}\epsilon_{\frac{\ell}{n}}(q^n,{\bm q}^n)}{n},
\label{Emc}
\end{align}
with $\epsilon_d(q,{\bm q})$ being the multi-covering component was proposed \cite{HMMO,FMS}, where ${\bm q}$ are the arguments of characters and identified with parameters of the curve.
Here in the multi-covering structure, $\E_\ell$ consists not only of the contribution $\epsilon_\ell(q,{\bm q})$ of degree $\ell$ but also those from lower degrees of divisors.

From the results obtained from the explicit analysis in \cite{FMS} for the $D_5$ curve, we expect that the multi-covering components are expressed as
\begin{align}
\epsilon_d(q,{\bm q})=\sum_{j,{\bf R}}m_j^{d,{\bf R}}(q^j+q^{-j})\chi_{\bf R}({\bm q}),
\label{emulticover}
\end{align}
which is given in terms of the group characters $\chi_{\bf R}({\bm q})$ with non-negative integers of multiplicities $m_j^{d,{\bf R}}$ of the representation ${\bf R}$ depending on degrees $d$ and spins $j$.
Note that $j$ should be interpreted as a suitable modification of spins, since the factor $q^j+q^{-j}$ is not the $\text{SU}(2)$ character itself, 
\begin{align}
\chi_j(q)=\frac{q^{j+\frac{1}{2}}-q^{-j-\frac{1}{2}}}{q^{\frac{1}{2}}-q^{-\frac{1}{2}}},
\label{su2}
\end{align}
but $q^j+q^{-j}=\chi_j(q)-\chi_{j-1}(q)$.
In \cite{FMS} it was found that the representations appearing in the mirror map match with those appearing in the BPS indices except for the trivial case of degree one.

After obtaining the quantum curves of $E_n$ $(n=6,7,8)$ in section \ref{e6qc}, section \ref{e7qc} and section \ref{e8qc}, we can embark on our analysis of the mirror maps for these curves by evaluating the quantum A-periods similarly.
Following the same steps, we reconfirm the group structure and the multi-covering structure proposed for the $D_5$ curve and work out the concrete expressions, which are given in table \ref{e6mirrormap}, table \ref{e7mirrormap} and table \ref{e8mirrormap} respectively for the curves of $E_n$ $(n=6,7,8)$.
Our strategies and the results are summarized as follows.

We find that the toric diagram in the so-called triangular realization as in figure \ref{e6tri}, figure \ref{e7tri} and figure \ref{e8tri} (respecting especially the Weyl subgroup $A_2\times A_1\times A_{n-4}$) seems to lead to the simplest Schr\"odinger equation for the mirror map.
For the $E_8$ curve we propose to introduce a slightly different-looking equation to study the mirror map as discussed at the end of section \ref{e8mirror}.

To express the results in terms of characters, typically we need characters of large representations for exceptional algebras, which can be very lengthy to generate.
Here for simplicity, instead of generating characters by ourselves, we construct characters from the decompositions of representations into subgroups listed in \cite{Y}.
However, since the representations are restricted in the list, our analysis is also restricted.
Here the arguments of characters are determined by parameters of the curves and given explicitly in \eqref{e6t}, \eqref{e7t} and \eqref{e8t} for the curves of $E_n$ $(n=6,7,8)$ respectively.
Importantly, we encounter again similar identifications of the overall factors $\alpha$ in \eqref{e6alpha}, \eqref{e7alpha} and \eqref{e8alpha} for the curves of $E_n$ $(n=6,7,8)$.
The determinations of the arguments of characters and the identifications of the overall factors $\alpha$ are obtained by studying the Weyl group actions in section \ref{e6wg}, section \ref{e7wg} and section \ref{e8wg}.

Note that the proposed multi-covering structure \eqref{Emc} and \eqref{emulticover} is not guaranteed for all the del Pezzo geometries a priori.
After computing the quantum mirror maps for the $E_n$ $(n=6,7,8)$ curves, we can proceed to confirm the validity of \eqref{Emc} and \eqref{emulticover}.
In fact, after adopting the same multi-covering structure \eqref{Emc} as the $D_5$ curve, we find that all the coefficients of characters $m_j^{d,{\bf R}}$ in \eqref{emulticover} are non-negative integers.
The explicit values of multiplicities for the mirror maps $m_j^{d,{\bf R}}$ for the $E_n$ $(n=6,7,8)$ curves are given respectively in table \ref{e6mirrormap}, table \ref{e7mirrormap} and table \ref{e8mirrormap}.
We continue to find that, for the $E_7$ case, by comparing table \ref{e7mirrormap} for the mirror map with the table for the BPS indices in \cite{MNY}, the representations appearing in both tables match exactly with each other except for the trivial case of degree one.

Furthermore, we observe that the mirror maps for the $E_n$ $(n=6,7,8)$ curves enjoy the decoupling relations in section \ref{decoupling}.
Namely starting with the mirror map for the $E_8$ curve of the highest rank, by decomposing the representations into $E_7\times A_1$ and choosing representations of the $\text{U}(1)_\text{diag}(\subset A_1)$ charge agreeing with the degree, we can reproduce the mirror map for the $E_7$ curve.
We can check this property by lowering the ranks down to the $D_5$ curve.
This relation is expected since curves of lower ranks are obtained by taking the limit for curves of higher ranks and is explicitly known for the BPS indices.

\subsubsection{Non-perturbative effects}

After the redefinition of the chemical potential with the quantum mirror map, the non-perturbative effects reduce to the well-established structure given in \cite{HMMO,MNY,KM}.
We do not have anything to add here and we repeat the explanation simply for completeness.

The reduced grand potential splits into the perturbative part, the worldsheet instanton part and the membrane instanton part as
\begin{align}
J_k(\mu)=J^\text{pert}_k(\mu_\text{eff})+J^\text{WS}_k(\mu_\text{eff})+J^\text{MB}_k(\mu_\text{eff}),
\end{align}
which are given explicitly by \cite{HMO2,HMO3,HMMO}
\begin{align}
J^\text{WS}_k(\mu_\text{eff})=\sum_{m=1}^\infty d_m(k,{\bm b})e^{-m\frac{\mu_\text{eff}}{k}},\quad
J^\text{MB}_k(\mu_\text{eff})=\sum_{\ell=1}^\infty(\widetilde b_\ell(k,{\bm b})\mu_\text{eff}+\widetilde c_\ell(k,{\bm b}))e^{-\ell\mu_\text{eff}},
\label{wsmb}
\end{align}
with ${\bm b}$ identified with the arguments of characters by ${\bm q}=e^{2\pi ik{\bm b}}$.
Here all of the instanton coefficients $d_m$, $\widetilde b_\ell$ and $\widetilde c_\ell$ are expressed with the multi-covering structure by
\cite{MNY,KM}
\begin{align}
d_m(k,{\bm b})=(-1)^m\sum_{n|m}\frac{\delta_{\frac{m}{n}}(k/n,n{\bm b})}{n},\quad
\widetilde b_\ell(k,{\bm b})=\sum_{n|\ell}\frac{\beta_{\frac{\ell}{n}}(nk,{\bm b})}{n},\quad
\widetilde c_\ell(k,{\bm b})=-k^2\frac{\partial}{\partial k}\frac{\widetilde b_\ell(k,{\bm b})}{\ell k},
\end{align}
and the multi-covering components are expressed by characters as in
\begin{align}
\delta_d(k,{\bm b})&=\frac{(-1)^{d-1}}{(2\sin\frac{\pi}{k})^2}\sum_{j_\text{L},j_\text{R}}\sum_{\bf R}
n^{d,{\bf R}}_{j_\text{L},j_\text{R}}\chi_{\bf R}(e^{2\pi i{\bm b}})\chi_{j_\text{L}}(e^{\frac{4\pi i}{k}})\chi_{j_\text{R}}(1),\nonumber\\
\beta_d(k,{\bm b})&=\frac{(-1)^dd}{4\pi\sin\pi k}\sum_{j_\text{L},j_\text{R}}\sum_{\bf R}
n^{d,{\bf R}}_{j_\text{L},j_\text{R}}\chi_{\bf R}(e^{2\pi ik{\bm b}})\chi_{j_\text{L}}(e^{2\pi ik})\chi_{j_\text{R}}(e^{2\pi ik}),
\label{deltabeta}
\end{align}
with $n^{d,{\bf R}}_{j_\text{L},j_\text{R}}$ being multiplicities of the representation ${\bf R}$ for each degree $d$ and each spin $(j_\text{L},j_\text{R})$.
The $\text{SU}(2)$ character $\chi_j(q)$ is given by \eqref{su2} and the arguments for the membrane instantons $\beta_d(k,{\bm b})$ match with those for the mirror map, $q=e^{2\pi ik}$, ${\bm q}=e^{2\pi ik{\bm b}}$.
To imitate the expression for the non-perturbative effects, we may want to rewrite the multi-covering structure for the mirror map \eqref{Emc}, \eqref{emulticover} as
\begin{align}
\E_\ell=\sum_{n|\ell}\frac{(-1)^{n+1}\epsilon_{\frac{\ell}{n}}(nk,{\bm b})}{n},\quad
\epsilon_d(k,{\bm b})=\sum_{j,{\bf R}}m^{d,{\bf R}}_j(e^{2\pi ikj}+e^{-2\pi ikj})\chi_{\bf R}(e^{2\pi ik{\bm b}}),
\end{align}
by changing the arguments of the multi-covering components from $\epsilon_d(q,{\bm q})$ to $\epsilon_d(k,{\bm b})$.
The multiplicities $n^{d,{\bf R}}_{j_\text{L},j_\text{R}}$ for the curves of $D_5$ and $E_7$ can be found in \cite{MNY} and for completeness we also list the first few multiplicities $n^{d,{\bf R}}_{j_\text{L},j_\text{R}}$ for the curves of $E_6$ and $E_8$ in appendix \ref{bpslist}.

Let us stress that in these expressions for the non-perturbative effects we have fully utilized the group structure.
Compared with the original proposal in \cite{HMMO,GHM1} or later computations in \cite{MNN} where the choice of the K\"ahler parameters and the split of BPS indices can be ambiguous, we stress that these expressions fully respecting the group structure are much explicit and unambiguous \cite{MNY}.
Namely, in the description of \cite{HMMO} we need to choose the K\"ahler parameters first and determine the BPS indices for the combination of the K\"ahler parameters.
As pointed out in \cite{MNN} the choice of the K\"ahler parameters can be ambiguous, so are the BPS indices.
This ambiguity may be resolved by complexifying the curve \cite{GM,EMR,Hu}.
Instead, after we fully utilize the group structure in the description of the spectral determinant by the free energy of topological strings, the ambiguity in the description does not exist any more and the description boils down to two sets of multiplicities, $m^{d,{\bf R}}_j$ for the mirror maps and $n^{d,{\bf R}}_{j_\text{L},j_\text{R}}$ for the BPS indices.

Before our work, without the group structure clarified, it was difficult to present the exact statement of the conjecture for the correspondence explicitly.
It is only after we exploit the group structure that we can write down the statement of the conjecture explicitly, where for spectral theories we have the quantum curves explicitly, while for topological strings we conjecture that the reduced grand potential $J_k(\mu)$ is given explicitly by \eqref{CB}, \eqref{emulticover} and \eqref{deltabeta} and reduce the conjecture to two sets of multiplicities, $m^{d,{\bf R}}_j$ and $n^{d,{\bf R}}_{j_\text{L},j_\text{R}}$.

\subsection{Weyl groups}\label{weyl}

So far we have stressed that the group structure is helpful in our analysis.
Now let us proceed to explain that the group structure, at the same time, clarifies some of our previous assumptions adopted in the analysis.
Especially, we shall see the three questions we have raised in the introduction are clarified from the group structure.\footnote{We are grateful to Yasuhiko Yamada for valuable discussions.}
Let us choose the $D_5$ curve to explain our understandings, though the same explanation works for all of the $E_n$ $(n=6,7,8)$ curves as well.

In the perturbative analysis of topological strings, the phase space area surrounded by the algebraic curve generally contains imaginary linear terms of $\mu$, which results in imaginary quadratic terms in the reduced grand potential.
As a consequence the perturbative expression is not given in group invariants.
It is only after we shift the chemical potential by $\mu=\mu'+(i/2)(\widetilde q_1-\widetilde q_2-\widetilde q_3-\widetilde q_4+\widetilde q_5)$ \eqref{d5muredef} that we can remove the imaginary linear terms.
In the non-perturbative analysis, to express the computational result in terms of group characters in \cite{FMS}, we have identified the overall factor $\alpha$ of the spectral operator $\H$ with the combination of parameters of the curve which transforms identically as $\alpha$, $\alpha=h_2^{\frac{1}{2}}e_1^{\frac{1}{4}}e_3^{-\frac{1}{2}}e_5^{-\frac{1}{4}}$ \eqref{d5alpha}.

To explain these processes from the group-theoretical viewpoint, let us first focus on the spectral determinant $\Det(1+z\H^{-1})$ in \eqref{sttseq}.
It is clear that the spectral determinant is invariant under simultaneous rescalings of the spectral operator $\H$ and the fugacity $z$.
It is then desirable to determine the overall factor so that both $\H$ and $z$ are invariant separately under the Weyl group transformation.
The overall factor $\alpha$ of the spectral operator $\H$, however, is defined simply from the top term of the exponentiated canonical operators $(\Q,\P)$, $\H/\alpha=q^{-\frac{1}{2}}\Q\P+\cdots$ \eqref{d5curve}, and there are no reasons to expect the coefficient of the top term to transform trivially.
Indeed, from the explicit analysis in \cite{FMS}, $\alpha$ transforms under the Weyl group action identically as the combination $h_2^{\frac{1}{2}}e_1^{\frac{1}{4}}e_3^{-\frac{1}{2}}e_5^{-\frac{1}{4}}$.
To obtain the spectral operator $\H$ with the overall factor $\alpha$ transforming trivially, it is the easiest to rescale $\H$ and $z$ simultaneously by this combination
\begin{align}
\H=(h_2^{\frac{1}{2}}e_1^{\frac{1}{4}}e_3^{-\frac{1}{2}}e_5^{-\frac{1}{4}})\H_\bullet,\quad
z=(h_2^{\frac{1}{2}}e_1^{\frac{1}{4}}e_3^{-\frac{1}{2}}e_5^{-\frac{1}{4}})z_\bullet,
\end{align}
where the subscript $\bullet$ denotes the invariance under the Weyl group actions.
We can regard the rescaling of the spectral operator $\H$ as the rescaling of its overall factor $\alpha$,
\begin{align}
\alpha=(h_2^{\frac{1}{2}}e_1^{\frac{1}{4}}e_3^{-\frac{1}{2}}e_5^{-\frac{1}{4}})\alpha_\bullet.
\end{align}
This is essentially what we have done in identifying $\alpha$ as the combination transforming identically as $\alpha$.
Namely, after factoring out $h_2^{\frac{1}{2}}e_1^{\frac{1}{4}}e_3^{-\frac{1}{2}}e_5^{-\frac{1}{4}}$, the remaining factor $\alpha_\bullet$ transforms trivially and serves only the same role of counting degrees as $z$.
Hence, we can easily absorb it in $z$ or simply set $\alpha_\bullet=1$.

Due to the above identification of $\alpha$, the variable $z'=z/\alpha$ introduced for \eqref{8points} reduces to
\begin{align}
z'=\frac{z}{\alpha}=(h_2^{\frac{1}{2}}e_1^{\frac{1}{4}}e_3^{-\frac{1}{2}}e_5^{-\frac{1}{4}})^{-1}z=\biggl(\frac{q_2q_3q_4}{q_1q_5}\biggr)^{\frac{1}{2}}z,
\end{align}
with the relation \eqref{h12e135} substituted.
This effectively shifts the chemical potential by
\begin{align}
\mu=\mu'+\frac{i}{2}(\widetilde q_1-\widetilde q_2-\widetilde q_3-\widetilde q_4+\widetilde q_5),
\end{align}
if we substitute $z=e^\mu$, $z'=e^{\mu'}$ and $q_i=e^{i\widetilde q_i}$.
This is nothing but the shift of the chemical potential \eqref{d5muredef} we encounter to remove the imaginary linear terms in evaluating the phase space area surrounded by the curve.

As a result, the classical phase space area is given by a quadratic polynomial in $\mu$ without linear terms, where the constant is a group invariant, the quadratic Casimir operator $\Omega(D_5)$ \eqref{d5area}.
After the integration, we obtain the reduced grand potential $J_k(\mu)$ with the perturbative contribution being a cubic polynomial in $\mu$ without quadratic terms.
For this reason, the absence of quadratic terms of $\mu$ in the reduced grand potential $J_k(\mu)$ is classically understood by the fact that the lowest group invariant is the quadratic Casimir operator $\Omega(D_5)$.
Namely, from the group structure on the spectral theory side, we also expect the group structure on the topological string side, though classically there are no linear group invariants to couple to quadratic terms of $\mu$ in $J_k(\mu)$.
(This argument does not apply for quantum corrections directly.
We can at least require the group invariance, however.)

Here we have chosen the $D_5$ curve for our explanation of the group-theoretical viewpoint.
The same arguments apply to all of the $E_n$ $(n=6,7,8)$ curves as well.
Especially, the identification of the overall factor $\alpha$ of the spectral operator $\H$ in \eqref{e6alpha}, \eqref{e7alpha} and \eqref{e8alpha} matches respectively well with the shift in the chemical potential $\mu$ in \eqref{e6muredef}, \eqref{e7muredef} and \eqref{e8muredef}.
Indeed the shifts in terms of the exponentiated fugacities $z'=e^{\mu'}$ read respectively
\begin{align}
z'((f_1f_2f_3)^2g_1g_2g_3)^{-\frac{1}{3}},\quad
z'\biggl(\frac{\prod_{i=1}^4f_i}{\prod_{i=1}^4h_i}\biggr)^{-\frac{1}{4}},\quad
z'\biggl(\frac{\prod_{i=1}^3f_i}{\prod_{i=1}^6h_i}\biggr)^{-\frac{1}{3}},
\end{align}
($f_i=e^{i\widetilde f_i}$, $g_i=e^{i\widetilde g_i}$ and $h_i=e^{i\widetilde h_i}$) for the $E_n$ $(n=6,7,8)$ curves.
After respectively substituting the gauge fixing conditions and the constraints in \eqref{e6gf}, \eqref{e7gf} and \eqref{e8gf}, all of them reduce to the original fugacity $z$.

As noted at the end of section \ref{perturbations}, for some examples of the del Pezzo geometries studied in \cite{GKMR}, the quadratic terms of $\mu$ exist.
These curves of lower ranks are special since the symmetries contain ``the Weyl group of $\text{U}(1)$'', whose meaning is of course ambiguous.
For this reason, it is unclear to us how to fix the overall factors $\alpha$ directly by requiring the Weyl group symmetries for these curves.
Nevertheless, we still believe that there should be arguments to fix the overall factors (such as considering the limit from curves of higher ranks) and finally we need to rescale the spectral operators in \cite{GKMR} by $m^{-\frac{1}{8}}$ and $(m_1m_2)^{-\frac{1}{7}}$ respectively for these curves.
These overall factors cause the shifts of the chemical potentials introduced previously.
It is an important future direction to explain the overall factors from the first principle.

To summarize, in the introduction we have raised three assumptions which need to be clarified.
From the group-theoretical viewpoint, we find that all of them are answered from the principle of manifesting the group structure.
By requiring both the spectral operators $\H$ and the fugacity $z$ to be invariant separately under the Weyl group transformations, we find that we need to rescale both of them, which results in the perturbative grand potential of a cubic polynomial without quadratic terms and simultaneously explains the shift of the chemical potential for the classical phase space area and the identification of the overall factor $\alpha$ in the quantum mirror map.

\section{Reviews: $D_5$ curve}\label{d5}

Let us now start reviewing the analysis for the $D_5$ curve, so that we can proceed to the curves of higher ranks in the following sections.
For our purpose, the quantum curve is defined to be a quantum spectral problem, where the spectral operator $\H$ is given as a Laurent polynomial of the exponentiated canonical operators $(\Q,\P)=(e^{\widehat q},e^{\widehat p})$ \eqref{QPexp}, satisfying the canonical commutation relation $[\widehat q,\widehat p]=i\hbar$ as in quantum mechanics.
The main observable is the spectral determinant \eqref{fredholmdet} of the operator.
We first review the $D_5$ quantum curve with a slightly different notation from \cite{KMN}.

We start by introducing an order, which we call the $q$-order.
We define the $q$-order for the monomial operators as
\begin{align}
\bigl[\Q^{\alpha_1}\P^{\beta_1}\Q^{\alpha_2}\P^{\beta_2}\cdots\bigr]_q
=q^{-\frac{\alpha\beta}{2}}\Q^\alpha\P^\beta
\bigl(=q^{\frac{\alpha\beta}{2}}\P^\beta\Q^\alpha\bigr),
\label{qorder}
\end{align}
with $\alpha=\sum_i\alpha_i$ and $\beta=\sum_i\beta_i$.
For the moment we regard the coefficients as unaffected by the $q$-order.
At this stage this is nothing but the Weyl order.
In the Weyl order we symmetrize the order by taking various combinations.
Here we symmetrize the order of operators in the exponent in the sense that we can express the $q$-order \eqref{qorder} by
\begin{align}
\bigl[\Q^{\alpha_1}\P^{\beta_1}\Q^{\alpha_2}\P^{\beta_2}\cdots\bigr]_q=e^{\alpha\log\Q+\beta\log\P},
\end{align}
where we can use the Baker-Campbell-Hausdorff formula and the canonical commutation relation to bring it back to the original definition \eqref{qorder}.
Later we shall generalize the $q$-order for degenerate curves.

With this notation, we can easily define the $D_5$ quantum curve by
\begin{align}
&\H/\alpha=\bigl[\Q^{-1}(\Q+e_3)(\Q+e_4)\P\nonumber\\
&\quad+(e_1^{-1}+e_2^{-1})\Q+E/\alpha+e_3e_4h_2^{-1}(e_5+e_6)\Q^{-1}\nonumber\\
&\quad+(e_1e_2)^{-1}\Q^{-1}(\Q+h_1e_7^{-1})(\Q+h_1e_8^{-1})\P^{-1}\bigr]_q,
\end{align}
with the parameters satisfying the constraint $h_1^2h_2^2=e_1e_2e_3e_4e_5e_6e_7e_8$ following its classical analysis given in \cite{KNY}.
This is called the $D_5$ curve because the curve enjoys the symmetry isomorphic to the $D_5$ Weyl group, as explained later in this section.
Note that the asymptotic values of the classical cousin of the curve is given by
\begin{align}
(Q,P)&=(\infty,-e_1^{-1}),(\infty,-e_2^{-1}),(-e_3,\infty),(-e_4,\infty),\nonumber\\
&\quad(0,-h_2^{-1}e_5),(0,-h_2^{-1}e_6),(-h_1e_7^{-1},0),(-h_1e_8^{-1},0),
\label{asymptotic}
\end{align}
which we simply refer to as
\begin{align}
\bigl\{e_1^{-1},e_2^{-1};e_3,e_4;h_2^{-1}e_5,h_2^{-1}e_6;h_1e_7^{-1},h_1e_8^{-1}\bigr\},
\label{asymptoticvalue}
\end{align}
by dropping minus signs and trivial coordinates.
We can alternatively regard the asymptotic values as the competitions among each of the eight terms except the constant $E$.
Roughly speaking, different terms dominate when crossing the asymptotic values.
If we follow the rule of the $q$-order, we can write down the quantum curve explicitly as
\begin{align}
&\H/\alpha=q^{-\frac{1}{2}}\Q^{-1}(\Q+q^{\frac{1}{2}}e_3)(\Q+q^{\frac{1}{2}}e_4)\P\nonumber\\
&\quad+(e_1^{-1}+e_2^{-1})\Q+E/\alpha+e_3e_4h_2^{-1}(e_5+e_6)\Q^{-1}\nonumber\\
&\quad+q^{\frac{1}{2}}(e_1e_2)^{-1}\Q^{-1}(\Q+q^{-\frac{1}{2}}h_1e_7^{-1})(\Q+q^{-\frac{1}{2}}h_1e_8^{-1})\P^{-1}.
\label{d5curve}
\end{align}
Although literally the asymptotic values of the quantum curve in this $q$-order are shifted by $q^{\pm\frac{1}{2}}$, we define the asymptotic values of the quantum curve to be the same as those of the classical one \eqref{asymptoticvalue}.

Let us consider the similarity transformation
\begin{align}
\Q'=\Q,\quad
\P'&=\bigl[(\Q+e_3)\P(\Q+h_1e_7^{-1})^{-1}\bigr]_q
=q^{-1}(\Q+q^{\frac{1}{2}}e_3)\P(\Q+q^{-\frac{1}{2}}h_1e_7^{-1})^{-1},
\label{s3}
\end{align}
for this curve.
Note that the transformation rule itself can also be presented in the $q$-order \eqref{qorder}.
It is not difficult to show that the transformation \eqref{s3} leads to
\begin{align}
(\Q+q^{\frac{1}{2}}e_3)\P=(\Q'+q^{\frac{1}{2}}h_1e_7^{-1})\P',\quad
(\Q+q^{-\frac{1}{2}}h_1e_7^{-1})\P^{-1}=(\Q'+q^{-\frac{1}{2}}e_3)\P'^{-1},
\end{align}
which is directly applicable to the quantum curve \eqref{d5curve}.
After the transformation, the asymptotic values \eqref{asymptoticvalue} change into
\begin{align}
&\bigl\{e_1'^{-1},e_2'^{-1};e_3',e_4';h_2'^{-1}e_5',h_2'^{-1}e_6';h_1'e_7'^{-1},h_1'e_8'^{-1}\bigr\}\nonumber\\
&\quad=\Bigl\{e_1^{-1},e_2^{-1};h_1e_7^{-1},e_4;\frac{e_3}{h_1e_7^{-1}}h_2^{-1}e_5,\frac{e_3}{h_1e_7^{-1}}h_2^{-1}e_6;e_3,h_1e_8^{-1}\Bigr\},
\end{align}
in the same notation as \eqref{asymptoticvalue}.
Note that the other combinations $e_1'^{-1}+e_2'^{-1}=e_1^{-1}+e_2^{-1}$, $e_3'e_4'h_2'^{-1}(e_5'+e_6')=e_3e_4h_2^{-1}(e_5+e_6)$, $(e_1'e_2')^{-1}=(e_1e_2)^{-1}$ appearing in the coefficients of $\Q$, $\Q^{-1}$ and $\P^{-1}$ are also unchanged in the transformation.
Similarly, the similarity transformation
\begin{align}
\Q'=\bigl[(\P+h_2^{-1}e_5)^{-1}\Q(\P+e_1^{-1})\bigr]_q=q^{-1}(\P+q^{-\frac{1}{2}}h_2^{-1}e_5)^{-1}\Q(\P+q^{\frac{1}{2}}e_1^{-1}),\quad
\P'=\P,
\end{align}
implies the change of asymptotic values
\begin{align}
\Bigl\{h_2^{-1}e_5,e_2^{-1};e_3,e_4;e_1^{-1},h_2^{-1}e_6;\frac{e_1^{-1}}{h_2^{-1}e_5}h_1e_7^{-1},\frac{e_1^{-1}}{h_2^{-1}e_5}h_1e_8^{-1}\Bigr\}.
\end{align}
As explained in \cite{KNY,KMN}, after combining the trivial symmetries switching asymptotic values,
\begin{align}
s_1:h_1e_7^{-1}\leftrightarrow h_1e_8^{-1},\quad
s_2:e_3\leftrightarrow e_4,\quad
s_5:e_1^{-1}\leftrightarrow e_2^{-1},
\end{align}
the $D_5$ Weyl group is generated.
With the gauge fixing
\begin{align}
e_2=e_4=e_6=e_8=1,
\label{gaugefixing}
\end{align}
the transformations are explicitly given by
\begin{align}
s_1:(h_1,h_2,e_1,e_3,e_5;\alpha)&\mapsto((h_1h_2^2)^{-1}e_1e_3e_5,h_2,e_1,e_3,e_5;\alpha),\nonumber\\
s_2:(h_1,h_2,e_1,e_3,e_5;\alpha)&\mapsto(h_1e_3^{-1},h_2,e_1,e_3^{-1},e_5;e_3\alpha),\nonumber\\
s_3:(h_1,h_2,e_1,e_3,e_5;\alpha)&\mapsto(h_1,(h_1h_2)^{-1}e_1e_5,e_1,(h_1h_2^2)^{-1}e_1e_3e_5,e_5;\alpha),\nonumber\\
s_4:(h_1,h_2,e_1,e_3,e_5;\alpha)&\mapsto(h_1h_2(e_1e_5)^{-1},h_2,h_2e_5^{-1},e_3,h_2e_1^{-1};\alpha),\nonumber\\
s_5:(h_1,h_2,e_1,e_3,e_5;\alpha)&\mapsto(h_1,h_2e_1^{-1},e_1^{-1},e_3,e_5;e_1^{-1}\alpha).
\end{align}
Note here that, by introducing the $q$-order \eqref{qorder}, various factors of $q$ in \cite{KMN} disappear and the computation is clarified largely.
From this viewpoint, it is clear that the original symmetry of the Weyl group for the classical curve is lifted quantum-mechanically.

Using the quantum curve and the Weyl group, in \cite{FMS} the mirror map for the $D_5$ curve was studied by solving the Schr\"odinger equation for the quantum curve following \cite{ACDKV},
\begin{align}
\biggl[\frac{\H}{\alpha}+\frac{z}{\alpha}\biggr]\Psi(x)=0.
\label{schrodinger}
\end{align}
Here we denote the wave function as $\Psi(x)$ and refer to the same function as $\Psi[X]$ when expressed in terms of $X=e^x$.
Then, the action of the canonical operators $\Q$ and $\P$ on the wave function $\Psi(x)$ is given by
\begin{align}
\Q\Psi(x)=e^x\Psi(x)=X\Psi[X],\quad
\P\Psi(x)=\Psi(x-i\hbar)=\Psi[q^{-1}X].
\end{align}
Motivated by the action of $\P$ if we define
\begin{align}
P[X]=\frac{\Psi[q^{-1}X]}{\Psi[X]},
\label{PX}
\end{align}
the Schr\"odinger equation \eqref{schrodinger} is given by
\begin{align}
&\frac{(X+q^{\frac{1}{2}}e_3)(X+q^{\frac{1}{2}})}{q^{\frac{1}{2}}X}P[X]
+(e_1^{-1}+1)X+\frac{e_3(e_5+1)}{h_2X}
+\frac{q^{\frac{1}{2}}(X+q^{-\frac{1}{2}}h_1e_7^{-1})(X+q^{-\frac{1}{2}}h_1)}{e_1XP[qX]}\nonumber\\
&\qquad\qquad\qquad\qquad\qquad\qquad+z_E/\alpha=0,
\label{d5S}
\end{align}
where we have introduced the constant shift of $z$ by
\begin{align}
z_E=z+E.
\label{zbarzE}
\end{align}

The A-periods and the B-periods are defined for classical curves by integrating the symplectic form
\begin{align}
\oint pdx=\oint\frac{1}{X}\log P[X]dX,
\end{align}
along the A-cycles and the B-cycles respectively.
Motivated by this the quantum A-period is defined as
\begin{align}
\Pi_A(z)=\Res_{X=0}\frac{1}{X}\log\frac{P[X]}{P_0[X]},
\label{PiA}
\end{align}
with the leading order in the large $z$ expansion being
\begin{align}
P_0[X]=\frac{-z}{\alpha}\frac{q^{\frac{1}{2}}X}{(X+q^{\frac{1}{2}}e_3)(X+q^{\frac{1}{2}})}.
\end{align}
Then the quantum mirror map is given by
\begin{align}
\log z_\text{eff}=\log z+\Pi_A(z).
\label{zeffmirror}
\end{align}

In \cite{FMS} it was found that the coefficient at each order of $z$ contains those of lower orders in the nest which implies that, if we redefine $z$ by $z_E$ \eqref{zbarzE}, the quantum mirror map is given by
\begin{align}
\log z_\text{eff}=\log z_E-\sum_{\ell=1}^\infty\frac{(-1)^\ell\A_\ell}{z_E^\ell},
\label{zeffzbar}
\end{align}
where $\A_\ell$ is a function of the parameters $q$, $(h_1,h_2,e_1,e_3,e_5)$ and $\alpha$.
Note that we can alternatively study the Schr\"odinger equation \eqref{d5S} with the constant term $E$ incorporated by $z_E$ \eqref{zbarzE} from the beginning to skip the process of resolving the nest structure in \eqref{zeffzbar}.
Then we can solve the mirror map \eqref{zeffzbar} inversely by
\begin{align}
\log z_E=\log z_\text{eff}+\sum_{\ell=1}^\infty(-1)^\ell\E_\ell z_\text{eff}^{-\ell},
\label{zbar}
\end{align}
with $\E_\ell$ being a function of the same set of parameters, which leads to the expression of \eqref{Eell}.
If we identify the overall factor $\alpha$ with the combination of the parameters $(h_1,h_2,e_1,e_3,e_5)$ which transforms identically as $\alpha$,
\begin{align}
\alpha=h_2^{\frac{1}{2}}e_1^{\frac{1}{4}}e_3^{-\frac{1}{2}}e_5^{-\frac{1}{4}},
\label{d5alpha}
\end{align}
we find that the results are given by the $D_5$ characters.
In section \ref{weyl} this identification is explained by requiring the invariance of the spectral operator $\H$ under the Weyl group.
In \cite{FMS} it was found that if we introduce the multi-covering structure \eqref{Emc}
\begin{align}
\E_\ell=\sum_{n|\ell}\frac{(-1)^{n+1}\epsilon_{\frac{\ell}{n}}(q^n,{\bm q}^n)}{n},
\label{multicover}
\end{align}
we can further simplify the expression with the multi-covering components $\epsilon_d(q,{\bm q})$ listed in \cite{FMS}.
This mirror map explains the redefinition of the chemical potential $\mu=\log z$ in the corresponding matrix models.

With the redefinition of the chemical potential in \eqref{zeffmirror} the worldsheet instantons take care of all the bound states.
This suggests a picture that each of the worldsheet instantons is accompanied by an infinite tower of BPS excitations of bound states, which are counted by these non-negative integers.
This viewpoint matches with the proposal of \cite{AKV} where the mirror map is interpreted as counting BPS excitations in the presence of domain walls.

\section{$E_6$ curve}\label{e6}

After reviewing for the $D_5$ curve, we now turn to the analysis for the three curves, $E_6$, $E_7$ and $E_8$.
We first start with the toric case $E_6$.
Using the $q$-order introduced in the previous section, we can write down the quantum curve, identify the Weyl group action and proceed to studying the mirror map without so much difficulty.
We also rewrite the $E_6$ curve into a non-toric one, which is helpful for later analysis for the curves of higher ranks.
The $E_6$ quantum curve in the non-toric realization was determined previously in \cite{T} in a slightly different guise, though we believe that the derivation here with the introduction of the $q$-order is more systematic and applicable to the curves of higher ranks.

\subsection{Quantum curve}\label{e6qc}

In this subsection, we first write down the $E_6$ quantum curve.

\begin{figure}[!t]
\centering\includegraphics[scale=0.4,angle=-90]{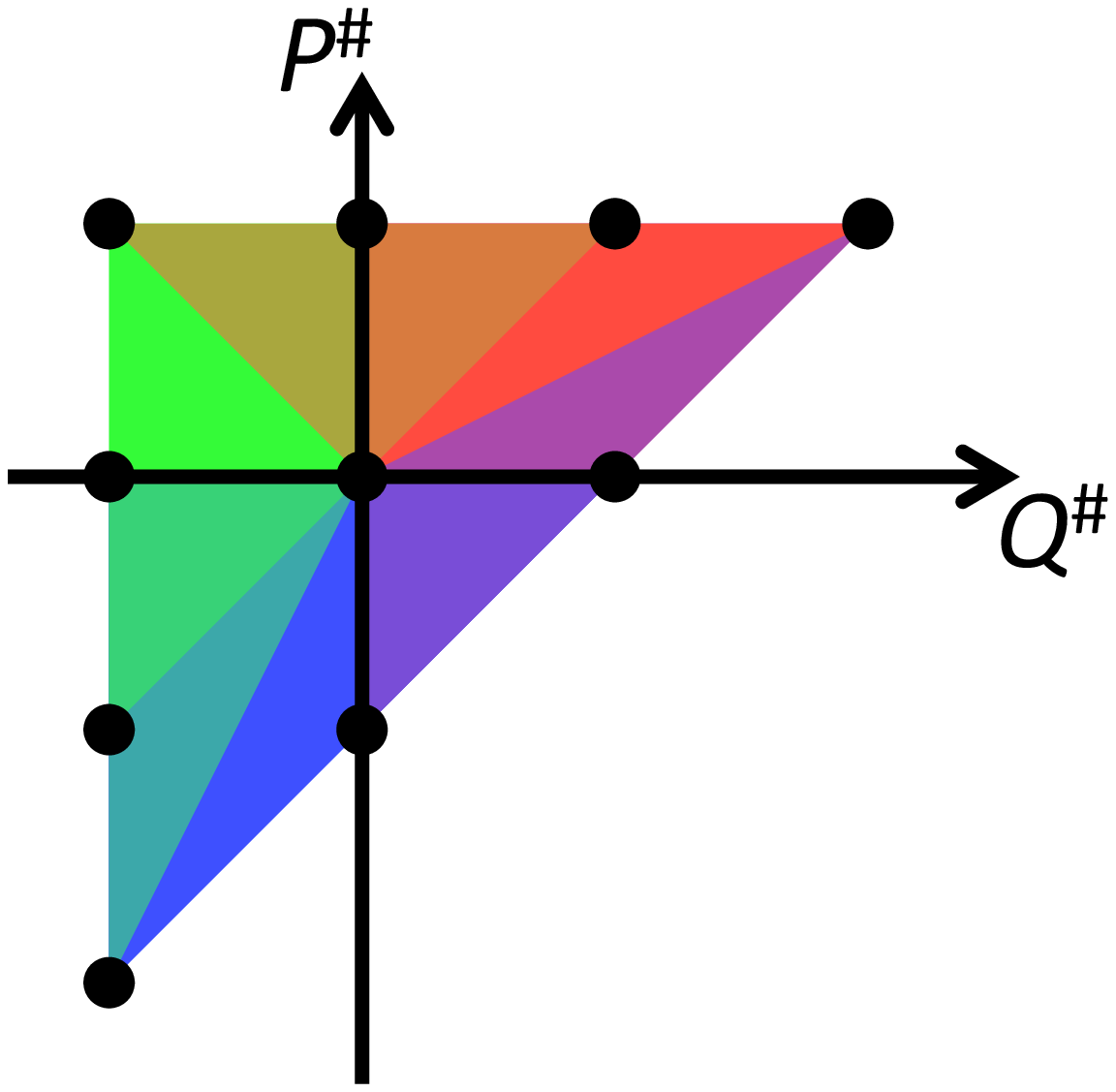}\qquad\qquad\includegraphics[scale=0.6,angle=-90]{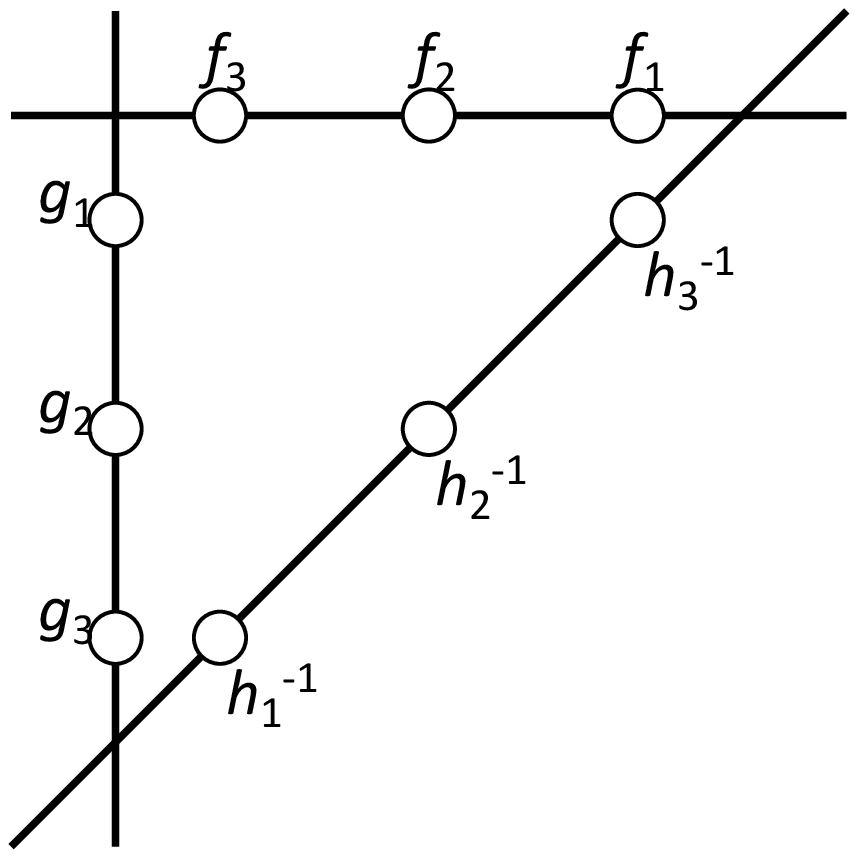}
\caption{A toric diagram (left) and a schematic diagram for asymptotic values (right) for the $E_6$ del Pezzo curve in the triangular realization.}
\label{e6tri}
\end{figure}

Since the $E_6$ curve is the del Pezzo geometry of the largest rank which enjoys a realization by the toric diagram, let us utilize the toric expression.
The toric diagram for the $E_6$ curve has a triangular shape (see figure \ref{e6tri}) and we refer to it as the triangular realization.
Using our $q$-order, the $E_6$ curve is given by
\begin{align}
&\H/\alpha
=\bigl[\Q^{-1}(\Q+f_1)(\Q+f_2)(\Q+f_3)\P\nonumber\\
&\quad+(h_1^{-1}+h_2^{-1}+h_3^{-1})\Q+E/\alpha+f_1f_2f_3(g_1+g_2+g_3)\Q^{-1}\nonumber\\
&\quad+(h_1h_2h_3)^{-1}\bigl((h_1+h_2+h_3)\Q+(g_1^{-1}+g_2^{-1}+g_3^{-1})\bigr)\Q^{-1}\P^{-1}\nonumber\\
&\quad+(h_1h_2h_3)^{-1}\Q^{-1}\P^{-2}\bigr]_q,
\end{align}
where the parameters enjoy the constraint
\begin{align}
f_1f_2f_3g_1g_2g_3h_1h_2h_3=1.
\label{e6constraint}
\end{align}

The classical cousin of the curve has the asymptotic values
\begin{align}
(Q,P)&=(-f_1,\infty),(-f_2,\infty),(-f_3,\infty),(0,-g_1),(0,-g_2),(0,-g_3),\nonumber\\
&\quad(\infty,-h_1^{-1}/\infty),(\infty,-h_2^{-1}/\infty),(\infty,-h_3^{-1}/\infty).
\end{align}
Namely, the asymptotic behaviors in the region $|\Q|\gg 1$ and $|\P|\ll 1$ are $\Q\P=-h_1^{-1}$, $\Q\P=-h_2^{-1}$, $\Q\P=-h_3^{-1}$.
As in the case of $D_5$ we often drop minus signs and ratios and refer to the asymptotic values simply as
\begin{align}
\bigl\{f_1,f_2,f_3;g_1,g_2,g_3;h_1^{-1},h_2^{-1},h_3^{-1}\bigr\}_\text{tri}.
\end{align}
From the definition of the $q$-order, the explicit expression of the quantum curve is given by
\begin{align}
&\H/\alpha
=q^{-1}\Q^{-1}(\Q+q^{\frac{1}{2}}f_1)(\Q+q^{\frac{1}{2}}f_2)(\Q+q^{\frac{1}{2}}f_3)\P\nonumber\\
&\quad+(h_1^{-1}+h_2^{-1}+h_3^{-1})\Q+E/\alpha+f_1f_2f_3(g_1+g_2+g_3)\Q^{-1}\nonumber\\
&\quad+(h_1h_2h_3)^{-1}
\bigl((h_1+h_2+h_3)\Q+q^{-\frac{1}{2}}(g_1^{-1}+g_2^{-1}+g_3^{-1})\bigr)\Q^{-1}\P^{-1}\nonumber\\
&\quad+q^{-1}(h_1h_2h_3)^{-1}\Q^{-1}\P^{-2},
\label{e6triP}
\end{align}
by collecting terms of the same order in $\P$, or 
\begin{align}
&\H/\alpha
=q^{-1}\Q^2\P\nonumber\\
&\quad+q^{-\frac{1}{2}}\Q\bigl((f_1+f_2+f_3)\P+q^{\frac{1}{2}}(h_1^{-1}+h_2^{-1}+h_3^{-1})\bigr)\nonumber\\
&\quad+f_1f_2f_3(f_1^{-1}+f_2^{-1}+f_3^{-1})\P+E/\alpha+(h_1h_2h_3)^{-1}(h_1+h_2+h_3)\P^{-1}\nonumber\\
&\quad+q^{\frac{1}{2}}f_1f_2f_3\Q^{-1}
(\P+q^{-\frac{1}{2}}g_1)(\P+q^{-\frac{1}{2}}g_2)(\P+q^{-\frac{1}{2}}g_3)\P^{-2},
\label{e6triQ}
\end{align}
by collecting terms of the same order in $\Q$ (where both expressions are useful for later discussions).
It is obvious that this curve enjoys the symmetries of exchanging two asymptotic values in each of the three subsets $\{f_1,f_2,f_3\}$, $\{g_1,g_2,g_3\}$ and $\{h_1^{-1},h_2^{-1},h_3^{-1}\}$.
Namely, the transformations
\begin{align}
&s_1:f_3\leftrightarrow f_2,\quad s_2:f_2\leftrightarrow f_1,\nonumber\\
&s_0:g_3\leftrightarrow g_2,\quad s_4:g_2\leftrightarrow g_1,\nonumber\\
&s_5:h_1\leftrightarrow h_2,\quad s_6:h_2\leftrightarrow h_3,
\label{s120456}
\end{align}
leave the curve invariant trivially.
(The numbering of the transformations is clarified later.)

Note that our toric diagram of the quantum curve in figure \ref{e6tri} contains a black dot at the origin.
We tessellate the diagram by connecting the origin with all of the outer black dots on the boundary of the toric diagram and paint each region with different colors.
This tessellation corresponds to the large $z$ limit, where we need to compare each term with the spectral parameter $z$.
As we see later, the coloring is helpful in determining the coefficients for the degenerate curves.

\begin{figure}[!t]
\centering\includegraphics[scale=0.4,angle=-90]{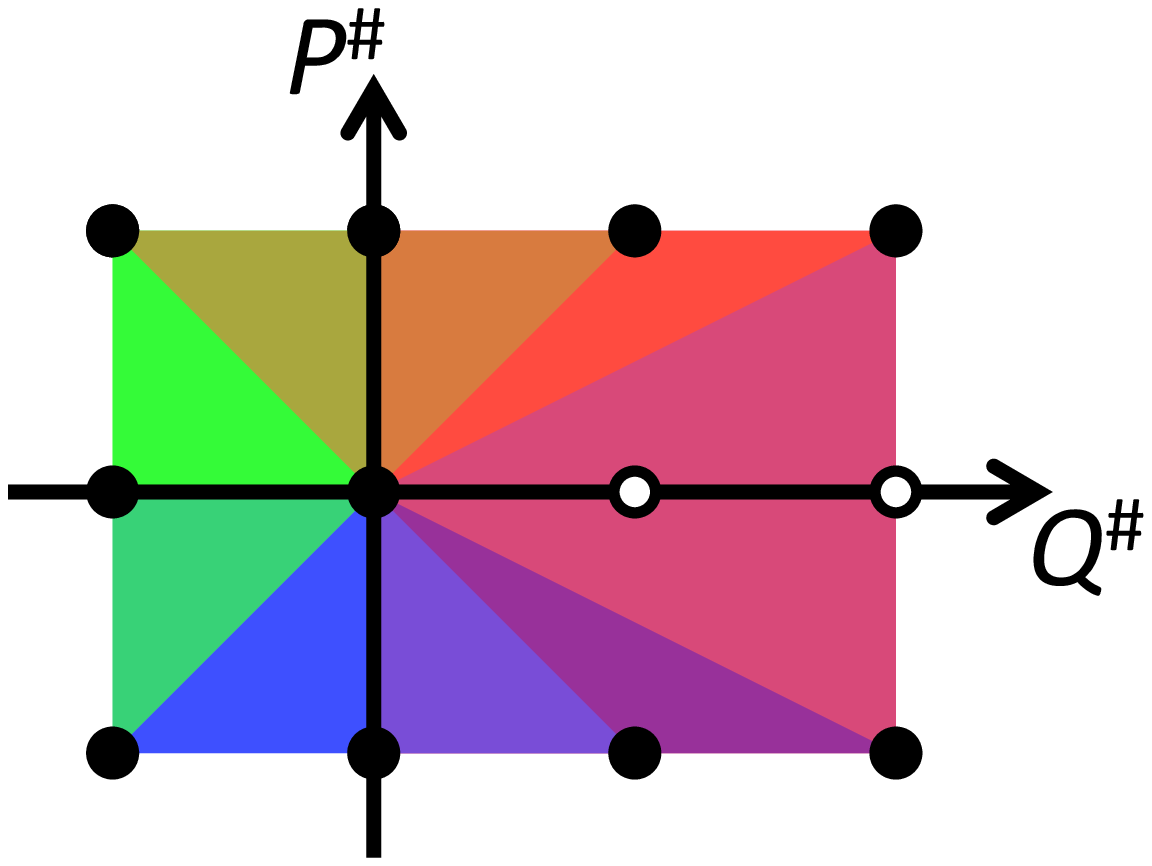}\qquad\qquad\includegraphics[scale=0.6,angle=-90]{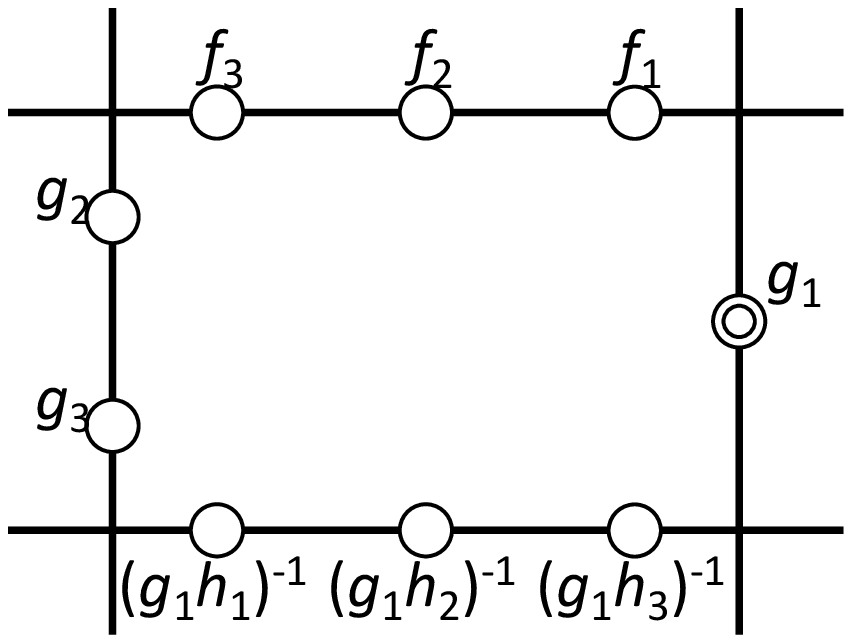}
\caption{A toric diagram (left) and a schematic diagram for asymptotic values (right) for the $E_6$ del Pezzo curve in the rectangular realization.}
\label{e6rect}
\end{figure}

To see the transformation mixing among the three subsets, it is more convenient to move to another realization.
Namely, we introduce the similarity transformation
\begin{align}
\Q'=\bigl[\P(\P+g_1)^{-1}\Q\bigr]_q=\P(\P+q^{-\frac{1}{2}}g_1)^{-1}\Q,\quad\P'=\P,
\label{trirect}
\end{align}
which implies, for example, that
\begin{align}
&\Q^3=\Q'^3(\P'+q^{\frac{5}{2}}g_1)(\P'+q^{\frac{3}{2}}g_1)(\P'+q^{\frac{1}{2}}g_1)\P'^{-3},\quad
\Q^{-1}(\P+q^{-\frac{1}{2}}g_1)\P^{-1}=\Q'^{-1},
\nonumber\\
&\Q^2=\Q'^2(\P'+q^{\frac{3}{2}}g_1)(\P'+q^{\frac{1}{2}}g_1)\P'^{-2},\quad
\Q^{-2}(\P+q^{-\frac{3}{2}}g_1)(\P+q^{-\frac{1}{2}}g_1)\P^{-2}=\Q'^{-2},
\nonumber\\
&\Q=\Q'(\P'+q^{\frac{1}{2}}g_1)\P'^{-1},\quad
\Q^{-3}(\P+q^{-\frac{5}{2}}g_1)(\P+q^{-\frac{3}{2}}g_1)(\P+q^{-\frac{1}{2}}g_1)\P^{-3}=\Q'^{-3}.
\label{trirect+}
\end{align}
After applying the transformation for \eqref{e6triQ} and removing the primes, we find
\begin{align}
&\H/\alpha
=q^{-1}\Q^2(\P+q^{\frac{3}{2}}g_1)(\P+q^{\frac{1}{2}}g_1)\P^{-1}\nonumber\\
&\quad+q^{-\frac{1}{2}}\Q(\P+q^{\frac{1}{2}}g_1)\bigl((f_1+f_2+f_3)\P+q^{\frac{1}{2}}(h_1^{-1}+h_2^{-1}+h_3^{-1})\bigr)\P^{-1}\nonumber\\
&\quad+f_1f_2f_3(f_1^{-1}+f_2^{-1}+f_3^{-1})\P+E/\alpha+(h_1h_2h_3)^{-1}(h_1+h_2+h_3)\P^{-1}\nonumber\\
&\quad+q^{\frac{1}{2}}f_1f_2f_3\Q^{-1}
(\P+q^{-\frac{1}{2}}g_2)(\P+q^{-\frac{1}{2}}g_3)\P^{-1},
\label{e6rectQ}
\end{align}
or in other words,
\begin{align}
&\H/\alpha
=q^{-1}\Q^{-1}(\Q+q^{\frac{1}{2}}f_1)(\Q+q^{\frac{1}{2}}f_2)(\Q+q^{\frac{1}{2}}f_3)\P
\nonumber\\
&\quad
+(q^{\frac{1}{2}}+q^{-\frac{1}{2}})g_1\Q^2+\bigl(g_1(f_1+f_2+f_3)+(h_1^{-1}+h_2^{-1}+h_3^{-1})\bigr)\Q
\nonumber\\
&\qquad
+E/\alpha+(f_1f_2f_3)(g_2+g_3)\Q^{-1}
\nonumber\\
&\quad
+qg_1^2\Q^{-1}(\Q+q^{-\frac{1}{2}}(g_1h_1)^{-1})
(\Q+q^{-\frac{1}{2}}(g_1h_2)^{-1})
(\Q+q^{-\frac{1}{2}}(g_1h_3)^{-1})\P^{-1}.
\label{e6rectP}
\end{align}
Since the toric diagram has a rectangular shape (see figure \ref{e6rect}), we refer to this as the rectangular realization.

Note that, compared with the classical curve, the degeneracy looks different in the quantum curve.
As in \eqref{e6rectP}, it seems that the binomial coefficient $2$ of $g_1\Q^2$ for the classical curve is changed into the $q$-integer $[2]_q=q^{\frac{1}{2}}+q^{-\frac{1}{2}}$, which is generally defined by
\begin{align}
[n]_q=\frac{q^{\frac{n}{2}}-q^{-\frac{n}{2}}}{q^{\frac{1}{2}}-q^{-\frac{1}{2}}}.
\end{align}
Also, by collecting terms of the same order in $\Q$ in \eqref{e6rectQ}, it turns out that the degeneracy appears as the step-by-step enhancement of powers of $q$ such as $(\P+q^{\frac{3}{2}}g_1)(\P+q^{\frac{1}{2}}g_1)$ in the $O(\Q^2)$ terms and $(\P+q^{\frac{1}{2}}g_1)$ in the $O(\Q)$ terms.
Hence, we may adopt the rule of the step-by-step enhancement of powers of $q$ for the degeneracy as part of the definition of the $q$-order in \eqref{qorder},
\begin{align}
\bigl[\Q^2(\P+g_1)^2\P^{-1}\bigr]_q=q^{-1}\Q^2(\P+q^{\frac{3}{2}}g_1)(\P+q^{\frac{1}{2}}g_1)\P^{-1}.
\label{quantumdegeneracy}
\end{align}
It seems that the step-by-step enhancement of powers of $q$ is mandatory for the similarity transformation to work as a symmetry for the quantum del Pezzo geometries.
These structures of degenerate curves were partially recognized in \cite{KMN}, though, with the $q$-order introduced in this paper, the structures appear more systematic.

In \cite{BBT,KY} white dots were introduced in the toric diagrams to express the brane configuration with parallel 5-branes ending on the same 7-brane, which corresponds to the degeneracy of curves.
Since the tessellations by the 5-branes and the colors (black/white) of the inner dots (located in the bulk of the toric diagram) are not unique due to the flop transitions, in our toric diagram in figure \ref{e6rect}, the white dots simply denote the dependence of the coefficients on those of the black ones due to the degeneracy of the curve.
As in figure \ref{e6tri}, we tessellate the toric diagram by connecting the origin with the outer black dots and paint each region with different colors in figure \ref{e6rect}.
Note that the information of the asymptotic values matches with the coloring.
Namely, the coefficients of the two white dots at $\Q$ and $\Q^2$ are both determined by the same asymptotic value $g_1$ and located in the region of the same color.

The asymptotic values of the quantum curve are defined to be those of their classical cousin,
\begin{align}
&(\infty,-g_1)\text{[double]},
(-f_1,\infty),(-f_2,\infty),(-f_3,\infty),(0,-g_2),(0,-g_3),\nonumber\\
&\qquad(-(g_1h_1)^{-1},0),(-(g_1h_2)^{-1},0),(-(g_1h_3)^{-1},0),
\end{align}
which we refer to as
\begin{align}
\bigl\{g_1\text{[double]};f_1,f_2,f_3;g_2,g_3;(g_1h_1)^{-1},(g_1h_2)^{-1},(g_1h_3)^{-1}\bigr\}_\text{rect}.
\label{e6double}
\end{align}
Note that, literally speaking, in the quantum curve \eqref{e6rectQ} the classical degenerate asymptotic values $(\infty,-g_1)$ seem to split into $(\infty,-q^{\frac{3}{2}}g_1)$ and $(\infty,-q^{\frac{1}{2}}g_1)$ \cite{KMN}, though we stick to the terminology of the double root.

Now let us consider the similarity transformation,
\begin{align}
\Q'=\Q,\quad
\P'=[(\Q+f_1)\P(\Q+(g_1h_1)^{-1})^{-1}]_q=q^{-1}(\Q+q^{\frac{1}{2}}f_1)\P(\Q+q^{-\frac{1}{2}}(g_1h_1)^{-1})^{-1},
\label{s3similarity}
\end{align}
which implies
\begin{align}
(\Q+q^{\frac{1}{2}}f_1)\P=(\Q'+q^{\frac{1}{2}}(g_1h_1)^{-1})\P',\quad
(\Q+q^{-\frac{1}{2}}(g_1h_1)^{-1})\P^{-1}=(\Q'+q^{-\frac{1}{2}}f_1)\P'^{-1}.
\label{s3similarity+}
\end{align}
By applying this transformation to the $E_6$ curve in the rectangular realization \eqref{e6rectP}, the asymptotic values \eqref{e6double} are changed into
\begin{align}
\bigl\{g_1\text{[double]};(g_1h_1)^{-1},f_2,f_3;(f_1g_1h_1)g_2,(f_1g_1h_1)g_3;f_1,(g_1h_2)^{-1},(g_1h_3)^{-1}\bigr\}_\text{rect}.
\label{e6s3}
\end{align}
Note that all of the coefficients $g_1$, $g_1(f_1+f_2+f_3)+h_1^{-1}+h_2^{-1}+h_3^{-1}$, $f_1f_2f_3(g_2+g_3)$ are invariant under the transformation.

\subsection{Weyl group}\label{e6wg}

\begin{figure}[!t]
\centering\includegraphics[scale=0.4,angle=-90]{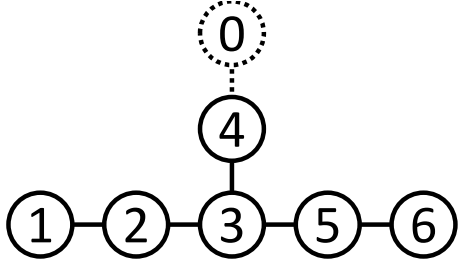}
\caption{Dynkin diagram of the $E_6$ algebra.}
\label{e6dynkin} 
\end{figure}

In this subsection we identify the symmetry of the curve as the $E_6$ Weyl group.
The computations in this subsection follow directly previous ones for the $D_5$ curve in \cite{KMN,KM,FMS}.
Various results are crucial later for the study of the mirror map in section \ref{e6mirror}.

Here let us choose the gauge $f_3=g_3=1$ and solve the constraint \eqref{e6constraint} by
\begin{align}
f_3=1,\quad g_3=1,\quad h_3=\frac{1}{f_1f_2g_1g_2h_1h_2}.
\label{e6gf}
\end{align}
Then the transformation from \eqref{e6double} to \eqref{e6s3} is given by
\begin{align}
s_3:(f_1,f_2,g_1,g_2,h_1,h_2;\alpha)\mapsto((g_1h_1)^{-1},f_2,(f_1h_1)^{-1},g_2,h_1,(f_1g_1h_1)h_2;(f_1g_1h_1)\alpha).
\end{align}
We can combine this with the other transformations in \eqref{s120456} as
\begin{align}
s_1:(f_1,f_2,g_1,g_2,h_1,h_2;\alpha)&\mapsto
(f_2^{-1}f_1,f_2^{-1},g_1,g_2,f_2h_1,f_2h_2;f_2^2\alpha),\nonumber\\
s_2:(f_1,f_2,g_1,g_2,h_1,h_2;\alpha)&\mapsto
(f_2,f_1,g_1,g_2,h_1,h_2;\alpha),\nonumber\\
s_3:(f_1,f_2,g_1,g_2,h_1,h_2;\alpha)&\mapsto
((g_1h_1)^{-1},f_2,(f_1h_1)^{-1},g_2,h_1,(f_1g_1h_1)h_2;(f_1g_1h_1)\alpha),\nonumber\\
s_4:(f_1,f_2,g_1,g_2,h_1,h_2;\alpha)&\mapsto
(f_1,f_2,g_2,g_1,h_1,h_2;\alpha),\nonumber\\
s_5:(f_1,f_2,g_1,g_2,h_1,h_2;\alpha)&\mapsto
(f_1,f_2,g_1,g_2,h_2,h_1;\alpha),\nonumber\\
s_6:(f_1,f_2,g_1,g_2,h_1,h_2;\alpha)&\mapsto
(f_1,f_2,g_1,g_2,h_1,(f_1f_2g_1g_2h_1h_2)^{-1};\alpha),\nonumber\\
s_0:(f_1,f_2,g_1,g_2,h_1,h_2;\alpha)&\mapsto
(f_1,f_2,g_2^{-1}g_1,g_2^{-1},g_2h_1,g_2h_2;g_2\alpha).
\label{e6weyl}
\end{align}
From the commutation relation of the transformations we find that the symmetries generate the $E_6$ Weyl group.
Namely, if we connect $i$ and $j$ for $(s_is_j)^3=1$ and disconnect them for $(s_is_j)^2=1$, we can reproduce exactly the $E_6$ Dynkin diagram (see figure \ref{e6dynkin}).
This is how we have numbered the Weyl action.

Although the elements of the Weyl group consist of the Weyl reflections along the simple roots in the orthogonal basis, the transformation rules in \eqref{e6weyl} are not interpreted directly as reflections.
Here, let us identify the simple roots and the fundamental weights as vectors in the linear space $(\log f_1,\log f_2,\log g_1,\log g_2,\log h_1,\log h_2)$ respecting the linear relation of the Weyl reflections and the Cartan matrix, as in \cite{KMN,FMS}.
Namely, we simultaneously require that the Weyl reflections $s_i$ along the simple roots $\alpha_i$ act on the fundamental weights $\omega_j$ as
\begin{align}
s_i(\omega_j)=\omega_j-\delta_{ij}\alpha_i,
\end{align}
(with no sum over $i$) and that the simple roots $\alpha_i$ are expanded by the fundamental weights $\omega_j$ with the Cartan matrix $A_{ij}$ as
\begin{align}
\alpha_i=A_{ij}\omega_j.
\end{align}
Then, from the transformation rule \eqref{e6weyl}, we can identify the simple roots and the fundamental weights as
\begin{align}
\alpha_1&=(-1,-2,0,0,1,1),&\omega_1&=(-1,-1,1,1,0,0),\nonumber\\
\alpha_2&=(-1,1,0,0,0,0),&\omega_2&=(-1,0,2,2,-1,-1),\nonumber\\
\alpha_3&=(1,0,1,0,0,-1),&\omega_3&=(0,0,3,3,-2,-2),\nonumber\\
\alpha_4&=(0,0,-1,1,0,0),&\omega_4&=(0,0,1,2,-1,-1),\nonumber\\
\alpha_5&=(0,0,0,0,-1,1),&\omega_5&=(0,0,2,2,-2,-1),\nonumber\\
\alpha_6&=(0,0,0,0,0,-1),&\omega_6&=(0,0,1,1,-1,-1),
\label{e6srfw}
\end{align}
in the linear space of $(\log f_1,\log f_2,\log g_1,\log g_2,\log h_1,\log h_2)$.
For example, by requiring the invariance under the transformations $s_i$ $(1\le i\le 5)$, we can parametrize the direction by $(f_1,f_2,g_1,g_2,h_1,h_2)=(t^0,t^0,t^1,t^1,t^{-1},t^{-1})$, which leads to $\omega_6=(0,0,1,1,-1,-1)$.
We can further apply $s_6$ to find out $\alpha_6=(0,0,0,0,0,-1)$ from $\omega_6-\alpha_6=(0,0,1,1,-1,0)$.
Then, the relation from the Cartan matrix $\alpha_6=2\omega_6-\omega_5$ implies $\omega_5=(0,0,2,2,-2,-1)$.
We can follow these steps to determine all of the simple roots and the fundamental weights.

In the next subsection, we shall compute the A-period for the $E_6$ curve.
As in \eqref{d5alpha} for the $D_5$ curve, to express the final results in terms of group characters, we need to identify the overall coefficient $\alpha$ as a product of $(f_1,f_2,g_1,g_2,h_1,h_2)$ which transforms exactly identically as $\alpha$, which is
\begin{align}
\alpha=(f_1f_2)^{-\frac{2}{3}}(g_1g_2)^{-\frac{1}{3}}.
\label{e6alpha}
\end{align}
The reason for this identification has been explained in section \ref{weyl}.

In identifying the mirror map as representations we need to prepare characters.
In general it is time-consuming to generate characters for exceptional algebras.
Instead of generating characters by ourselves by applying various group-theoretical formulas, we simply pick up results from \cite{Y}.
For example, since the adjoint representation ${\bf 78}$ of $E_6$ is decomposed as ${\bf 45}_0+{\bf 16}_3+\overline{\bf 16}_{-3}+{\bf 1}_0$ in the subgroup $[D_5]_{\text{U}(1)}$ due to the list in \cite{Y}, we can reconstruct the character by
\begin{align}
\chi^{E_6}_{\bf 78}=\chi^{D_5}_{\bf 45}+t_6^3\chi^{D_5}_{\bf 16}+t_6^{-3}\chi^{D_5}_{\overline{\bf 16}}+\chi^{D_5}_{\bf 1},
\label{e6chi78}
\end{align}
as long as we know the $D_5$ characters 
\begin{align}
&\chi^{D_5}_{\bf 45}=5+\sum_{i<j}(t_i^2+t_i^{-2})(t_j^2+t_j^{-2}),\quad\chi^{D_5}_{\bf 1}=1,\nonumber\\
&\chi^{D_5}_{\bf 16}=\frac{\sum_it_i^2+\sum_{i<j<k}t_i^2t_j^2t_k^2+t_1^2t_2^2t_3^2t_4^2t_5^2}{t_1t_2t_3t_4t_5},\quad
\chi^{D_5}_{\overline{\bf 16}}=\frac{1+\sum_{i<j}t_i^2t_j^2+\sum_{i<j<k<l}t_i^2t_j^2t_k^2t_l^2}{t_1t_2t_3t_4t_5},
\end{align}
which can be generated by the Weyl character formula without difficulty.

Later we need to express our results of the mirror map (which are given by the parameters $(f_1,f_2,g_1,g_2,h_1,h_2)$) in terms of characters (given by the variables $(t_1,t_2,t_3,t_4,t_5,t_6)$).
The variables $(t_1,t_2,t_3,t_4,t_5,t_6)$ are associated to the orthogonal basis, where the simple roots and the fundamental weights are given by
\begin{align}
\alpha^\perp_1&=\textstyle{(-1,1,0,0,0,0)},&
\omega^\perp_1&=\textstyle{(-1,0,0,0,0,-\frac{\sqrt{3}}{3})},\nonumber\\
\alpha^\perp_2&=\textstyle{(0,-1,1,0,0,0)},&
\omega^\perp_2&=\textstyle{(-1,-1,0,0,0,-\frac{2\sqrt{3}}{3})},\nonumber\\
\alpha^\perp_3&=\textstyle{(0,0,-1,1,0,0)},&
\omega^\perp_3&=\textstyle{(-1,-1,-1,0,0,-\sqrt{3})},\nonumber\\
\alpha^\perp_4&=\textstyle{(0,0,0,-1,1,0)},&
\omega^\perp_4&=\textstyle{(-\frac{1}{2},-\frac{1}{2},-\frac{1}{2},-\frac{1}{2},\frac{1}{2},-\frac{\sqrt{3}}{2})},\nonumber\\
\alpha^\perp_5&=\textstyle{(0,0,0,-1,-1,0)},&
\omega^\perp_5&=\textstyle{(-\frac{1}{2},-\frac{1}{2},-\frac{1}{2},-\frac{1}{2},-\frac{1}{2},-\frac{5\sqrt{3}}{6})},\nonumber\\
\alpha^\perp_6&=\textstyle{(\frac{1}{2},\frac{1}{2},\frac{1}{2},\frac{1}{2},\frac{1}{2},-\frac{\sqrt{3}}{2})},&
\omega^\perp_6&=\textstyle{(0,0,0,0,0,-\frac{2\sqrt{3}}{3})}.
\label{e6ortho}
\end{align}
To identify our variables $(f_1,f_2,g_1,g_2,h_1,h_2)$ with the orthogonal ones $(t_1,t_2,t_3,t_4,t_5,t_6)$, we need to compare two sets of the fundamental weights in \eqref{e6srfw} and \eqref{e6ortho} by
\begin{align}
&(f_1,f_2,g_1,g_2,h_1,h_2)
=f_1^{(1,0,0,0,0,0)}f_2^{(0,1,0,0,0,0)}g_1^{(0,0,1,0,0,0)}g_2^{(0,0,0,1,0,0)}h_1^{(0,0,0,0,1,0)}h_2^{(0,0,0,0,0,1)}\nonumber\\
&=f_1^{-\omega_2+\omega_3-\omega_6}f_2^{-\omega_1+\omega_2-\omega_6}
g_1^{\omega_3-\omega_4-\omega_6}g_2^{\omega_4-\omega_6}
h_1^{\omega_3-\omega_5-\omega_6}h_2^{\omega_5-2\omega_6}\nonumber\\
&\leftrightarrow
f_1^{-\omega^\perp_2+\omega^\perp_3-\omega^\perp_6}f_2^{-\omega^\perp_1+\omega^\perp_2-\omega^\perp_6}
g_1^{\omega^\perp_3-\omega^\perp_4-\omega^\perp_6}g_2^{\omega^\perp_4-\omega^\perp_6}
h_1^{\omega^\perp_3-\omega^\perp_5-\omega^\perp_6}h_2^{\omega^\perp_5-2\omega^\perp_6}\nonumber\\
&=f_1^{(0,0,-1,0,0,\frac{\sqrt{3}}{3})}f_2^{(0,-1,0,0,0,\frac{\sqrt{3}}{3})}
g_1^{(-\frac{1}{2},-\frac{1}{2},-\frac{1}{2},\frac{1}{2},-\frac{1}{2},\frac{\sqrt{3}}{6})}
g_2^{(-\frac{1}{2},-\frac{1}{2},-\frac{1}{2},-\frac{1}{2},\frac{1}{2},\frac{\sqrt{3}}{6})}\nonumber\\
&\qquad\qquad\times h_1^{(-\frac{1}{2},-\frac{1}{2},-\frac{1}{2},\frac{1}{2},\frac{1}{2},\frac{\sqrt{3}}{2})}
h_2^{(-\frac{1}{2},-\frac{1}{2},-\frac{1}{2},-\frac{1}{2},-\frac{1}{2},\frac{\sqrt{3}}{2})}\nonumber\\
&=\bigl((g_1g_2h_1h_2)^{-\frac{1}{2}},(f_2^2g_1g_2h_1h_2)^{-\frac{1}{2}},(f_1^2g_1g_2h_1h_2)^{-\frac{1}{2}},
(g_1g_2^{-1}h_1h_2^{-1})^{\frac{1}{2}},(g_1^{-1}g_2h_1h_2^{-1})^{\frac{1}{2}},\nonumber\\
&\qquad\qquad(f_1^2f_2^2g_1g_2h_1^3h_2^3)^{\frac{\sqrt{3}}{6}}\bigr),
\end{align}
as in \cite{KM,FMS}.
By identifying this set of variables as
\begin{align}
{\bm q}=(q_1,q_2,q_3,q_4,q_5,q_6)=(t_1^2,t_2^2,t_3^2,t_4^2,t_5^2,t_6^{-2\sqrt{3}}),
\label{e6t}
\end{align}
we can solve the relation inversely and find
\begin{align}
f_1=\frac{t_1^2}{t_3^2},\quad
f_2=\frac{t_1^2}{t_2^2},\quad
g_1=\frac{t_4t_6^3}{t_1t_2t_3t_5},\quad
g_2=\frac{t_5t_6^3}{t_1t_2t_3t_4},\quad
h_1=\frac{t_2t_3t_4t_5}{t_1t_6^3},\quad
h_2=\frac{t_2t_3}{t_1t_4t_5t_6^3}.
\label{e6fght}
\end{align}
Namely, after obtaining the results for the mirror map, we can substitute \eqref{e6fght} to compare with the known characters.

\subsection{Mirror map}\label{e6mirror}

\begin{table}[t!]
\begin{align*}
\epsilon_1&=0,\\
\epsilon_2&=\chi_{\bf 27},\\
\epsilon_3&=(q^{\frac{3}{2}}+q^{-\frac{3}{2}})\chi_{\bf 1}
+(q^{\frac{1}{2}}+q^{-\frac{1}{2}})(\chi_{\bf 78}+3\chi_{\bf 1}),\\
\epsilon_4&=(q^2+q^{-2})\chi_{\overline{\bf 27}}
+(q+q^{-1})(\chi_{\overline{\bf 351}}+3\chi_{\overline{\bf 27}})
+4\chi_{\overline{\bf 27}},\\
\epsilon_5&=(q^{\frac{7}{2}}+q^{-\frac{7}{2}})\chi_{\bf 27}
+(q^{\frac{5}{2}}+q^{-\frac{5}{2}})(\chi_{\bf 351}+4\chi_{\bf 27})
+(q^{\frac{3}{2}}+q^{-\frac{3}{2}})(\chi_{\bf 1728}+3\chi_{\bf 351}+9\chi_{\bf 27})\\
&\quad+(q^{\frac{1}{2}}+q^{-\frac{1}{2}})(3\chi_{\bf 351}+9\chi_{\bf 27}),\\
\epsilon_6&=(q^6+q^{-6})\chi_{\bf 1}
+(q^5+q^{-5})(\chi_{\bf 78}+3\chi_{\bf 1})
+(q^4+q^{-4})(\chi_{\bf 650}+4\chi_{\bf 78}+10\chi_{\bf 1})\\
&\quad+(q^3+q^{-3})(\chi_{\bf 2925}+4\chi_{\bf 650}+13\chi_{\bf 78}+23\chi_{\bf 1})\\
&\quad+(q^2+q^{-2})(\chi_{\bf 5824}+3\chi_{\bf 2925}+\chi_{\bf 2430}+9\chi_{\bf 650}+25\chi_{\bf 78}+38\chi_{\bf 1})\\
&\quad+(q+q^{-1})(2\chi_{\bf 2925}+\chi_{\bf 2430}+9\chi_{\bf 650}+27\chi_{\bf 78}+45\chi_{\bf 1})\\
&\quad+\chi_{\bf 5824}+2\chi_{\bf 2925}+9\chi_{\bf 650}+27\chi_{\bf 78}+44\chi_{\bf 1},\\
\epsilon_7&=(q^{\frac{15}{2}}+q^{-\frac{15}{2}})\chi_{\overline{\bf 27}}
+(q^{\frac{13}{2}}+q^{-\frac{13}{2}})(\chi_{\overline{\bf 351}}+4\chi_{\overline{\bf 27}})\\
&\quad+(q^{\frac{11}{2}}+q^{-\frac{11}{2}})(\chi_{\overline{\bf 1728}}+\chi_{\overline{\bf 351'}}+4\chi_{\overline{\bf 351}}+13\chi_{\overline{\bf 27}})\\
&\quad+(q^{\frac{9}{2}}+q^{-\frac{9}{2}})(\chi_{\overline{\bf 7371}}+4\chi_{\overline{\bf 1728}}+4\chi_{\overline{\bf 351'}}+14\chi_{\overline{\bf 351}}+35\chi_{\overline{\bf 27}})\\
&\quad+(q^{\frac{7}{2}}+q^{-\frac{7}{2}})(\chi_{\overline{\bf 17550}}+4\chi_{\overline{\bf 7371}}+13\chi_{\overline{\bf 1728}}+10\chi_{\overline
{\bf 351'}}+35\chi_{\overline{\bf 351}}+76\chi_{\overline{\bf 27}})\\
&\quad+(q^{\frac{5}{2}}+q^{-\frac{5}{2}})(\chi_{\overline{\bf 19305}}+4\chi_{\overline{\bf 17550}}+9\chi_{\overline{\bf 7371}}+25\chi_{\overline{\bf 1728}}+19\chi_{\overline{\bf 351'}}+62\chi_{\overline{\bf 351}}+122\chi_{\overline{\bf 27}})\\
&\quad+(q^{\frac{3}{2}}+q^{-\frac{3}{2}})(3\chi_{\overline{\bf 17550}}+8\chi_{\overline{\bf 7371}}+27\chi_{\overline{\bf 1728}}+19\chi_{\overline{\bf 351'}}+69\chi_{\overline{\bf 351}}+143\chi_{\overline{\bf 27}})\\
&\quad+(q^{\frac{1}{2}}+q^{-\frac{1}{2}})(\chi_{\overline{\bf 19305}}+3\chi_{\overline{\bf 17550}}+9\chi_{\overline{\bf 7371}}+27\chi_{\overline{\bf 1728}}+22\chi_{\overline{\bf 351'}}+72\chi_{\overline{\bf 351}}+149\chi_{\overline{\bf 27}}).
\end{align*}
\caption{Multi-covering components of the quantum mirror map for the $E_6$ curve.\label{e6mirrormap}}
\end{table}

After determining the $E_6$ quantum curve and the Weyl group, we can move to the study of the quantum mirror map.
We can either adopt \eqref{e6triP} in the triangular realization of figure \ref{e6tri} or \eqref{e6rectP} in the rectangular realization of figure \ref{e6rect}, both of which have their advantages and disadvantages.
For the triangular realization \eqref{e6triP}, the Schr\"odinger equation reads
\begin{align}
&\biggl[q^{-1}X^2+q^{-\frac{1}{2}}F_1X+F_2+q^{\frac{1}{2}}\frac{F_3}{X}\biggr]P[X]
+F_3G_3H_2X
+\frac{F_3G_1}{X}\nonumber\\
&\quad
+\frac{F_3G_3H_1X+q^{-\frac{1}{2}}F_3G_2}{XP[qX]}
+\frac{q^{-1}F_3G_3}{XP[qX]P[q^2X]}
+\frac{z_E}{\alpha}=0,
\label{e6triSeq}
\end{align}
($z_E=z+E$) with the leading term in the large $z_E$ expansion being
\begin{align}
P_0[X]=\frac{-z_E}{\alpha}\frac{1}{q^{-1}X^2+q^{-\frac{1}{2}}F_1X+F_2+q^{\frac{1}{2}}F_3X^{-1}}.
\end{align}
Here since the expression in the triangular realization enjoys explicitly the Weyl group of the subalgebra $(A_2)^3$, it is convenient to introduce the elementary symmetric polynomials
\begin{align}
&F_1=f_1+f_2+f_3,\quad F_2=f_1f_2+f_1f_3+f_2f_3,\quad F_3=f_1f_2f_3,\nonumber\\
&G_1=g_1+g_2+g_3,\quad G_2=g_1g_2+g_1g_3+g_2g_3,\quad G_3=g_1g_2g_3,\nonumber\\
&H_1=h_1+h_2+h_3,\quad H_2=h_1h_2+h_1h_3+h_2h_3,\quad H_3=h_1h_2h_3,
\label{e6elementary}
\end{align}
subject to the constraint \eqref{e6constraint}
\begin{align}
F_3G_3H_3=1.
\end{align}
One disadvantage for the computation is the appearance of the product $(P[qX]P[q^2X])^{-1}$ in \eqref{e6triSeq}.
When we solve for the function $P[X]$ order by order in the large $z_E$ expansion, we encounter a rather complicated nested structure.
On the other hand, for the rectangular realization \eqref{e6rectP}, since $\P^{-2}$ does not appears in the spectral operator, the complication of the nested structure does not happen.
Instead, since now we need to break the symmetry into $(A_2)^2\times A_1$, the algebra is more complicated
\begin{align}
&\Bigl[q^{-1}X^2+q^{-\frac{1}{2}}F_1X+F_2+q^{\frac{1}{2}}F_3X^{-1}\Bigr]P[X]\nonumber\\
&+[2]_qg_1X^2+(g_1F_1+F_3G_3H_2)X+(g_2+g_3)F_3X^{-1}\nonumber\\
&+\Bigl[qg_1^2X^2+q^{\frac{1}{2}}g_1F_3G_3H_2+F_3G_3H_1+q^{-\frac{1}{2}}g_1^{-1}F_3G_3X^{-1}\Bigr]\frac{1}{P[qX]}
+\frac{z_E}{\alpha}=0.
\end{align}
We find that, as a result, it is more convenient to utilize the triangular realization \eqref{e6triSeq} to compute the mirror map \eqref{PiA}.

The quantum mirror map for the $E_6$ curve obtained by computing the quantum A-period \eqref{PiA} satisfies the same group structure and the same multi-covering structure found for the $D_5$ curve.
Namely we find that, after identifying the overall parameter $\alpha$ with the combination of the parameters $(f_1,f_2,g_1,g_2,h_1,h_2)$ transforming identically as $\alpha$ as in \eqref{e6alpha} and rewrite the parameters in terms of $(t_1,t_2,t_3,t_4,t_5,t_6)$ using \eqref{e6fght}, the results are expressed by the $E_6$ characters constructed as in \eqref{e6chi78}.
Furthermore by adopting exactly the same multi-covering structure \eqref{multicover} for the $D_5$ curve \cite{FMS}, we find that the multi-covering component $\epsilon_d(q,{\bm q})$ is expressed by the $E_6$ characters $\chi_{\bf R}({\bm q})$ with non-negative integral coefficients for each degree $d$.
The results are listed in table \ref{e6mirrormap}.

\section{$E_7$ curve}\label{e7}

Next let us move to the $E_7$ case.
The non-toric expression in the rectangular realization for the $E_6$ case is helpful for our determination of the $E_7$ curve.
The $E_7$ quantum curve (in the rectangular realization) was determined previously in \cite{T,KMN} in a slightly different notation.
We repeat the analysis because the derivation here is more systematic after introducing the $q$-order and applicable directly to the $E_8$ curve in the next section.
The resulting quantum curve is, hence, simpler and also suitable for our studies of the quantum mirror map.

\subsection{Quantum curve}\label{e7qc}

The triangular realization of the $E_7$ curve is given by
\begin{align}
&\H/\alpha=q^{-1}\widehat Q^2\widehat P\nonumber\\
&\quad+q^{-\frac{1}{2}}\widehat Q((f_1+f_2+f_3+f_4)\widehat P+q^{\frac{1}{2}}(h_1^{-1}+h_2^{-1}+h_3^{-1}+h_4^{-1}))\nonumber\\
&\quad+(f_1f_2+\cdots+f_3f_4)\widehat P+E/\alpha+(h_1^{-1}h_2^{-1}+\cdots+h_3^{-1}h_4^{-1})\widehat P^{-1}\nonumber\\
&\quad+q^{\frac{1}{2}}f_1f_2f_3f_4\widehat Q^{-1}(\widehat P+q^{-\frac{1}{2}}g_1)(\widehat P+q^{-\frac{1}{2}}g_2)\nonumber\\
&\qquad\times((f_1^{-1}+f_2^{-1}+f_3^{-1}+f_4^{-1})\widehat P+q^{-\frac{1}{2}}g_1g_2(h_1+h_2+h_3+h_4))\P^{-2}\nonumber\\
&\quad+q f_1f_2f_3f_4\widehat Q^{-2}
(\widehat P+q^{-\frac{3}{2}}g_1)(\widehat P+q^{-\frac{1}{2}}g_1)
(\widehat P+q^{-\frac{3}{2}}g_2)(\widehat P+q^{-\frac{1}{2}}g_2)\widehat P^{-3},
\label{e7tricurve}
\end{align}
with the parameters subject to the constraint
\begin{align}
f_1f_2f_3f_4(g_1g_2)^2h_1h_2h_3h_4=1.
\label{e7constraint}
\end{align}
Here the abbreviation stands for $f_1f_2+\cdots+f_3f_4=\sum_{i<j}f_if_j$, $h_1^{-1}h_2^{-1}+\cdots+h_3^{-1}h_4^{-1}=\sum_{i<j}h_i^{-1}h_j^{-1}$.

\begin{figure}[!t]
\centering\includegraphics[scale=0.4,angle=-90]{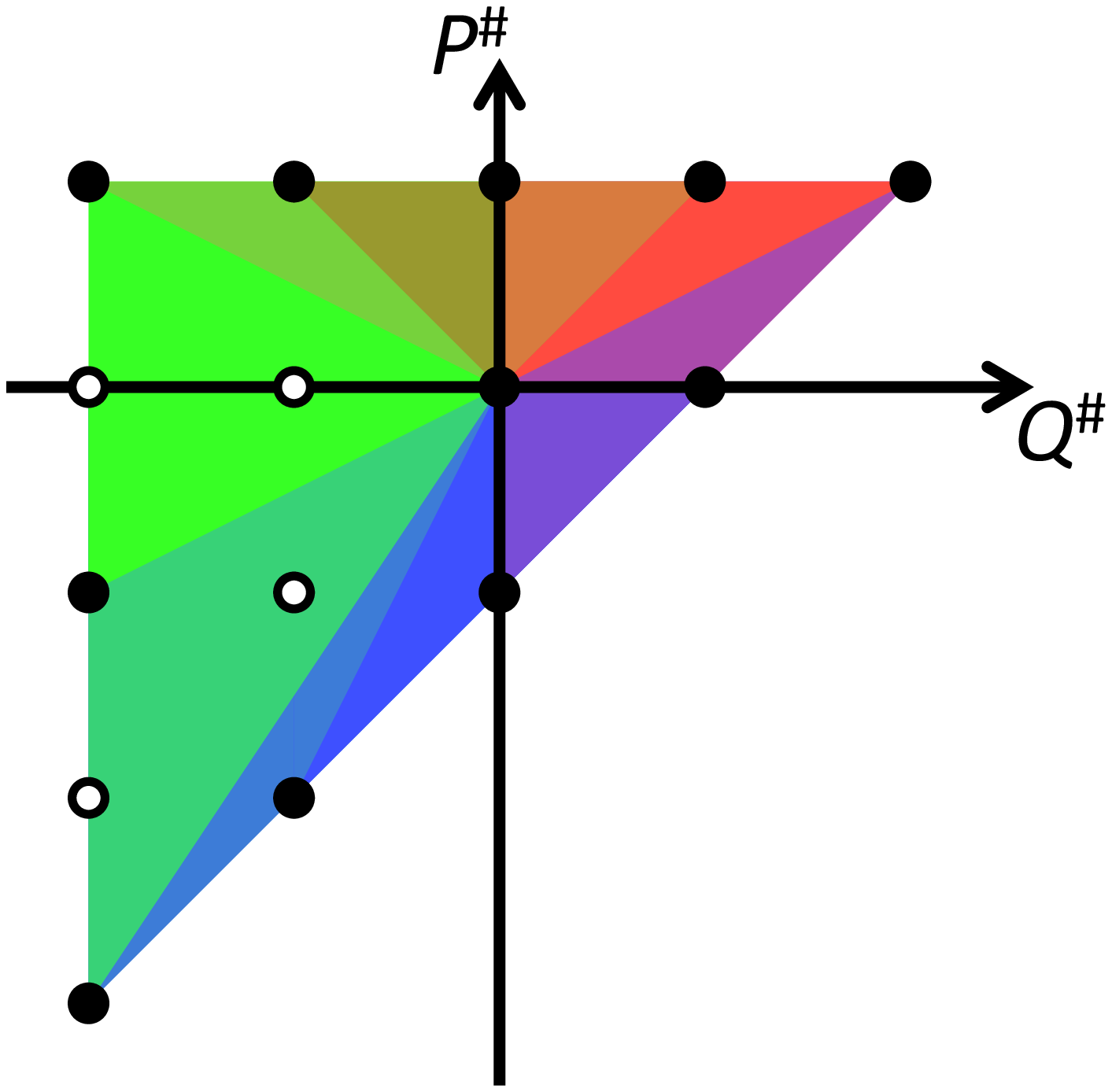}\qquad\qquad\includegraphics[scale=0.6,angle=-90]{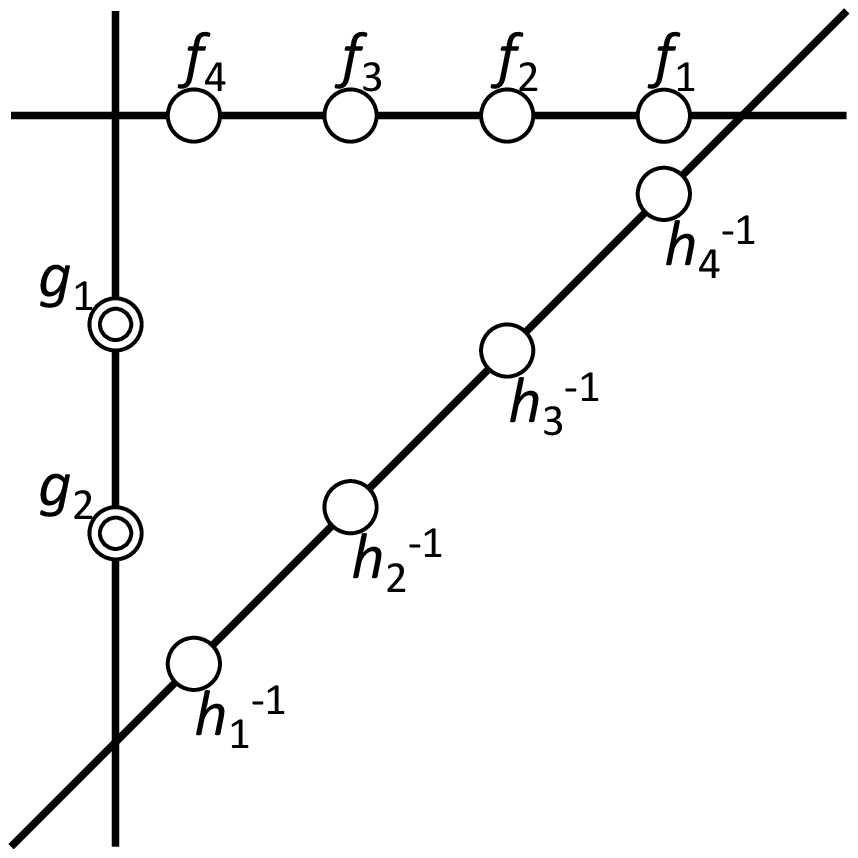}
\caption{A toric diagram (left) and a schematic diagram for asymptotic values (right) for the $E_7$ del Pezzo curve in the triangular realization.}
\label{e7tri}
\end{figure}

This is determined as follows (see figure \ref{e7tri}).
We expect that the degenerate $E_7$ curve behaves exactly in the same way as we have experienced in the degenerate $E_6$ curve with the step-by-step enhancement of powers of $q$ \eqref{quantumdegeneracy}.
For this reason, the coefficients of the terms on the boundary $\widehat Q^\alpha\widehat P (2\ge\alpha\ge -2)$, $\widehat Q^{-2}\widehat P^\beta (1\ge\beta\ge-3)$ and $\widehat Q^{\gamma+1}\widehat P^\gamma (-3\le\gamma\le 1)$ are determined from the asymptotic values of the curve by
\begin{align}
&\bigl[\Q^{-2}(\Q+f_1)(\Q+f_2)(\Q+f_3)\P\bigr]_q\nonumber\\
&\quad=q^{-1}\widehat Q^{-2}
(\widehat Q+q^{\frac{1}{2}}f_1)(\widehat Q+q^{\frac{1}{2}}f_2)(\widehat Q+q^{\frac{1}{2}}f_3)(\widehat Q+q^{\frac{1}{2}}f_4)\P,\nonumber\\
&\bigl[f_1f_2f_3f_4\Q^{-2}(\P+g_1)^2(\P+g_2)^2\P^{-3}\bigr]_q\nonumber\\
&\quad=qf_1f_2f_3f_4\widehat Q^{-2}
(\widehat P+q^{-\frac{3}{2}}g_1)(\widehat P+q^{-\frac{1}{2}}g_1)
(\widehat P+q^{-\frac{3}{2}}g_2)(\widehat P+q^{-\frac{1}{2}}g_2)\widehat P^{-3},\nonumber\\
&\bigl[\Q^{-2}(\Q\P+h_1^{-1})(\Q\P+h_2^{-1})(\Q\P+h_3^{-1})(\Q\P+h_4^{-1})\P^{-3}\bigr]_q\nonumber\\
&\quad=q^{-1}\widehat Q^2\widehat P+(h_1^{-1}+h_2^{-1}+h_3^{-1}+h_4^{-1})\widehat Q
+(h_1^{-1}h_2^{-1}+\cdots+h_3^{-1}h_4^{-1})\widehat P^{-1}\nonumber\\
&\qquad
+q^{-1}(h_1h_2h_3h_4)^{-1}(h_1+h_2+h_3+h_4)\widehat Q^{-1}\widehat P^{-2}+q^{-3}(h_1h_2h_3h_4)^{-1}\widehat Q^{-2}\widehat P^{-3}.
\end{align}
Here in the first and third equations we simply incorporate the $q$-order, while in the second equation we adopt the rule of the step-by-step enhancement of powers of $q$ for degenerate curves as in \eqref{quantumdegeneracy}.
After determining the boundary terms, the coefficients of $\widehat Q^{-1}$ and $\widehat Q^{-1}\widehat P^{-1}$ are determined by requiring that the $O(\widehat Q^{-1})$ terms have the two factors
\begin{align}
(\widehat P+q^{-\frac{1}{2}}g_1)(\widehat P+q^{-\frac{1}{2}}g_2),
\label{g1g2}
\end{align}
due to the degeneracy condition.

As explained for the $E_6$ curve, the process of determining the coefficients is clarified by the colors in figure \ref{e7tri}.
After we have determined the coefficients of the dots on the boundary, it remains to determine the coefficients of the dots $\Q^{-1}$ and $\Q^{-1}\P^{-1}$ (the two inner white dots in the regions of different colors in figure \ref{e7tri}).
For this purpose we require the $O(\widehat Q^{-1})$ terms to have the two factors \eqref{g1g2} where the two asymptotic values $g_1$ and $g_2$ correspond to the two different colors in figure \ref{e7tri}.
As we find later, this observation would be helpful when we proceed to the $E_8$ curve.

In the triangular realization, we refer to the asymptotic values
\begin{align}
&(-f_1,\infty),(-f_2,\infty),(-f_3,\infty),(-f_4,\infty),(0,-g_1)\text{[double]},(0,-g_2)\text{[double]},\nonumber\\
&\quad(\infty,-h_1^{-1}/\infty),(\infty,-h_2^{-1}/\infty),(\infty,-h_3^{-1}/\infty),(\infty,-h_4^{-1}/\infty),
\end{align}
as
\begin{align}
\bigl\{f_1,f_2,f_3,f_4;g_1,g_2\text{[double]};h_1^{-1},h_2^{-1},h_3^{-1},h_4^{-1}\bigr\}_\text{tri}.
\label{e7asymptoticvalues}
\end{align}
As in the $E_6$ case, clearly the curve enjoys the symmetries of exchanging two asymptotic values in each of the three subsets $\{f_1,f_2,f_3,f_4\}$, $\{g_1,g_2\}$ and $\{h_1^{-1},h_2^{-1},h_3^{-1},h_4^{-1}\}$,
\begin{align}
&s_0: f_4\leftrightarrow f_3,\quad s_1:f_3\leftrightarrow f_2,\quad s_2:f_2\leftrightarrow f_1,\nonumber\\
&s_4:g_2\leftrightarrow g_1,\nonumber\\
&s_5: h_1\leftrightarrow h_2,\quad s_6:h_2\leftrightarrow h_3,\quad s_7:h_3\leftrightarrow h_4.
\label{s0124567}
\end{align}

\begin{figure}[!t]
\centering\includegraphics[scale=0.4,angle=-90]{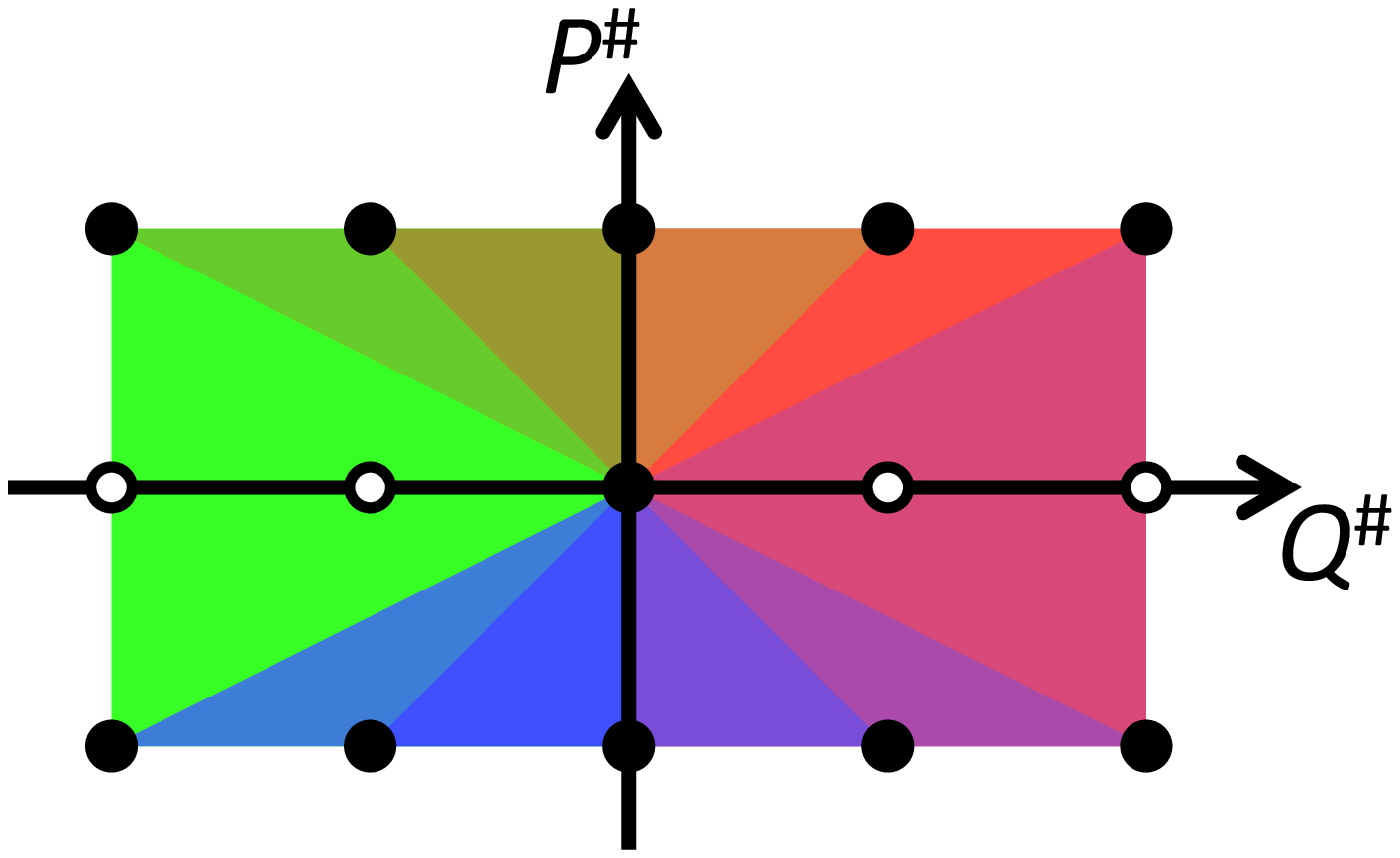}\qquad\qquad\includegraphics[scale=0.6,angle=-90]{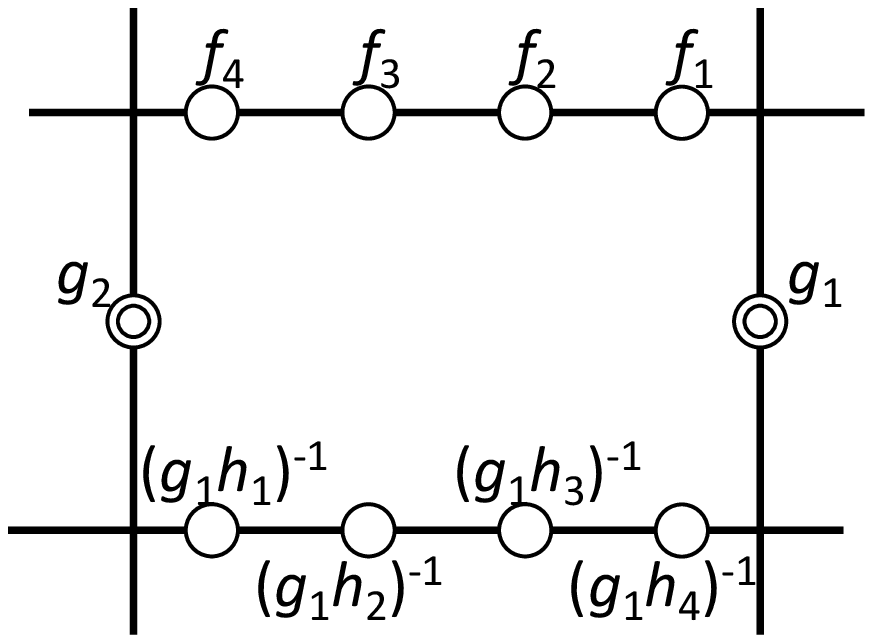}
\caption{A toric diagram (left) and a schematic diagram for asymptotic values (right) for the $E_7$ del Pezzo curve in the rectangular realization.}
\label{e7rect}
\end{figure}

We can then apply exactly the same transformation \eqref{trirect} along with \eqref{trirect+} (constructed originally for the $E_6$ curve) in moving to the rectangular realization
\begin{align}
&\H/\alpha=q^{-1}\widehat Q^2(\widehat P+q^{\frac{3}{2}}g_1)(\widehat P+q^{\frac{1}{2}}g_1)\widehat P^{-1}\nonumber\\
&\;+q^{-\frac{1}{2}}\widehat Q(\widehat P+q^{\frac{1}{2}}g_1)
((f_1+f_2+f_3+f_4)\widehat P+q^{\frac{1}{2}}(h_1^{-1}+h_2^{-1}+h_3^{-1}+h_4^{-1}))\widehat P^{-1}\nonumber\\
&\;+(f_1f_2+\cdots+f_3f_4)\widehat P+E/\alpha+(h_1^{-1}h_2^{-1}+\cdots+h_3^{-1}h_4^{-1})\widehat P^{-1}\nonumber\\
&\;+q^{\frac{1}{2}}f_1f_2f_3f_4\widehat Q^{-1}(\widehat P+q^{-\frac{1}{2}}g_2)
((f_1^{-1}+f_2^{-1}+f_3^{-1}+f_4^{-1})\widehat P+q^{-\frac{1}{2}}g_1g_2(h_1+h_2+h_3+h_4))\widehat P^{-1}\nonumber\\
&\;+qf_1f_2f_3f_4\widehat Q^{-2}(\widehat P+q^{-\frac{3}{2}}g_2)(\widehat P+q^{-\frac{1}{2}}g_2)\widehat P^{-1},
\label{e7rectcurve}
\end{align}
(see figure \ref{e7rect}).
Here we refer to the asymptotic values
\begin{align}
&(\infty,-g_1)\text{[double]},(-f_1,\infty),(-f_2,\infty),(-f_3,\infty),(-f_4,\infty),\nonumber\\
&\quad(0,-g_2)\text{[double]},
(-(g_1h_1)^{-1},0),(-(g_1h_2)^{-1},0),(-(g_1h_3)^{-1},0),(-(g_1h_4)^{-1},0),
\end{align}
as
\begin{align}
\bigl\{g_1\text{[double]};f_1,f_2,f_3,f_4;g_2\text{[double]};(g_1h_1)^{-1},(g_1h_2)^{-1},(g_1h_3)^{-1},(g_1h_4)^{-1}\bigr\}_\text{rect}.
\label{e7double}
\end{align}
Then, as in the $E_6$ case, we can apply the similarity transformation \eqref{s3similarity} along with \eqref{s3similarity+} (again constructed originally for the $E_6$ curve).
This transformation takes the asymptotic values \eqref{e7double} into
\begin{align}
\bigl\{g_1\text{[double]};(g_1h_1)^{-1},f_2,f_3,f_4;(f_1g_1h_1)g_2\text{[double]};f_1,(g_1h_2)^{-1},(g_1h_3)^{-1},(g_1h_4)^{-1}\bigr\}_\text{rect},
\label{e7s3}
\end{align}
and generates the $s_3$ transformation for the $E_7$ curve.
Note that all the other coefficients are also invariant under the $s_3$ transformation.

\subsection{Weyl group}\label{e7wg}

\begin{figure}[!t]
\centering\includegraphics[scale=0.4,angle=-90]{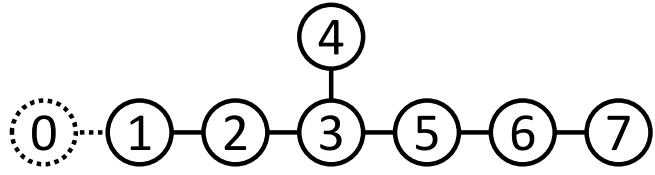}
\caption{Dynkin diagram of the $E_7$ algebra.}
\label{e7dynkin}
\end{figure}

Before proceeding to the analysis of the mirror map in the next subsection, we need to clarify the symmetries of the curve.
After fixing the gauge,
\begin{align}
f_4=g_2=1,\quad h_4=(f_1f_2f_3g_1^2h_1h_2h_3)^{-1},
\label{e7gf}
\end{align}
the transformations \eqref{s0124567} and \eqref{e7s3} are given by
\begin{align}
s_1&:(f_1,f_2,f_3,g_1,h_1,h_2,h_3;\alpha)
\mapsto(f_1,f_3,f_2,g_1,h_1,h_2,h_3;\alpha),\nonumber\\
s_2&:(f_1,f_2,f_3,g_1,h_1,h_2,h_3;\alpha)
\mapsto(f_2,f_1,f_3,g_1,h_1,h_2,h_3;\alpha),\nonumber\\
s_3&:(f_1,f_2,f_3,g_1,h_1,h_2,h_3;\alpha)\nonumber\\
&\hspace{10mm}\mapsto((g_1h_1)^{-1},f_2,f_3,(f_1h_1)^{-1},h_1,(f_1g_1h_1)h_2,(f_1g_1h_1)h_3;(f_1g_1h_1)\alpha),\nonumber\\
s_4&:(f_1,f_2,f_3,g_1,h_1,h_2,h_3;\alpha)
\mapsto(f_1,f_2,f_3,g_1^{-1},g_1h_1,g_1h_2,g_1h_3;g_1\alpha),\nonumber\\
s_5&:(f_1,f_2,f_3,g_1,h_1,h_2,h_3;\alpha)
\mapsto(f_1,f_2,f_3,g_1,h_2,h_1,h_3;\alpha),\nonumber\\
s_6&:(f_1,f_2,f_3,g_1,h_1,h_2,h_3;\alpha)
\mapsto(f_1,f_2,f_3,g_1,h_1,h_3,h_2;\alpha),\nonumber\\
s_7&:(f_1,f_2,f_3,g_1,h_1,h_2,h_3;\alpha)
\mapsto(f_1,f_2,f_3,g_1,h_1,h_2,(f_1f_2f_3g_1^2h_1h_2h_3)^{-1};\alpha),\nonumber\\
s_0&:(f_1,f_2,f_3,g_1,h_1,h_2,h_3;\alpha)
\mapsto(f_3^{-1}f_1,f_3^{-1}f_2,f_3^{-1},g_1,f_3h_1,f_3h_2,f_3h_3;f_3^2\alpha).
\end{align}
From the commutation relations of the transformations $s_i$, we reproduce the $E_7$ Dynkin diagram with the numbering given in figure \ref{e7dynkin}.
The combination of the parameters which transforms identically as $\alpha$ is $(f_1f_2f_3g_1)^{-\frac{1}{2}}$ and we shall identify them when expressing the mirror map in terms of characters,
\begin{align}
\alpha=(f_1f_2f_3g_1)^{-\frac{1}{2}}.
\label{e7alpha}
\end{align}
From the transformations, as in the $D_5$ case \cite{KM,FMS} or the previous $E_6$ case \eqref{e6srfw}, we can identify the simple roots and the fundamental weights as
\begin{align}
\alpha_1&=(0,-1,1,0,0,0,0),&\omega_1&=(1,1,2,0,-1,-1,-1),\nonumber\\
\alpha_2&=(-1,1,0,0,0,0,0),&\omega_2&=(2,3,3,0,-2,-2,-2),\nonumber\\
\alpha_3&=(1,0,0,1,0,-1,-1),&\omega_3&=(4,4,4,0,-3,-3,-3),\nonumber\\
\alpha_4&=(0,0,0,-2,1,1,1),&\omega_4&=(2,2,2,-1,-1,-1,-1),\nonumber\\
\alpha_5&=(0,0,0,0,-1,1,0),&\omega_5&=(3,3,3,0,-3,-2,-2),\nonumber\\
\alpha_6&=(0,0,0,0,0,-1,1),&\omega_6&=(2,2,2,0,-2,-2,-1),\nonumber\\
\alpha_7&=(0,0,0,0,0,0,-1),&\omega_7&=(1,1,1,0,-1,-1,-1),
\label{e7srfw}
\end{align}
in the linear space of $(\log f_1,\log f_2,\log f_3,\log g_1,\log h_1,\log h_2,\log h_3)$ spanned by the parameters of the curve.

In identifying the mirror map for the $E_7$ curve as characters later, we need to compare with the $E_7$ characters constructed from the orthogonal basis, where the simple roots and the fundamental weights are chosen to be
\begin{align}
\alpha^\perp_1&=\textstyle{(-1,1,0,0,0,0,0)},&
\omega^\perp_1&=\textstyle{(-1,0,0,0,0,-\frac{\sqrt{3}}{3},-\frac{\sqrt{6}}{3})},\nonumber\\
\alpha^\perp_2&=\textstyle{(0,-1,1,0,0,0,0)},&
\omega^\perp_2&=\textstyle{(-1,-1,0,0,0,-\frac{2\sqrt{3}}{3},-\frac{2\sqrt{6}}{3})},\nonumber\\
\alpha^\perp_3&=\textstyle{(0,0,-1,1,0,0,0)},&
\omega^\perp_3&=\textstyle{(-1,-1,-1,0,0,-\sqrt{3},-\sqrt{6})},\nonumber\\
\alpha^\perp_4&=\textstyle{(0,0,0,-1,1,0,0)},&
\omega^\perp_4&=\textstyle{(-\frac{1}{2},-\frac{1}{2},-\frac{1}{2},-\frac{1}{2},\frac{1}{2},-\frac{\sqrt{3}}{2},-\frac{\sqrt{6}}{2})},\nonumber\\
\alpha^\perp_5&=\textstyle{(0,0,0,-1,-1,0,0)},&
\omega^\perp_5&=\textstyle{(-\frac{1}{2},-\frac{1}{2},-\frac{1}{2},-\frac{1}{2},-\frac{1}{2},-\frac{5\sqrt{3}}{6},-\frac{5\sqrt{6}}{6})},\nonumber\\
\alpha^\perp_6&=\textstyle{(\frac{1}{2},\frac{1}{2},\frac{1}{2},\frac{1}{2},\frac{1}{2},-\frac{\sqrt{3}}{2},0)},&
\omega^\perp_6&=\textstyle{(0,0,0,0,0,-\frac{2\sqrt{3}}{3},-\frac{2\sqrt{6}}{3})},\nonumber\\
\alpha^\perp_7&=\textstyle{(0,0,0,0,0,\frac{2\sqrt{3}}{3},-\frac{\sqrt{6}}{3})},&
\omega^\perp_7&=\textstyle{(0,0,0,0,0,0,-\frac{\sqrt{6}}{2})}.
\label{e7ortho}
\end{align}
We can then relate the fundamental weights obtained from the parameters of the $E_7$ curve \eqref{e7srfw} with those in the orthogonal basis \eqref{e7ortho} as
\begin{align}
&(f_1,f_2,f_3,g_1,h_1,h_2,h_3)\nonumber\\
&=f_1^{(1,0,0,0,0,0,0)}f_2^{(0,1,0,0,0,0,0)}f_3^{(0,0,1,0,0,0,0)}g_1^{(0,0,0,1,0,0,0)}h_1^{(0,0,0,0,1,0,0)}h_2^{(0,0,0,0,0,1,0)}h_3^{(0,0,0,0,0,0,1)}\nonumber\\
&=f_1^{-\omega_2+\omega_3-\omega_7}f_2^{-\omega_1+\omega_2-\omega_7}f_3^{\omega_1-\omega_7}
g_1^{\omega_3-\omega_4-2\omega_7}
h_1^{\omega_3-\omega_5-\omega_7}h_2^{\omega_5-\omega_6-\omega_7}h_3^{\omega_6-2\omega_7}\nonumber\\
&\leftrightarrow
f_1^{-\omega^\perp_2+\omega^\perp_3-\omega^\perp_7}
f_2^{-\omega^\perp_1+\omega^\perp_2-\omega^\perp_7}
f_3^{\omega^\perp_1-\omega^\perp_7}
g_1^{\omega^\perp_3-\omega^\perp_4-2\omega^\perp_7}
h_1^{\omega^\perp_3-\omega^\perp_5-\omega^\perp_7}
h_2^{\omega^\perp_5-\omega^\perp_6-\omega^\perp_7}
h_3^{\omega^\perp_6-2\omega^\perp_7}\nonumber\\
&=f_1^{(0,0,-1,0,0,-\frac{\sqrt{3}}{3},\frac{\sqrt{6}}{6})}
f_2^{(0,-1,0,0,0,-\frac{\sqrt{3}}{3},\frac{\sqrt{6}}{6})}
f_3^{(-1,0,0,0,0,-\frac{\sqrt{3}}{3},\frac{\sqrt{6}}{6})}
g_1^{(-\frac{1}{2},-\frac{1}{2},-\frac{1}{2},\frac{1}{2},-\frac{1}{2},-\frac{\sqrt{3}}{2},\frac{\sqrt{6}}{2})}\nonumber\\
&\quad\times
h_1^{(-\frac{1}{2},-\frac{1}{2},-\frac{1}{2},\frac{1}{2},\frac{1}{2},-\frac{\sqrt{3}}{6},\frac{\sqrt{6}}{3})}
h_2^{(-\frac{1}{2},-\frac{1}{2},-\frac{1}{2},-\frac{1}{2},-\frac{1}{2},-\frac{\sqrt{3}}{6},\frac{\sqrt{6}}{3})}
h_3^{(0,0,0,0,0,-\frac{2\sqrt{3}}{3},\frac{\sqrt{6}}{3})}\nonumber\\
&=\bigl((f_3^2g_1h_1h_2)^{-\frac{1}{2}},(f_2^2g_1h_1h_2)^{-\frac{1}{2}},(f_1^2g_1h_1h_2)^{-\frac{1}{2}},
(g_1h_1h_2^{-1})^{\frac{1}{2}},(g_1^{-1}h_1h_2^{-1})^{\frac{1}{2}},\nonumber\\
&\qquad(f_1^2f_2^2f_3^2g_1^3h_1h_2h_3^4)^{-\frac{\sqrt{3}}{6}},
(f_1f_2f_3g_1^3h_1^2h_2^2h_3^2)^{\frac{\sqrt{6}}{6}}\bigr).
\end{align}
If we identify this set of variables as
\begin{align}
{\bm q}=(q_1,q_2,q_3,q_4,q_5,q_6,q_7)=(t_1^2,t_2^2,t_3^2,t_4^2,t_5^2,t_6^{-2\sqrt{3}},t_7^{-\sqrt{6}}),
\label{e7t}
\end{align}
we can solve the relations reversely and find
\begin{align}
f_1=\frac{t_6^2t_7^2}{t_3^2},\quad
f_2=\frac{t_6^2t_7^2}{t_2^2},\quad
f_3=\frac{t_6^2t_7^2}{t_1^2},\quad
g_1=\frac{t_4^2}{t_5^2},\quad
h_1=\frac{t_5^2}{t_6^2t_7^2},\quad
h_2=\frac{1}{t_4^2t_6^2t_7^2},\quad
h_3=\frac{t_1t_2t_3t_5t_6}{t_4t_7^2}.
\label{e7fght}
\end{align}
We shall use this identification in the next subsection to express the results of the mirror map in terms of the $E_7$ characters.

\subsection{Mirror map}\label{e7mirror}

\begin{table}[t!]
\begin{align*}
\epsilon_1&=0,\\
\epsilon_2&=(q+q^{-1})\chi_{\bf 1}+\chi_{\bf 133}+3\chi_{\bf 1},\\
\epsilon_3&=(q^{\frac{3}{2}}+q^{-\frac{3}{2}})\chi_{\bf 56}
+(q^{\frac{1}{2}}+q^{-\frac{1}{2}})(\chi_{\bf 912}+3\chi_{\bf 56}),\\
\epsilon_4&=(q^4+q^{-4})\chi_{\bf 1}
+(q^3+q^{-3})(\chi_{\bf 133}+3\chi_{\bf 1})
+(q^2+q^{-2})(\chi_{\bf 1539}+4\chi_{\bf 133}+9\chi_{\bf 1})\\
&\quad
+(q+q^{-1})(\chi_{\bf 8645}+3\chi_{\bf 1539}+9\chi_{\bf 133}+13\chi_{\bf 1})
+(4\chi_{\bf 1539}+8\chi_{\bf 133}+16\chi_{\bf 1}),\\
\epsilon_5&=(q^{\frac{11}{2}}+q^{-\frac{11}{2}})\chi_{\bf 56}
+(q^{\frac{9}{2}}+q^{-\frac{9}{2}})(\chi_{\bf 912}+4\chi_{\bf 56})
+(q^{\frac{7}{2}}+q^{-\frac{7}{2}})(\chi_{\bf 6480}+4\chi_{\bf 912}+13\chi_{\bf 56})\\
&\quad
+(q^{\frac{5}{2}}+q^{-\frac{5}{2}})(\chi_{\bf 27664}+4\chi_{\bf 6480}+13\chi_{\bf 912}+31\chi_{\bf 56})\\
&\quad
+(q^{\frac{3}{2}}+q^{-\frac{3}{2}})(\chi_{\bf 86184}+3\chi_{\bf 27664}+9\chi_{\bf 6480}+25\chi_{\bf 912}+51\chi_{\bf 56})\\
&\quad
+(q^{\frac{1}{2}}+q^{-\frac{1}{2}})(3\chi_{\bf 27664}+9\chi_{\bf 6480}+27\chi_{\bf 912}+57\chi_{\bf 56}).
\end{align*}
\caption{Multi-covering components of the quantum mirror map for the $E_7$ curve.\label{e7mirrormap}}
\end{table}

As in the previous case of the $E_6$ curve, let us consider the associated Schr\"odinger equation for the $E_7$ curve in the triangular realization \eqref{e7tricurve}
\begin{align}
&\Bigl[q^{-1}X^2+q^{-\frac{1}{2}}F_1X+F_2+q^{\frac{1}{2}}F_3X^{-1}+qF_4X^{-2}\Bigr]P[X]\nonumber\\
&+F_4G_2^2H_3X+(F_3G_1+F_4G_2H_1)X^{-1}+(q^{\frac{1}{2}}+q^{-\frac{1}{2}})F_4G_1X^{-2}\nonumber\\
&+\Bigl[F_4G_2^2H_2+q^{-\frac{1}{2}}(F_3G_2+F_4G_2G_1H_1)X^{-1}+q^{-1}F_4(G_1^2+(q+q^{-1})G_2)X^{-2}\Bigr]
\frac{1}{P[qX]}\nonumber\\
&+\Bigl[q^{-1}F_4G_2^2H_1X^{-1}+q^{-2}(q^{\frac{1}{2}}+q^{-\frac{1}{2}})F_4G_2G_1X^{-2}\Bigr]\frac{1}{P[qX]P[q^2X]}\nonumber\\
&+q^{-3}F_4G_2^2X^{-2}\frac{1}{P[qX]P[q^2X]P[q^3X]}+\frac{z_E}{\alpha}=0,
\label{e7schrodinger}
\end{align}
where the elementary symmetric polynomials are given by
\begin{align}
&F_1=\sum_{i}f_i,\quad F_2=\sum_{i<j}f_if_j,\quad F_3=\sum_{i<j<k}f_if_jf_k,\quad F_4=f_1f_2f_3f_4,\nonumber\\
&G_1=g_1+g_2,\quad G_2=g_1g_2,\nonumber\\
&H_1=\sum_{i}h_i,\quad H_2=\sum_{i<j}h_ih_j,\quad H_3=\sum_{i<j<k}h_ih_jh_k,\quad H_4=h_1h_2h_3h_4,
\label{e7elemetary}
\end{align}
subject to the constraint $F_4G_2^2H_4=1$ \eqref{e7constraint}.
Then, the large $z_E$ expansion of $P[X]$ is given by
\begin{align}
P_0[X]=\frac{-z_E}{\alpha}\frac{1}{q^{-1}X^2+q^{-\frac{1}{2}}F_1X+F_2+q^{\frac{1}{2}}F_3X^{-1}+qF_4X^{-2}}.
\end{align}
Technically for solving for $P[X]$ the simplest way is probably to first solve the Schr\"odinger equation with general coefficients
\begin{align}
&\sum_{n=-2}^2c_{n,1}q^{-\frac{n}{2}}X^nP[X]+\sum_{n=-2}^1c_{n,0}X^n\nonumber\\
&+\sum_{n=-2}^0\frac{c_{n,-1}q^{\frac{n}{2}}X^n}{P[qX]}+\sum_{n=-2}^{-1}\frac{c_{n,-2}q^nX^n}{P[qX]P[q^2X]}+\frac{c_{-2,-3}q^{-3}X^{-2}}{P[qX]P[q^2X]P[q^3X]}+\frac{z_E}{\alpha}=0,
\end{align}
and then specify the coefficients to be those in \eqref{e7schrodinger} afterwards.
The analysis is parallel to the previous $E_6$ case.
We find again that the mirror map enjoys the group structure and the multi-covering structure.
Namely after picking up the residue for the A-period and substituting the identification of $\alpha$ \eqref{e7alpha} and the change of parameters \eqref{e7fght}, the results are given by the $E_7$ characters.
After solving reversely and assuming the same multi-covering structure as in \eqref{multicover}, we find out the multi-covering components with non-negative integral coefficients.
We list the results in table \ref{e7mirrormap}.
It is interesting to note that, as in the $D_5$ case \cite{FMS}, by comparing with the table of the BPS indices (see table 5 in \cite{MNY}), the representations appearing in the mirror map match exactly with those appearing in the BPS indices except for the trivial case of degree one.

\section{$E_8$ curve}\label{e8}

Equipped with all the techniques and the observations obtained so far such as the introduction of the $q$-order, the coloring of the toric diagram and the appearance of the $q$-integer for the degenerate curves, we can now proceed to the study of the $E_8$ case.
To the best of our knowledge, the $E_8$ quantum curve was not given previously.
After writing down the $E_8$ quantum curve we can proceed to study the mirror maps for the $E_8$ curve.

\subsection{Quantum curve}\label{e8qc}

The $E_8$ quantum curve is given as follows.
In the triangular realization, it is given by
\begin{align}
&\H/\alpha=q^{-3}\Q^3\P^2\nonumber\\
&\;+q^{-2}\Q^2\P\bigl\{[2]_qF_1\P+qF_3^2G_2^3H_5\bigr\}\nonumber\\
&\;+q^{-1}\Q\bigl\{\bigl(([3]_q-1)F_2+F_1^2\bigr)\P^2+q^{\frac{1}{2}}\bigl(F_3^2F_1G_2^3H_5+F_2G_1+F_3G_2H_1\bigr)\P+qF_3^2G_2^3H_4\bigr\}\nonumber\\
&\;+\bigl([2]_qF_2F_1+([4]_q-[2]_q)F_3\bigr)\P^2+\bigl(F_3^2F_2G_2^3H_5+[3]_qF_3G_1+F_2F_1G_1+F_3F_1G_2H_1\bigr)\P
\nonumber\\
&\;\quad+E/\alpha+F_3^2G_2^3H_3\P^{-1}
\nonumber\\
&\;+q\Q^{-1}\P^{-2}(\P+q^{-\frac{1}{2}}g_1)(\P+q^{-\frac{1}{2}}g_2)\nonumber\\
&\quad\times\bigl\{\bigl(([3]_q-1)F_3F_1+F_2^2\bigr)\P^2
+q^{-\frac{1}{2}}\bigl(F_3^3G_2^3H_5+F_3F_1G_1+F_3F_2G_2H_1\bigr)\P+q^{-1}F_3^2G_2^2H_2\bigr\}\nonumber\\
&\;+q^2\Q^{-2}\P^{-3}(\P+q^{-\frac{3}{2}}g_1)(\P+q^{-\frac{1}{2}}g_1)(\P+q^{-\frac{3}{2}}g_2)(\P+q^{-\frac{1}{2}}g_2)
\bigl\{[2]_qF_3F_2\P+q^{-1}F_3^2G_2H_1\bigr\}\nonumber\\
&\;+q^3\Q^{-3}\P^{-4}(\P+q^{-\frac{5}{2}}g_1)(\P+q^{-\frac{3}{2}}g_1)(\P+q^{-\frac{1}{2}}g_1)
(\P+q^{-\frac{5}{2}}g_2)(\P+q^{-\frac{3}{2}}g_2)(\P+q^{-\frac{1}{2}}g_2)F_3^2.
\label{e8trieq}
\end{align}
Here we have introduced the elementary symmetric polynomials
\begin{align}
&F_1=f_1+f_2+f_3,\quad F_2=f_1f_2+f_1f_3+f_2f_3,\quad F_3=f_1f_2f_3,\quad G_1=g_1+g_2,\quad G_2=g_1g_2,\nonumber\\
&H_1=h_1+\cdots+h_6,\quad H_2=h_1h_2+\cdots+h_5h_6,\quad\cdots,\quad H_6=h_1\cdots h_6,
\label{e8elemetary}
\end{align}
from the beginning, with the variables subject to the constraint
\begin{align}
F_3^2G_2^3H_6=1.
\end{align}

\begin{figure}[!t]
\centering\includegraphics[scale=0.4,angle=-90]{e8tri.eps}\qquad\qquad\includegraphics[scale=0.6,angle=-90]{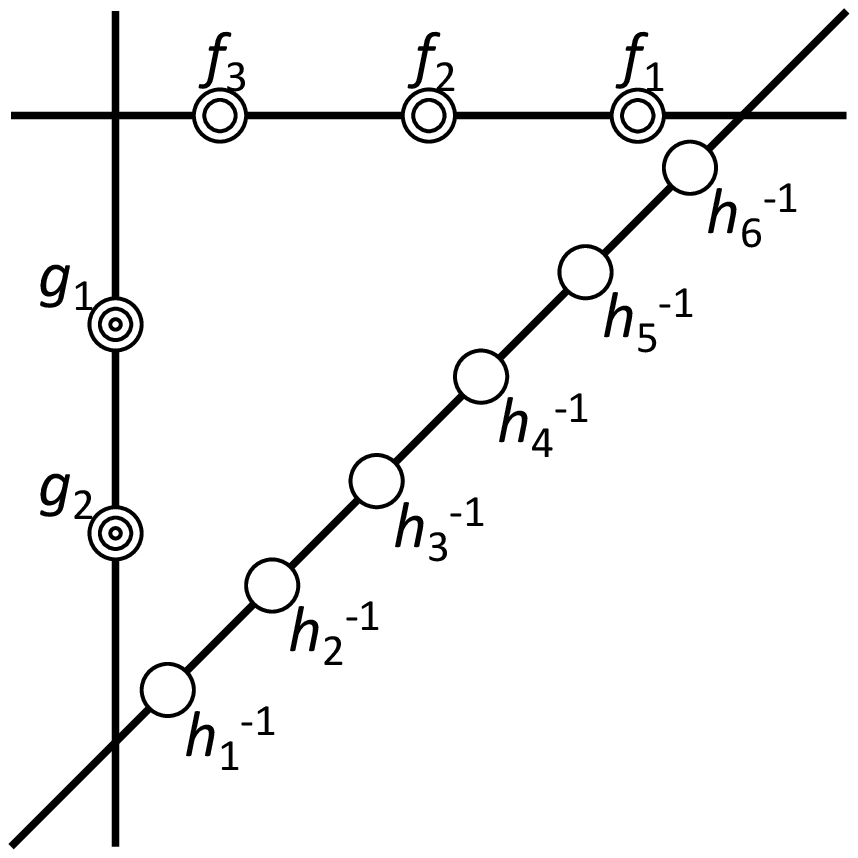}
\caption{A toric diagram (left) and a schematic diagram for asymptotic values (right) for the $E_8$ del Pezzo curve in the triangular realization.}
\label{e8tri}
\end{figure}

The expression is determined by requiring the curve to have asymptotic values
\begin{align}
\bigl\{f_1,f_2,f_3\text{[double]};g_1,g_2\text{[triple]};h_1^{-1},h_2^{-1},h_3^{-1},h_4^{-1},h_5^{-1},h_6^{-1}\bigr\}_\text{tri},
\label{e8asymptoticvalues}
\end{align}
where the notation is the same as \eqref{e7asymptoticvalues} (see figure \ref{e8tri}).
Namely, we require that the coefficients of the boundary terms $\Q^\alpha\P^2 (3\ge\alpha\ge-3)$, $\Q^{-3}\P^\beta (2\ge\beta\ge-4)$ and $\Q^{\gamma+1}\P^\gamma (-4\le\gamma\le 2)$ are given by
\begin{align}
&\bigl[\Q^{-3}(\Q+f_1)^2(\Q+f_2)^2(\Q+f_3)^2\P^2\bigr]_q\nonumber\\
&\quad=q^{-3}\Q^{-3}(\Q+q^{\frac{3}{2}}f_1)(\Q+q^{\frac{1}{2}}f_1)(\Q+q^{\frac{3}{2}}f_2)(\Q+q^{\frac{1}{2}}f_2)(\Q+q^{\frac{3}{2}}f_3)(\Q+q^{\frac{1}{2}}f_3)\P^2,\nonumber\\
&\bigl[(f_1f_2f_3)^2\Q^{-3}(\P+g_1)^3(\P+g_2)^3\P^{-4}\bigr]_q\nonumber\\
&\quad=q^3F_3^2\Q^{-3}(\P+q^{-\frac{5}{2}}g_1)(\P+q^{-\frac{3}{2}}g_1)(\P+q^{-\frac{1}{2}}g_1)
(\P+q^{-\frac{5}{2}}g_2)(\P+q^{-\frac{3}{2}}g_2)(\P+q^{-\frac{1}{2}}g_2)\P^{-4},\nonumber\\
&\bigl[\Q^{-3}(\Q\P+h_1^{-1})(\Q\P+h_2^{-1})(\Q\P+h_3^{-1})(\Q\P+h_4^{-1})(\Q\P+h_5^{-1})(\Q\P+h_6^{-1})\P^{-4}\bigr]_q\nonumber\\
&\quad=H_6^{-1}(q^{-3}H_6\Q^3\P^2+q^{-1}H_5\Q^2\P+H_4\Q\nonumber\\
&\qquad\qquad+H_3\P^{-1}+q^{-1}H_2\Q^{-1}\P^{-2}+q^{-3}H_1\Q^{-2}\P^{-3}+q^{-6}\Q^{-3}\P^{-4}).
\end{align}
After determining the coefficients of $\Q^{-2}\P^2$ and $\Q^{-2}\P^{-3}$, since there are exactly four terms $\Q^{-2}\P^\beta (1\ge\beta\ge-2)$ whose coefficients remains to be determined, we can proceed to fixing the coefficients of the $O(\Q^{-2})$ terms by requiring the expression to contain the four factors
\begin{align}
(\P+q^{-\frac{3}{2}}g_1)(\P+q^{-\frac{1}{2}}g_1)(\P+q^{-\frac{3}{2}}g_2)(\P+q^{-\frac{1}{2}}g_2).
\label{Pg1Pg2}
\end{align}
If we try to proceed to the coefficients of the $O(\Q^{-1})$ terms by requiring the expression to contain the two factors $(\P+q^{-\frac{1}{2}}g_1)(\P+q^{-\frac{1}{2}}g_2)$, we find that the requiremnt is not enough since we have three unknown coefficients of $\Q^{-1}\P$, $\Q^{-1}$ and $\Q^{-1}\P^{-1}$.
Instead, we can first fix the coefficients of the $O(\P)$ terms by requiring the expression to contain the three factors, $(\Q+q^{\frac{1}{2}}f_1)(\Q+q^{\frac{1}{2}}f_2)(\Q+q^{\frac{1}{2}}f_3)$, since the coefficients of $\Q^2\P$, $\Q^{-2}\P$ and $\Q^{-3}\P$ have already been fixed and there are only three coefficients of $\Q\P$, $\P$ and $\Q^{-1}\P$ remaining.
Using the result we can now return to fix the coefficients of the $O(Q^{-1})$ terms by requiring the expression to contain the two factors $(\P+q^{\frac{1}{2}}g_1)(\P+q^{\frac{1}{2}}g_2)$ since now the coefficients of $\Q^{-1}\P^2$, $\Q^{-1}\P^{-2}$ and especially $\Q^{-1}\P$ have already been fixed and there are only two unknown coefficients of $\Q^{-1}$ and $\Q^{-1}\P^{-1}$ remaining.

As noticed around \eqref{g1g2} for the $E_7$ case, the process of determining the coefficient is more transparent from the viewpoint of the toric diagrams with colors assigned (see again figure \ref{e8tri}).
Namely, we consider the white dots in the region of the same color to be taken care of by the same asymptotic value.
After determining terms on the boundary, we have proceeded to determine the $O(\Q^{-2})$ terms from the four factors \eqref{Pg1Pg2}.
Here the four missing coefficients are located in the two regions of different colors for $g_1$ and $g_2$ with multiplicity $2$.
It is, however, not possible to proceed to the $O(\Q^{-1})$ terms since there is one white dot $\Q^{-1}\P$ located in the regions of the colors for $f_1,f_2,f_3$ which should be determined from $f_1,f_2,f_3$.
Hence we move to the $O(\P)$ terms first where the unknown coefficients are all determined from $f_1,f_2,f_3$ and then return to the $O(\Q^{-1})$ terms.
Our coloring of the toric diagram has largely clarified the process of determining the coefficients from the asymptotic values.

The expression of the $E_8$ curve in the triangular realization clearly enjoys the symmetries of exchanging two asymptotic values in each of the three sets $\{f_1,f_2,f_3\}$, $\{g_1,g_2\}$ and $\{h_1^{-1},h_2^{-1},\cdots,h_6^{-1}\}$,
\begin{align}
&s_1:f_3\leftrightarrow f_2,\quad s_2:f_2\leftrightarrow f_1,\nonumber\\
&s_4:g_2\leftrightarrow g_1,\nonumber\\
&s_5:h_1\leftrightarrow h_2,\quad s_6:h_2\leftrightarrow h_3,\quad s_7:h_3\leftrightarrow h_4,\quad s_8:h_4\leftrightarrow h_5,\quad
s_0:h_5\leftrightarrow h_6.
\end{align}

\begin{figure}[!t]
\centering\includegraphics[scale=0.4,angle=-90]{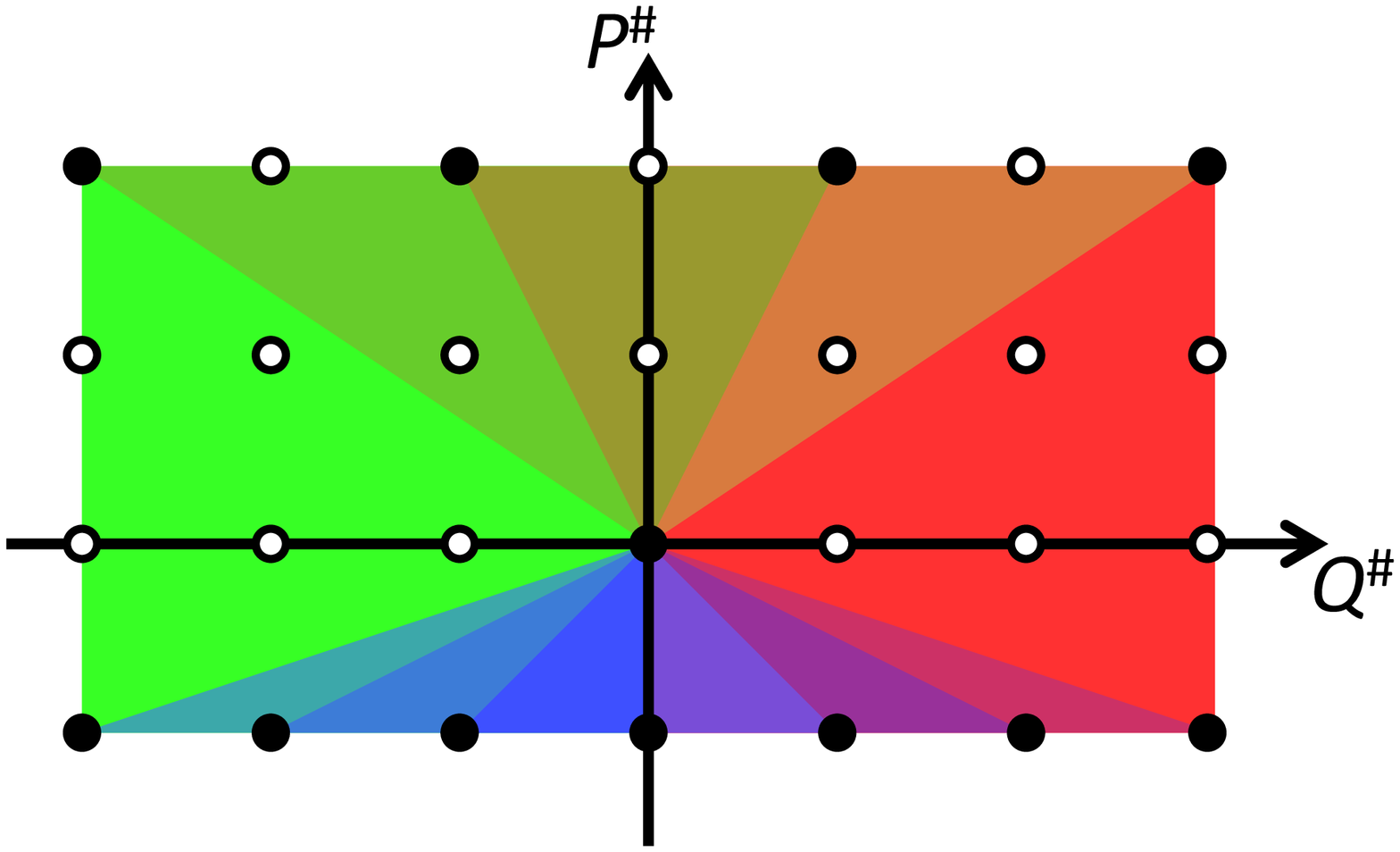}\qquad\qquad\includegraphics[scale=0.6,angle=-90]{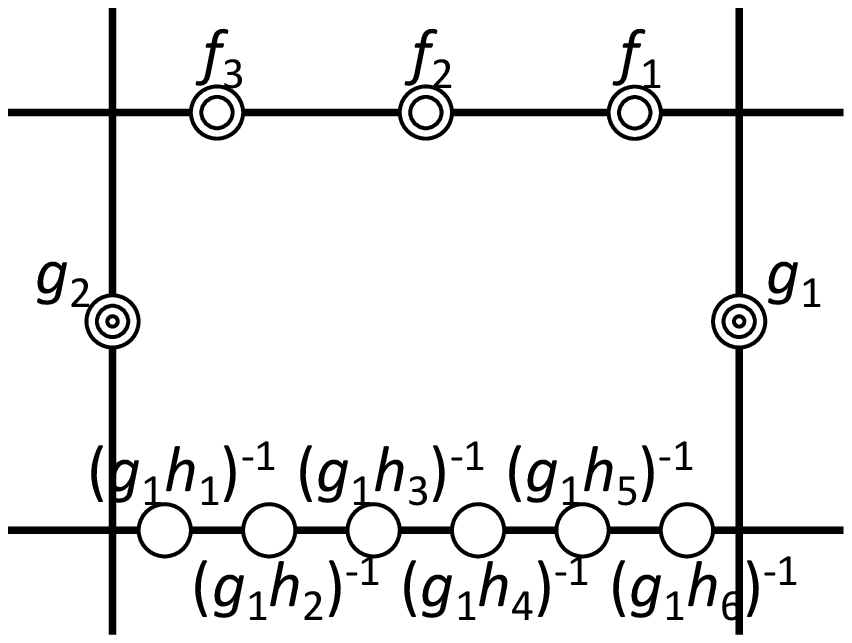}
\caption{A toric diagram (left) and a schematic diagram for asymptotic values (right) for the $E_8$ curve in the rectangular realization.}
\label{e8rect}
\end{figure}

We can move to the rectangular realization (see figure \ref{e8rect}) by using the similarity transformation \eqref{trirect} and \eqref{trirect+} (constructed, of course, originally for the $E_6$ curve).
Then we find, by collecting terms with the same power of $\P$, it is given by
\begin{align}
&\H/\alpha=q^{-3}(\Q+q^{\frac{3}{2}}f_1)(\Q+q^{\frac{1}{2}}f_1)(\Q+q^{\frac{3}{2}}f_2)(\Q+q^{\frac{1}{2}}f_2)(\Q+q^{\frac{3}{2}}f_3)(\Q+q^{\frac{1}{2}}f_3)\Q^{-3}\P^2
\nonumber\\
&\;
+q^{-\frac{3}{2}}(\Q+q^{\frac{1}{2}}f_1)(\Q+q^{\frac{1}{2}}f_2)(\Q+q^{\frac{1}{2}}f_3)\nonumber\\
&\;\quad\times\bigl\{[3]_qg_1\Q^3
+q^{\frac{1}{2}}\bigl(g_1F_1+F_3^2G_2^3H_5\bigr)\Q^2
+q\bigl(F_3G_2H_1+g_1^{-1}F_2G_2\bigr)\Q
+q^{\frac{3}{2}}[3]_qg_1^{-1}F_3G_2\bigr\}\Q^{-3}\P\nonumber\\
&\;
+[3]_qg_1^2\Q^3
+[2]_q\bigl(g_1^2F_1+g_1F_3^2G_2^3H_5\bigr)\Q^2
+\bigl(g_1(F_3^2F_1G_2^3H_5+F_2G_1+F_3G_2H_1)+F_3^2G_2^3H_4\bigr)\Q\nonumber\\
&\quad
+E/\alpha
+\bigl(F_3^2G_2^2H_2+g_1^{-1}(F_3F_1G_2G_1+F_3^3G_2^4H_5+F_3F_2G_2^2H_1)\bigr)\Q^{-1}\nonumber\\
&\quad
+[2]_q\bigl(g_1^{-1}F_3^2G_2^2H_1+g_1^{-2}F_3F_2G_2^2\bigr)\Q^{-2}
+[3]_qg_1^{-2}F_3^2G_2^2\Q^{-3}\nonumber\\
&\;
+q^{\frac{3}{2}}g_1^3(\Q+q^{-\frac{1}{2}}(g_1h_1)^{-1})(\Q+q^{-\frac{1}{2}}(g_1h_2)^{-1})
\cdots(\Q+q^{-\frac{1}{2}}(g_1h_6)^{-1})\Q^{-3}\P^{-1}.
\label{e8recteq}
\end{align}
With the viewpoint of the toric diagram with colors in mind as explained around \eqref{Pg1Pg2} we can alternatively obtain the expression in the rectangular realization directly (instead of from the similarity transformation \eqref{trirect}) and recognize that the result matches with that obtained from the similarity transformation.

From the similarity transformation \eqref{s3similarity} (constructed originally for the $E_6$ curve), which implies \eqref{s3similarity+} and
\begin{align}
&(\Q'+q^{\frac{1}{2}}(g_1h_1)^{-1})(\Q'+q^{\frac{3}{2}}(g_1h_1)^{-1})\P'^2
=q(\Q'+q^{\frac{1}{2}}(g_1h_1)^{-1})\P'(\Q'+q^{\frac{1}{2}}(g_1h_1)^{-1})\P'\nonumber\\
&=q(\Q+q^{\frac{1}{2}}f_1)\P(\Q+q^{\frac{1}{2}}f_1)\P=(\Q+q^{\frac{1}{2}}f_1)(\Q+q^{\frac{3}{2}}f_1)\P^2,
\end{align}
as well, we find that the asymptotic values of \eqref{e8recteq}
\begin{align}
&\bigl\{g_1\text{[triple]};f_1,f_2,f_3\text{[double]};g_2\text{[triple]};\nonumber\\
&\quad(g_1h_1)^{-1},(g_1h_2)^{-1},(g_1h_3)^{-1},(g_1h_4)^{-1},(g_1h_5)^{-1},(g_1h_6)^{-1}\bigr\}_\text{rect},
\end{align}
are transformed into
\begin{align}
&\bigl\{g_1\text{[triple]};(g_1h_1)^{-1},f_2,f_3\text{[double]};(f_1g_1h_1)g_2\text{[triple]};\nonumber\\
&\quad f_1,(g_1h_2)^{-1},(g_1h_3)^{-1},(g_1h_4)^{-1},(g_1h_5)^{-1},(g_1h_6)^{-1}\bigr\}_\text{rect}.
\end{align}
This generates the $s_3$ transformation for the $E_8$ curve.
We have checked that all the other coefficients are also invariant under the $s_3$ transformation.

\subsection{Weyl group}\label{e8wg}

\begin{figure}[!t]
\centering\includegraphics[scale=0.4,angle=-90]{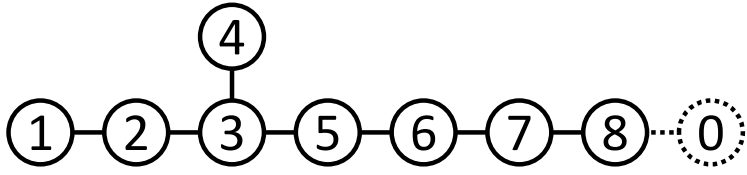}
\caption{Dynkin diagram of the $E_8$ algebra.}
\label{e8dynkin} 
\end{figure}

After fixing the gauge by choosing
\begin{align}
f_3=g_2=1,\quad h_6=(f_1^2f_2^2g_1^3h_1h_2h_3h_4h_5)^{-1},
\label{e8gf}
\end{align}
we find that all of the transformations are given by
\begin{align}
s_1&:(f_1,f_2,g_1,h_1,h_2,h_3,h_4,h_5;\alpha)
\mapsto(f_2^{-1}f_1,f_2^{-1},g_1,f_2h_1,f_2h_2,f_2h_3,f_2h_4,f_2h_5;f_2^3\alpha),\nonumber\\
s_2&:(f_1,f_2,g_1,h_1,h_2,h_3,h_4,h_5;\alpha)
\mapsto(f_2,f_1,g_1,h_1,h_2,h_3,h_4,h_5;\alpha),\nonumber\\
s_3&:(f_1,f_2,g_1,h_1,h_2,h_3,h_4,h_5;\alpha)\nonumber\\
&\hspace{5mm}\mapsto((g_1h_1)^{-1},f_2,(f_1h_1)^{-1},h_1,(f_1g_1h_1)h_2,(f_1g_1h_1)h_3,(f_1g_1h_1)h_4,(f_1g_1h_1)h_5;(f_1g_1h_1)^2\alpha),\nonumber\\
s_4&:(f_1,f_2,g_1,h_1,h_2,h_3,h_4,h_5;\alpha)\mapsto(f_1,f_2,g_1^{-1},g_1h_1,g_1h_2,g_1h_3,g_1h_4,g_1h_5;g_1^2\alpha),\nonumber\\
s_5&:(f_1,f_2,g_1,h_1,h_2,h_3,h_4,h_5;\alpha)\mapsto(f_1,f_2,g_1,h_2,h_1,h_3,h_4,h_5;\alpha),\nonumber\\
s_6&:(f_1,f_2,g_1,h_1,h_2,h_3,h_4,h_5;\alpha)\mapsto(f_1,f_2,g_1,h_1,h_3,h_2,h_4,h_5;\alpha),\nonumber\\
s_7&:(f_1,f_2,g_1,h_1,h_2,h_3,h_4,h_5;\alpha)\mapsto(f_1,f_2,g_1,h_1,h_2,h_4,h_3,h_5;\alpha),\nonumber\\
s_8&:(f_1,f_2,g_1,h_1,h_2,h_3,h_4,h_5;\alpha)\mapsto(f_1,f_2,g_1,h_1,h_2,h_3,h_5,h_4;\alpha),\nonumber\\
s_0&:(f_1,f_2,g_1,h_1,h_2,h_3,h_4,h_5;\alpha)\mapsto(f_1,f_2,g_1,h_1,h_2,h_3,h_4,(f_1^2f_2^2g_1^3h_1h_2h_3h_4h_5)^{-1};\alpha).
\end{align}
From the commutation relations of the transformations, we reproduce the $E_8$ Dynkin diagram in figure \ref{e8dynkin} .
Here the combination of parameters which transforms identically as $\alpha$ is
\begin{align}
\alpha=(f_1f_2g_1)^{-1}.
\label{e8alpha}
\end{align}
As previously, from these transformations, we can identify the simple roots and the fundamental weights as
\begin{align}
\alpha_1&=(-1,-2,0,1,1,1,1,1),&\omega_1&=(-1,-1,0,1,1,1,1,1),\nonumber\\
\alpha_2&=(-1,1,0,0,0,0,0,0),&\omega_2&=(-1,0,0,1,1,1,1,1),\nonumber\\
\alpha_3&=(1,0,1,0,-1,-1,-1,-1),&\omega_3&=(0,0,0,1,1,1,1,1),\nonumber\\
\alpha_4&=(0,0,-2,1,1,1,1,1),&\omega_4&=(0,0,-1,1,1,1,1,1),\nonumber\\
\alpha_5&=(0,0,0,-1,1,0,0,0),&\omega_5&=(0,0,0,0,1,1,1,1),\nonumber\\
\alpha_6&=(0,0,0,0,-1,1,0,0),&\omega_6&=(0,0,0,0,0,1,1,1),\nonumber\\
\alpha_7&=(0,0,0,0,0,-1,1,0),&\omega_7&=(0,0,0,0,0,0,1,1),\nonumber\\
\alpha_8&=(0,0,0,0,0,0,-1,1),&\omega_8&=(0,0,0,0,0,0,0,1),
\label{e8ao}
\end{align}
in the linear space of $(\log f_1,\log f_2,\log g_1,\log h_1,\log h_2,\log h_3,\log h_4,\log h_5)$ spanned by the parameters of the curve.

Comparing the simple roots and the fundamental weights \eqref{e8ao} identified from the symmetries of the $E_8$ curve with those in the orthogonal basis
\begin{align}
\alpha^\perp_1&=\textstyle{(-1,1,0,0,0,0,0,0)},&
\omega^\perp_1&=\textstyle{(-1,0,0,0,0,-\frac{\sqrt{3}}{3},-\frac{\sqrt{6}}{3},-\sqrt{2})},\nonumber\\
\alpha^\perp_2&=\textstyle{(0,-1,1,0,0,0,0,0)},&
\omega^\perp_2&=\textstyle{(-1,-1,0,0,0,-\frac{2\sqrt{3}}{3},-\frac{2\sqrt{6}}{3},-2\sqrt{2})},\nonumber\\
\alpha^\perp_3&=\textstyle{(0,0,-1,1,0,0,0,0)},&
\omega^\perp_3&=\textstyle{(-1,-1,-1,0,0,-\sqrt{3},-\sqrt{6},-3\sqrt{2})},\nonumber\\
\alpha^\perp_4&=\textstyle{(0,0,0,-1,1,0,0,0)},&
\omega^\perp_4&=\textstyle{(-\frac{1}{2},-\frac{1}{2},-\frac{1}{2},-\frac{1}{2},\frac{1}{2},-\frac{\sqrt{3}}{2},-\frac{\sqrt{6}}{2},-\frac{3\sqrt{2}}{2})},\nonumber\\
\alpha^\perp_5&=\textstyle{(0,0,0,-1,-1,0,0,0)},&
\omega^\perp_5&=\textstyle{(-\frac{1}{2},-\frac{1}{2},-\frac{1}{2},-\frac{1}{2},-\frac{1}{2},-\frac{5\sqrt{3}}{6},-\frac{5\sqrt{6}}{6},-\frac{5\sqrt{2}}{2})},\nonumber\\
\alpha^\perp_6&=\textstyle{(\frac{1}{2},\frac{1}{2},\frac{1}{2},\frac{1}{2},\frac{1}{2},-\frac{\sqrt{3}}{2},0,0)},&
\omega^\perp_6&=\textstyle{(0,0,0,0,0,-\frac{2\sqrt{3}}{3},-\frac{2\sqrt{6}}{3},-2\sqrt{2})},\nonumber\\
\alpha^\perp_7&=\textstyle{(0,0,0,0,0,\frac{2\sqrt{3}}{3},-\frac{\sqrt{6}}{3},0)},&
\omega^\perp_7&=\textstyle{(0,0,0,0,0,0,-\frac{\sqrt{6}}{2},-\frac{3\sqrt{2}}{2})},\nonumber\\
\alpha^\perp_8&=\textstyle{(0,0,0,0,0,0,\frac{\sqrt{6}}{2},-\frac{\sqrt{2}}{2})},&
\omega^\perp_8&=\textstyle{(0,0,0,0,0,0,0,-\sqrt{2})},
\end{align}
which are utilized in constructing the $E_8$ characters, we find the relation
\begin{align}
&(f_1,f_2,g_1,h_1,h_2,h_3,h_4,h_5)\nonumber\\
&=f_1^{-\omega_2+\omega_3}f_2^{-\omega_1+\omega_2}g_1^{\omega_3-\omega_4}
h_1^{\omega_3-\omega_5}h_2^{\omega_5-\omega_6}h_3^{\omega_6-\omega_7}h_4^{\omega_7-\omega_8}h_5^{\omega_8}\nonumber\\
&\leftrightarrow
f_1^{-\omega^\perp_2+\omega^\perp_3}
f_2^{-\omega^\perp_1+\omega^\perp_2}
g_1^{\omega^\perp_3-\omega^\perp_4}
h_1^{\omega^\perp_3-\omega^\perp_5}
h_2^{\omega^\perp_5-\omega^\perp_6}
h_3^{\omega^\perp_6-\omega^\perp_7}
h_4^{\omega^\perp_7-\omega^\perp_8}
h_5^{\omega^\perp_8}\nonumber\\
&=f_1^{(0,0,-1,0,0,-\frac{\sqrt{3}}{3},-\frac{\sqrt{6}}{3},-\sqrt{2})}
f_2^{(0,-1,0,0,0,-\frac{\sqrt{3}}{3},-\frac{\sqrt{6}}{3},-\sqrt{2})}
g_1^{(-\frac{1}{2},-\frac{1}{2},-\frac{1}{2},\frac{1}{2},-\frac{1}{2},-\frac{\sqrt{3}}{2},-\frac{\sqrt{6}}{2},-\frac{3\sqrt{2}}{2})}\nonumber\\
&\quad\times
h_1^{(-\frac{1}{2},-\frac{1}{2},-\frac{1}{2},\frac{1}{2},\frac{1}{2},-\frac{\sqrt{3}}{6},-\frac{\sqrt{6}}{6},-\frac{\sqrt{2}}{2})}
h_2^{(-\frac{1}{2},-\frac{1}{2},-\frac{1}{2},-\frac{1}{2},-\frac{1}{2},-\frac{\sqrt{3}}{6},-\frac{\sqrt{6}}{6},-\frac{\sqrt{2}}{2})}\nonumber\\
&\quad\times
h_3^{(0,0,0,0,0,-\frac{2\sqrt{3}}{3},-\frac{\sqrt{6}}{6},-\frac{\sqrt{2}}{2})}
h_4^{(0,0,0,0,0,0,-\frac{\sqrt{6}}{2},-\frac{\sqrt{2}}{2})}
h_5^{(0,0,0,0,0,0,0,-\sqrt{2})}\nonumber\\
&=\bigl((g_1h_1h_2)^{-\frac{1}{2}},(f_2^2g_1h_1h_2)^{-\frac{1}{2}},(f_1^2g_1h_1h_2)^{-\frac{1}{2}},
(g_1h_1h_2^{-1})^{\frac{1}{2}},(g_1^{-1}h_1h_2^{-1})^{\frac{1}{2}},\nonumber\\
&\qquad(f_1^2f_2^2g_1^3h_1h_2h_3^4)^{-\frac{\sqrt{3}}{6}},(f_1^2f_2^2g_1^3h_1h_2h_3h_4^3)^{-\frac{\sqrt{6}}{6}},
(f_1^2f_2^2g_1^3h_1h_2h_3h_4h_5^2)^{-\frac{\sqrt{2}}{2}}\bigr).
\end{align}
If we identify this as
\begin{align}
{\bm q}=(q_1,q_2,q_3,q_4,q_5,q_6,q_7,q_8)
=(t_1^2,t_2^2,t_3^2,t_4^2,t_5^2,t_6^{-2\sqrt{3}},t_7^{-\sqrt{6}},t_8^{-\sqrt{2}}),
\label{e8t}
\end{align}
we can solve the relations reversely and find
\begin{align}
&f_1=\frac{t_1^2}{t_3^2},\quad
f_2=\frac{t_1^2}{t_2^2},\quad
g_1=\frac{t_4^2}{t_5^2},\nonumber\\
&h_1=\frac{t_5^2}{t_1^2},\quad
h_2=\frac{1}{t_1^2t_4^2},\quad
h_3=\frac{t_2t_3t_5t_6^3}{t_1t_4},\quad
h_4=\frac{t_2t_3t_5t_7^2}{t_1t_4t_6},\quad
h_5=\frac{t_2t_3t_5t_8}{t_1t_4t_6t_7}.
\label{e8fght}
\end{align}

\subsection{Mirror map}\label{e8mirror}

The analysis of the mirror map for the $E_8$ curve is not as simple as the previous two cases.
If we start from the triangular realization \eqref{e8trieq}, the Schr\"odinger equation reads
\begin{align}
&q^{-3}X^{-3}
(X+q^{\frac{1}{2}}f_1)(X+q^{\frac{3}{2}}f_1)(X+q^{\frac{1}{2}}f_2)(X+q^{\frac{3}{2}}f_2)(X+q^{\frac{1}{2}}f_3)(X+q^{\frac{3}{2}}f_3)
P[X]P[q^{-1}X]\nonumber\\
&\quad+q^{-1}X^{-3}(X+q^{\frac{1}{2}}f_1)(X+q^{\frac{1}{2}}f_2)(X+q^{\frac{1}{2}}f_3)(aX^2+bX+c)P[X]+\cdots
+\frac{z_E}{\alpha}=0,
\end{align}
where the coefficients $a$, $b$, $c$ are given by various combinations of $F$, $G$, $H$ and $q$.
Then, in the large $z_E$ limit, the Schr\"odinger equation is solved by
\begin{align}
P_0[X]=\sqrt{\frac{-z_E}{\alpha}}\frac{q^{\frac{3}{4}}X^{\frac{3}{2}}}{(X+q^{\frac{1}{2}}f_1)(X+q^{\frac{1}{2}}f_2)(X+q^{\frac{1}{2}}f_3)},
\label{fractionalz}
\end{align}
and the behavior of the large $z_E$ expansion is completely different from the previous two cases.
Especially, it is unclear for us how to deal with the half-integral power of $z_E$ in \eqref{fractionalz}.

\begin{table}[t!]
\begin{align*}
\epsilon_1&=0,\\
\epsilon_2&=(q^2+q^{-2})\chi_{\bf 1}+
(q+q^{-1})(\chi_{\bf 248}+3\chi_{\bf 1})
+(\chi_{\bf 3875}+3\chi_{\bf 248}+7\chi_{\bf 1}),\\
\epsilon_3&=(q^{\frac{9}{2}}+q^{-\frac{9}{2}})\chi_{\bf 1}
+(q^{\frac{7}{2}}+q^{-\frac{7}{2}})(\chi_{\bf 248}+3\chi_{\bf 1})
+(q^{\frac{5}{2}}+q^{-\frac{5}{2}})(\chi_{\bf 3875}+4\chi_{\bf 248}+9\chi_{\bf 1})\\
&\quad+(q^{\frac{3}{2}}+q^{-\frac{3}{2}})(\chi_{\bf 30380}+4\chi_{\bf 3875}+12\chi_{\bf 248}+20\chi_{\bf 1})\\
&\quad+(q^{\frac{1}{2}}+q^{-\frac{1}{2}})(\chi_{\bf 147250}+3\chi_{\bf 30380}+10\chi_{\bf 3875}+23\chi_{\bf 248}+32\chi_{\bf 1}).
\end{align*}
\caption{Multi-covering components of the quantum mirror map for the $E_8$ curve.\label{e8mirrormap}}
\end{table}

For this purpose, we choose the rectangular realization \eqref{e8recteq} instead.
This time the Schr\"odinger equation reads
\begin{align}
&q^{-3}(X+q^{\frac{3}{2}}f_1)(X+q^{\frac{1}{2}}f_1)(X+q^{\frac{3}{2}}f_2)(X+q^{\frac{1}{2}}f_2)(X+q^{\frac{3}{2}}f_3)(X+q^{\frac{1}{2}}f_3)
/(X^3R[q^{-1}X]R[q^{-2}X])
\nonumber\\
&\;
+q^{-\frac{3}{2}}\bigl\{[3]_qg_1X^3
+q^{\frac{1}{2}}\bigl(g_1F_1+F_3^2G_2^3H_5\bigr)X^2
+q\bigl(F_3G_2H_1+g_1^{-1}F_2G_2\bigr)X
+q^{\frac{3}{2}}[3]_qg_1^{-1}F_3G_2\bigr\}
\nonumber\\
&\;\quad\times
(X+q^{\frac{1}{2}}f_1)(X+q^{\frac{1}{2}}f_2)(X+q^{\frac{1}{2}}f_3)/(X^3R[q^{-1}X])\nonumber\\
&\;
+[3]_qg_1^2X^3
+[2]_q\bigl(g_1^2F_1+g_1F_3^2G_2^3H_5\bigr)X^2
+\bigl(g_1(F_3^2F_1G_2^3H_5+F_2G_1+F_3G_2H_1)+F_3^2G_2^3H_4\bigr)X\nonumber\\
&\quad
+\bigl(F_3^2G_2^2H_2+g_1^{-1}(F_3F_1G_2G_1+F_3^3G_2^4H_5+F_3F_2G_2^2H_1)\bigr)X^{-1}\nonumber\\
&\quad
+[2]_q\bigl(g_1^{-1}F_3^2G_2^2H_1+g_1^{-2}F_3F_2G_2^2\bigr)X^{-2}
+[3]_qg_1^{-2}F_3^2G_2^2X^{-3}\nonumber\\
&\;
+q^{\frac{3}{2}}g_1^3(X+q^{-\frac{1}{2}}(g_1h_1)^{-1})(X+q^{-\frac{1}{2}}(g_1h_2)^{-1})
\cdots(X+q^{-\frac{1}{2}}(g_1h_6)^{-1})X^{-3}R[X]+z_E/\alpha=0,
\end{align}
where instead of using $P[X]$ defined in \eqref{PX} we introduce the ``inverse'' of it
\begin{align}
R[X]=\frac{\Psi[qX]}{\Psi[X]}.
\label{RX}
\end{align}
Then the leading term in the large $z_E$ expansion is
\begin{align}
R_0[X]=\frac{-z_E}{\alpha}
\frac{q^{-\frac{3}{2}}g_1^{-3}X^3}{(X+q^{-\frac{1}{2}}(g_1h_1)^{-1})(X+q^{-\frac{1}{2}}(g_1h_2)^{-1})\cdots(X+q^{-\frac{1}{2}}(g_1h_6)^{-1})},
\end{align}
and we can follow our previous analysis.
As in \eqref{PiA}, this time we can pick up the residue in
\begin{align}
\Pi_A(z)=\Res_{X=0}\frac{1}{X}\log\frac{R[X]}{R_0[X]},
\end{align}
for the mirror map and continue our analysis.
We find again that, by substituting the identification of $\alpha$ \eqref{e8alpha} and the change of parameters \eqref{e8fght}, we can confirm the group structure, in the sense that the results are given by the $E_8$ characters.
Furthermore by adopting the same multi-covering structure \eqref{multicover} for the $D_5$ case, we find non-negative integral coefficients.
The resulting multi-covering components are listed in table \ref{e8mirrormap}.

If we wish to continue our study of the mirror map for the $E_8$ curve to higher degrees, it should be more convenient to move back to the triangular realization due to its simplicity in algebra.
To avoid the fractional powers of $z$ appearing in $P_0[X]$ as in \eqref{fractionalz}, we need to apply the similarity transformation, 
\begin{align}
\widehat O\to\widehat G\widehat O\widehat G^{-1},\quad\widehat G=e^{-\frac{i}{2\hbar}\widehat p^2}
\end{align}
for the triangular realization.
With this similarity transformation all terms on the diagonal boundary $\Q^{\gamma+1}\P^\gamma (-4\le\gamma\le 2)$ are transformed into $\Q^{\gamma+1}\P^{-1}$ and it is possible to perform the analysis of \eqref{RX}.
The reader may wonder why we do not start with this expression of the $E_8$ curve from the beginning of our analysis.
We stress that the prescription of the $q$-order is difficult to incorporate when the degeneracy appears on the diagonal boundary and it is important for us to start with our $E_8$ curve in \eqref{e8trieq}.

\section{Observation}\label{decoupling}

In the previous three sections, after determining the $E_n$ $(n=6,7,8)$ quantum curves and fixing the Weyl groups, we have proceeded to study the mirror maps for these curves.
For all of these cases, we find that, by identifying the overall factor $\alpha$ as the combination of the asymptotic values which transforms identically as $\alpha$ (which is explained in section \ref{weyl}), we can express the result in terms of characters.
Furthermore, by adopting the same expression of the multi-covering structure for the $D_5$ curve \eqref{multicover}, we find that all of the multi-covering components listed in table \ref{e6mirrormap}, table \ref{e7mirrormap} and table \ref{e8mirrormap} have non-negative integral coefficients for characters.
Note also that, as we have stressed at the end of section \ref{e7}, for both the cases of $D_5$ \cite{FMS} and $E_7$, the representations appearing in the mirror maps match exactly with those appearing in the BPS indices except for the trivial case of degree one.

We can further make an interesting observation for the relation among the mirror maps for different curves.
Let us take the case of degree $d=2$ for explanation.
The multi-covering components of the mirror map for the $E_8$ curve in table \ref{e8mirrormap} contain the representations ${\bf 3875}$, ${\bf 248}$ and ${\bf 1}$.
The decomposition of these representations into the $E_7\times A_1$ subalgebra reads \cite{Y}
\begin{align}
{\bf 3875}&\to({\bf 1539},{\bf 1})+({\bf 912},{\bf 2})+({\bf 133},{\bf 3})+({\bf 56},{\bf 2})+({\bf 1},{\bf 1}),\nonumber\\
{\bf 248}&\to({\bf 133},{\bf 1})+({\bf 56},{\bf 2})+({\bf 1},{\bf 3}),\nonumber\\
{\bf 1}&\to({\bf 1},{\bf 1}).
\end{align}
In the decoupling limit we pick up representations with the $\text{U}(1)$ charge $2$ (for degree $d=2$) by breaking the $A_1$ algebra further into the diagonal $\text{U}(1)$ subalgebra normalized so that the fundamental representation of $A_1$ reduces to the $\text{U}(1)$ charge $\pm 1$.
This means picking up representations of $A_1$ with odd dimensions larger than $2$ effectively.
Then we find that only the $E_8$ representations ${\bf 3875}$ and ${\bf 248}$ remain and reduce to the $E_7$ representations ${\bf 133}$ and ${\bf 1}$ respectively.
Thus we obtain the multi-covering components for the $E_7$ curve as in table \ref{e7mirrormap}.

Furthermore if we decompose these representations of $E_7$ into $[E_6]_{\text{U}(1)}$ we find
\begin{align}
{\bf 133}&\to{\bf 27}_2+{\bf 78}_0+{\bf 1}_0+\overline{\bf 27}_{-2},\nonumber\\
{\bf 1}&\to{\bf 1}_0.
\end{align}
Therefore, by picking up the representations with the $\text{U}(1)$ charge 2, we find that only the $E_7$ representation ${\bf 133}$ reduces to the $E_6$ representation ${\bf 27}$, which is indeed the case in table \ref{e6mirrormap}.
We can proceed this down to the $D_5$ case in table 2 of \cite{FMS}.
Although we have chosen the degree $d=2$ for explanation, this works for all other degrees as well.

The property we find in this section is called the decoupling relation and known explicitly for the BPS indices.\footnote{We are grateful to Kazuhiro Sakai for valuable discussions.}
This relation originates from the fact that curves of lower degrees are obtained by taking the decoupling limit for the parameters in curves of higher degrees.
The confirmation of the decoupling relation here can probably be regarded as another evidence that we have identified the multi-covering structure correctly in \eqref{multicover}.

\section{Conclusion and discussions}\label{conclusion}

Originally it was proposed that multiple M2-branes on various orbifold backgrounds are described by supersymmetric Chern-Simons theories.
From the localization technique, the partition functions reduce to matrix models.
In the grand canonical ensemble these matrix models can be put further into the expression of spectral determinants.
The spectral operators take the form of algebraic curves, which enjoy the Weyl group symmetries of exceptional algebras for those known as the del Pezzo geometries.
This leads to the notion of quantum curves.
For some parameters the spectral operator is realized by the matrix models, while for others not \cite{KM}.
Still these curves enjoy the degrees of freedom $N^{\frac{3}{2}}$ and the perturbative completion by the Airy function for large $N$ as all other supersymmetric Chern-Simons theories.
In view of the full-fledged group structure of the curves we are naturally led to the idea that it is no more the matrix models or the original supersymmetric Chern-Simons theories that describes M2-branes.
Instead, we propose that the partition functions of multiple M2-branes are described by the quantum curves themselves.

From this viewpoint we investigate the correspondence between spectral theories and topological strings on the del Pezzo geometries of higher ranks fully in the group-theoretical language.
For spectral theories, we are able to write down all the quantum curves of $E_n$ $(n=6,7,8)$.
For topological strings, we conjecture the coefficients of the perturbative part and propose that the non-perturbative part boils down to two sets of multiplicities of representations.
We further identify the multiplicities for the mirror maps, since the multiplicities for the BPS indices were studied previously in \cite{HKP,MNY}.
We find that the multiplicities for the mirror maps enjoy the decoupling relation relating curves of lower ranks to those of higher ranks.
Namely starting from the $E_8$ curve, we find that the results for all the other curves are obtained by picking up the representations of the $\text{U}(1)$ charge corresponding to the degree.
Remarkably, as explained in section \ref{weyl}, with the group structure, we can resolve some unsatisfactory assumptions made in previous works: the shift of the chemical potential in the classical phase space area and the identification of the overall factor $\alpha$ in the quantum mirror map.
We shall list some further directions we would like to pursue in the future.

Firstly, motivated by understanding M2-branes, we have studied quantum curves of genus one and higher ranks.
For the $D_5$ curve it was known that the three-dimensional space of relative ranks in the brane configuration with two NS5-branes and two $(1,k)$5-branes are embedded into the five-dimensional space of the curve \cite{KM}.
We also know for a specific choice of six 5-branes, the matrix model is described by the $E_7$ curve \cite{MNY,KMN}.
It is unclear, however, how other configurations of these curves or other curves of higher ranks describe the backgrounds for M2-branes, even though these curves give perfectly the $N^{\frac{3}{2}}$ behavior and the Airy function.
Especially, we would like to identify the geometrical backgrounds for M2-branes for these quantum curves.

Secondly, in this paper we have adopted the rule of \cite{ACDKV} by picking up the residue to compute the quantum mirror map.
Classically the symplectic form is invariant under the symplectic transformation.
It is unclear to us, however, when lifting quantum-mechanically, the computation is still invariant under the similarity transformation.
We would like to clarify symmetries of the quantum periods rigorously to justify the computations using $R[X]$ in \eqref{RX}.

Thirdly, since the multiplicities of the mirror maps for all the del Pezzo geometries are governed by those of the $E_8$ curve from the decoupling relation, it is important to proceed to as high degrees as possible.
Although in section \ref{e8mirror} we have struggled to avoid the technical difficulties for the $E_8$ curve, it is desirable to establish a clean method to proceed to higher degrees.

Fourthly, after stating the conjecture for the correspondence between spectral theories and topological strings explicitly, it is important to prove it or collect evidences for it.
We would like to perform numerical computations for the $E_n$ $(n=6,7,8)$ curves to support the conjecture and understand the results for the curves studied in \cite{GKMR} in the same framework.

Fifthly, for our perturbative analysis of the reduced grand potential in the classical limit for the $E_n$ $(n=6,7,8)$ curves, we have concentrated on the expansion from the large $\mu$ limit.
The subleading group-invariant behavior was also obtained recently in \cite{HKLY}.
We would like to clarify its relation to our analysis.

Sixthly, we concentrate on the del Pezzo geometries of genus one since the Weyl group symmetries are well-known from their classical cousins.
It is interesting to explore curves of higher genus from the symmetry viewpoint \cite{HM,MN1,MN3,HHO,CGM,CGuM}.
In \cite{KM} after identifying most symmetries as the Hanany-Witten transition, a new brane transition was observed.
We would like to clarify the whole symmetry structure for the quantum curves or the brane systems.

Seventhly, as an application, we would like to further proceed to the study of the matrix model with the orthosymplectic supergroup \cite{MS1,H,O,MS2,MN5} which is expected to be interpreted as an orientifold.
As shown in \cite{H,MS2,MN5} these matrix models are interpreted as the chiral projections of the spectral operators.
We would like to study the corresponding quantum curves and determine the two sets of multiplicities from the chiral projections.

Finally, it is interesting to clarify the relation to Painlev\'e equations \cite{Sakai,BGT}.
In fact similar computations were performed for the $E_6$ and $E_7$ curves in the rectangular realization \cite{T} and its generalizations \cite{NRY} from the viewpoint of Painlev\'e equations.
Here with the $q$-order of quantum curves and the coloring of toric diagrams, our computation should be clearer.
We hope that this clarification is helpful in studying the relation to the Painlev\'e equations.
Integrabilities for the ABJM matrix model \cite{HHMO,GHM2,G,JT,2PT,2DTL,CK} should be a clue.

\section*{Acknowledgements}
We are grateful to Tomohiro Furukawa, Yasuyuki Hatsuda, Hirotaka Hayashi, Masazumi Honda, Hiroaki Kanno, Sung-Soo Kim, Naotaka Kubo, Kimyeong Lee, Marcos Marino, Tomoki Nosaka, Kazuhiro Sakai, Masaki Shigemori, Yuji Sugimoto, Futoshi Yagi and Yasuhiko Yamada for valuable discussions in various stages of this work.
The work of S.M.\ is supported by JSPS Grant-in-Aid for Scientific Research (C) \#19K03829.

\appendix
\section{Phase space area}\label{area}

This appendix is devoted to the classical computation of the area of the dual diagram in the phase space in the large $\mu=\log z$ limit.\footnote{We are grateful to Sung-Soo Kim, Kimyeong Lee and Futoshi Yagi for valuable discussions.}
In this limit, the tessellations of the toric diagrams for the $E_n$ $(n=6,7,8)$ curves are given by connecting the origin with the black dots on the boundary.
We stress that it is only after we shift the chemical potential $\mu$ \cite{AHK,BPTY,MPTY} that the area becomes real and expressed in the group invariants.
The shift of the chemical potential $\mu$ is explained in section \ref{weyl} from the group-theoretical viewpoint.

\subsection{$D_5$ curve}

\begin{figure}[!t]
\centering\includegraphics[scale=0.6,angle=-90]{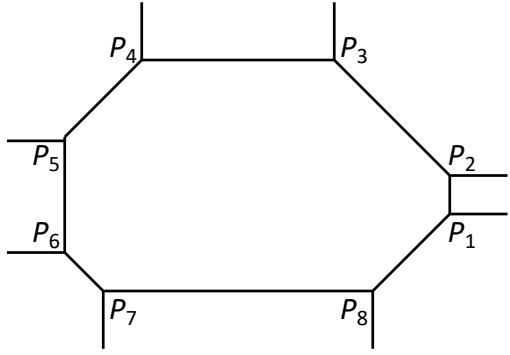}
\caption{Dual diagram of the toric diagram for the $D_5$ curve.}
\label{d5dual}
\end{figure}

The dual diagram of the toric diagram for the $D_5$ curve is given in figure \ref{d5dual}.
Note that the parameters $(h_1,h_2,e_1,e_2,\cdots,e_7,e_8)$ are in general phase functions in our application to the super Chern-Simons matrix models with rank deformations \cite{MNY,KMN,KM}.
For our computations, however, we formally regard them as satisfying inequalities $e_1\ge e_2$, $e_3\ge e_4$, $e_5\ge e_6$ and $e_7\ge e_8$ in depicting the dual diagram, though the resulting area does not depend on the inequalities.
We also consider that the fugacity $z$ is large enough so that the constant term $E$ is neglected. 
The coordinates of the dual diagram are given by
\begin{align}
&P_1:\biggl(e_2z',\frac{1}{e_1}\bigg),\quad\!
P_2:\biggl(e_2z',\frac{1}{e_2}\bigg),\quad\!
P_3:\biggl(e_3,\frac{z'}{e_3}\bigg),\quad\!
P_4:\biggl(e_4,\frac{z'}{e_3}\bigg),\nonumber\\
&P_5:\biggl(\frac{e_3e_4e_5}{h_2z'},\frac{e_5}{h_2}\bigg),\quad\!
P_6:\biggl(\frac{e_3e_4e_5}{h_2z'},\frac{e_6}{h_2}\bigg),\quad\!
P_7:\biggl(\frac{h_1}{e_7},\frac{h_1}{e_1e_2e_8z'}\bigg),\quad\!
P_8:\biggl(\frac{h_1}{e_8},\frac{h_1}{e_1e_2e_8z'}\bigg),
\label{8points}
\end{align}
where we have introduced a variable
\begin{align}
z'=z/\alpha,
\label{zprime}
\end{align}
since the curve is solved classically by equating the defining equation of the spectral operator $\H$ with the fugacity $z$.

Then we can compute the area without difficulty.
After substituting the gauge fixing condition \eqref{gaugefixing}, the constraint $e_7=(h_1h_2)^2/(e_1e_3e_5)$ and the relation to the orthogonal basis
\begin{align}
h_1=q_3q_4,\quad
h_2=\frac{1}{q_2q_3},\quad
e_1=\frac{q_1}{q_2},\quad
e_3=\frac{q_4}{q_5},\quad
e_5=\frac{1}{q_1q_2},
\label{h12e135}
\end{align}
obtained in \cite{FMS}, we find that the area is given by
\begin{align}
&\text{Area}(D_5)=2\mu'^2+2i(\widetilde q_1-\widetilde q_2-\widetilde q_3-\widetilde q_4+\widetilde q_5)\mu'\nonumber\\
&\quad+\widetilde q_1\widetilde q_2
+\widetilde q_1\widetilde q_3
+\widetilde q_1\widetilde q_4
-\widetilde q_1\widetilde q_5
-\widetilde q_2\widetilde q_3
-\widetilde q_2\widetilde q_4
+\widetilde q_2\widetilde q_5
-\widetilde q_3\widetilde q_4
+\widetilde q_3\widetilde q_5
+\widetilde q_4\widetilde q_5,
\label{d5area}
\end{align}
where we have introduced $z'=e^{\mu'}$ and $q_i=e^{i\widetilde q_i}$.
If we redefine the chemical potential by
\begin{align}
\mu=\mu'+\frac{i}{2}(\widetilde q_1-\widetilde q_2-\widetilde q_3-\widetilde q_4+\widetilde q_5),
\label{d5muredef}
\end{align}
to remove the imaginary linear terms in $\mu'$, we find that the area is given by
\begin{align}
\text{Area}(D_5)=2\mu^2+\frac{\Omega(D_5)}{2}.
\end{align}
with the quadratic Casimir element for $D_5$ being
\begin{align}
\Omega(D_5)=\widetilde q_1^2+\widetilde q_2^2+\widetilde q_3^2+\widetilde q_4^2+\widetilde q_5^2.
\end{align}
This leads to the general expression \eqref{classicalCB}.
The reason why the chemical potential with shifts \eqref{d5muredef} is interpreted as the original chemical potential $\mu=\log z$ is explained in section \ref{weyl}.

\subsection{$E_6$ curve}

\begin{figure}[!t]
\centering\includegraphics[scale=0.6,angle=-90]{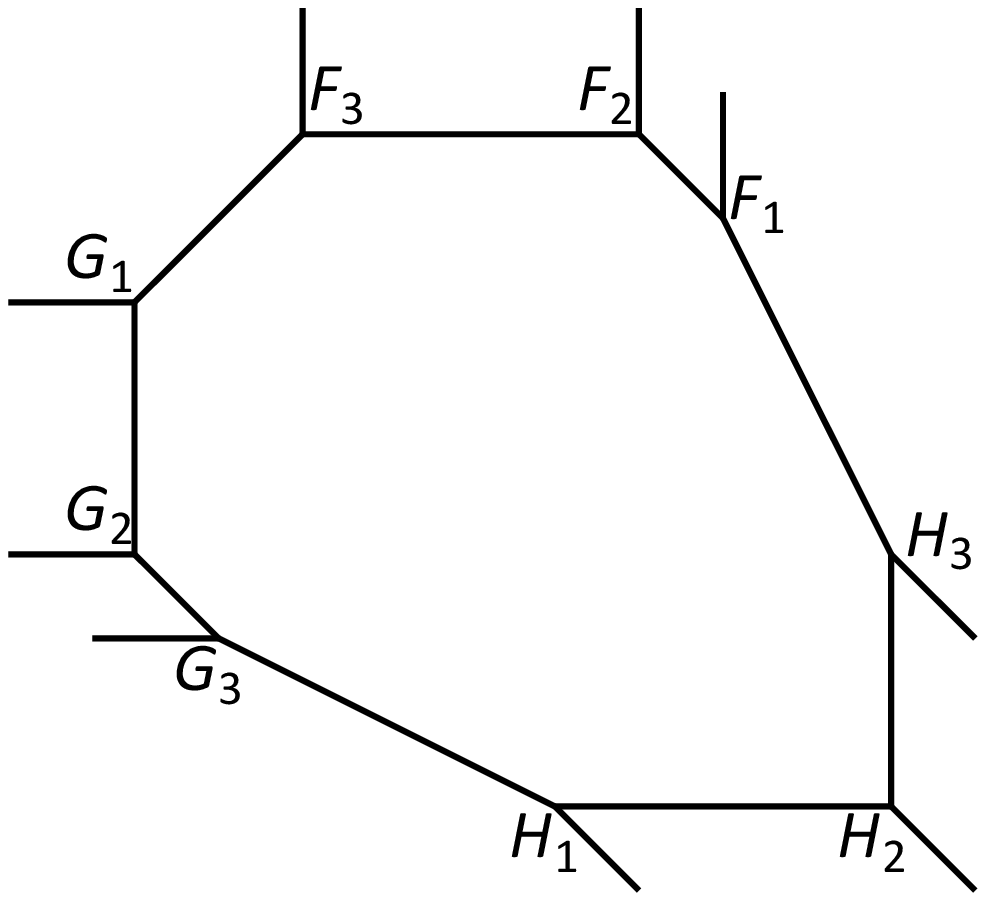}\qquad\qquad\qquad\includegraphics[scale=0.6,angle=-90]{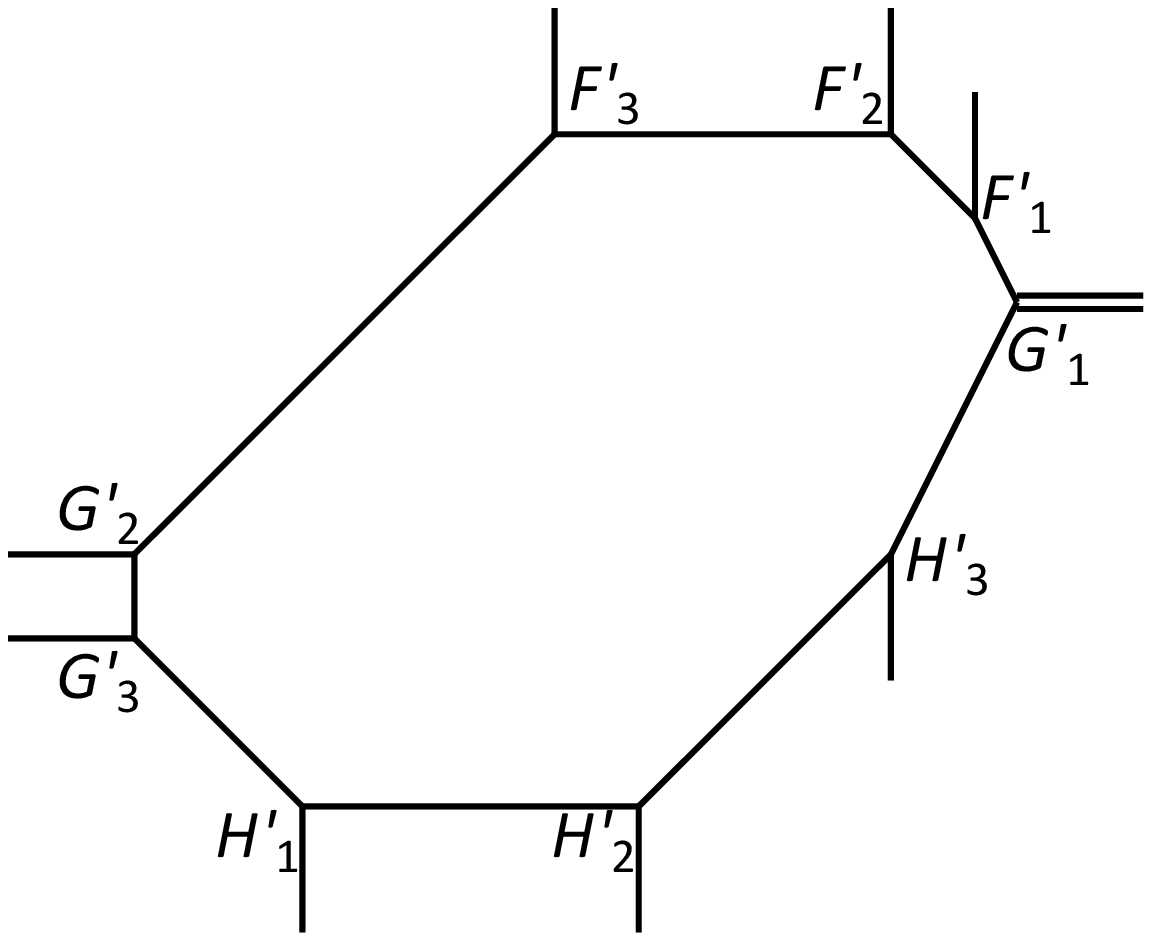}
\caption{Dual diagrams of the toric diagrams for the $E_6$ curve.
The left one is that in the triangular realization while the right one is that in the rectangular realization.}
\label{e6dual}
\end{figure}

The dual diagrams of the toric diagrams for the $E_6$ curve in the triangular realization (figure \ref{e6tri}) and the rectangular realization (figure \ref{e6rect}) are given explicitly in figure \ref{e6dual}.
It should be clear that the vertices of the diagrams $F_i$, $G_i$ and $H_i$ have nothing to do with the elementary symmetric polynomials introduced in \eqref{e6elementary}, \eqref{e7elemetary} or \eqref{e8elemetary}.
Note that $f_i$, $g_i$ and $h_i$ are in general expected to be imaginary phase functions.
Nevertheless we formally assume $f_1\ge f_2\ge f_3$, $g_1\ge g_2\ge g_3$ and $h_1\ge h_2\ge h_3$ in depicting the dual diagrams.
The coordinates of the vertices for the $E_6$ curve in the triangular realization are given by
\begin{align}
&F_1:\biggl(f_1,\frac{z'}{f_1^2}\biggr),\quad
F_2:\biggl(f_2,\frac{z'}{f_1f_2}\biggr),\quad
F_3:\biggl(f_3,\frac{z'}{f_1f_2}\biggr),\nonumber\\
&G_1:\biggl(\frac{f_1f_2f_3g_1}{z'},g_1\biggr),\quad
G_2:\biggl(\frac{f_1f_2f_3g_1}{z'},g_2\biggr),\quad
G_3:\biggl(\frac{f_1f_2f_3g_1g_2}{g_3z'},g_3\biggr),\nonumber\\
&H_1:\biggl(\frac{h_2h_3z'}{h_1},\frac{1}{h_2h_3z'}\biggr),\quad
H_2:\biggl(h_3z',\frac{1}{h_2h_3z'}\biggr),\quad
H_3:\biggl(h_3z',\frac{1}{h_3^2z'}\biggr),
\end{align}
while those in the rectangular realization are given by
\begin{align}
&F'_1:\biggl(f_1,\frac{z'}{f_1^2}\biggr),\quad
F'_2:\biggl(f_2,\frac{z'}{f_1f_2}\biggr),\quad
F'_3:\biggl(f_3,\frac{z'}{f_1f_2}\biggr)\nonumber\\
&G'_1:\biggl(\sqrt{\frac{z'}{g_1}},g_1\biggr),\quad
G'_2:\biggl(\frac{f_1f_2f_3g_2}{z'},g_2\biggr),\quad
G'_3:\biggl(\frac{f_1f_2f_3g_2}{z'},g_3\biggr),\nonumber\\
&H'_1:\biggl(\frac{1}{g_1h_1},\frac{1}{h_2h_3z'}\biggr),\quad
H'_2:\biggl(\frac{1}{g_1h_2},\frac{1}{h_2h_3z'}\biggr),\quad
H'_3:\biggl(\frac{1}{g_1h_3},\frac{1}{h_3^2z'}\biggr).
\end{align}

In either case, we can compute the area without difficulty,
\begin{align}
&\text{Area}(E_6)=\frac{3}{2}\mu'^2
-i(2\widetilde f_1+2\widetilde f_2+2\widetilde f_3+\widetilde g_1+\widetilde g_2+\widetilde g_3)\mu'
\nonumber\\
&\quad-(\widetilde f_1\widetilde f_2+\widetilde f_1\widetilde f_3+\widetilde f_2\widetilde f_3)
+\frac{1}{2}(\widetilde g_1^2+\widetilde g_2^2+\widetilde g_3^2)
-(\widetilde h_1\widetilde h_2+\widetilde h_1\widetilde h_3+\widetilde h_2\widetilde h_3),
\label{e6area}
\end{align}
where we have introduced the phases $f_i=e^{i\widetilde f_i}$, $g_i=e^{i\widetilde g_i}$ and $h_i=e^{i\widetilde h_i}$ and improved the expression using the constraint \eqref{e6constraint} so that the cross terms among variables from different sets of $\{f_i\}_{i=1}^3$, $\{g_i\}_{i=1}^3$ and $\{h_i\}_{i=1}^3$ are missing and the area is symmetric under the exchange of the variables in the same set.
If we redefine the chemical potential by
\begin{align}
\mu=\mu'-\frac{i}{3}(2\widetilde f_1+2\widetilde f_2+2\widetilde f_3+\widetilde g_1+\widetilde g_2+\widetilde g_3),
\label{e6muredef}
\end{align}
to remove the imaginary linear terms, we find that the area is given by
\begin{align}
\text{Area}(E_6)=\frac{3}{2}\mu^2+\frac{\Omega(E_6)}{2},
\end{align}
where $\Omega(E_6)$ is the quadratic Casimir element of $E_6$.
Using the variables ${\bm t}$ in \eqref{e6t}, $\Omega(E_6)$ is expressed as $(t_i=e^{i\widetilde t_i}$)
\begin{align}
\Omega(E_6)=4(\widetilde t_1^2+\widetilde t_2^2+\widetilde t_3^2+\widetilde t_4^2+\widetilde t_5^2+3\widetilde t_6^2).
\end{align}

\subsection{$E_7$ curve}

\begin{figure}[!t]
\centering\includegraphics[scale=0.6,angle=-90]{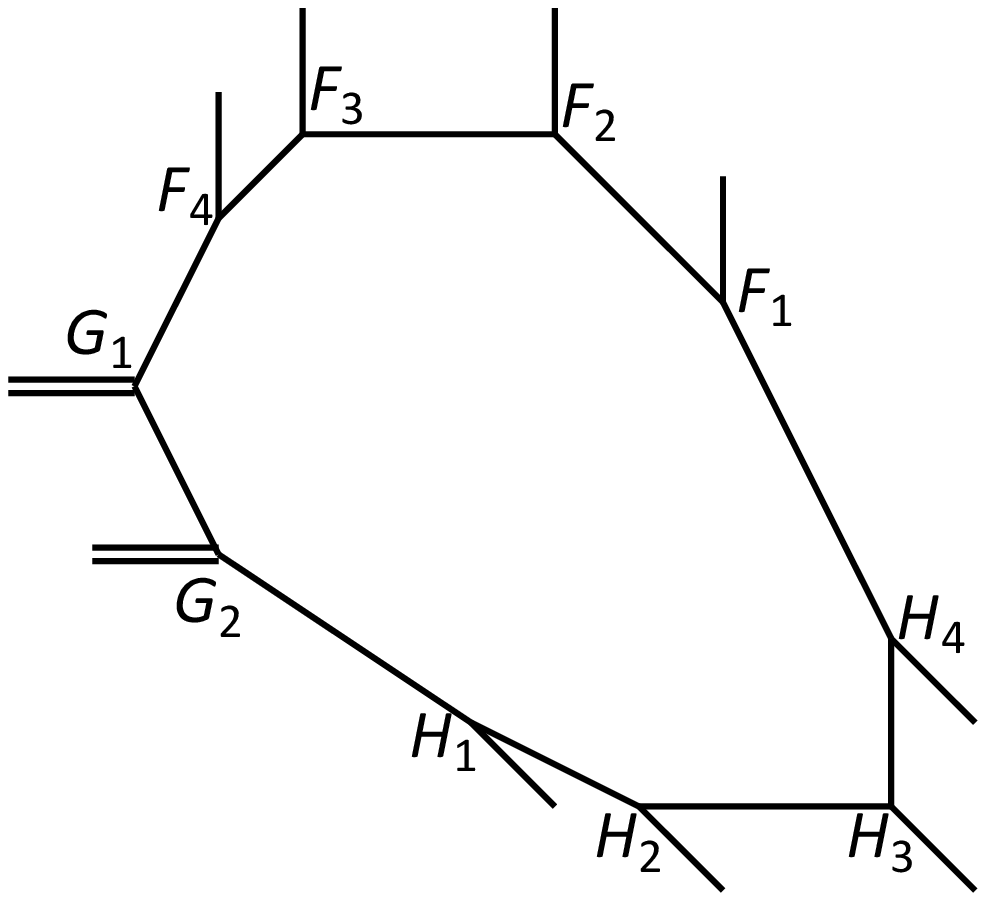}\qquad\qquad\qquad\includegraphics[scale=0.6,angle=-90]{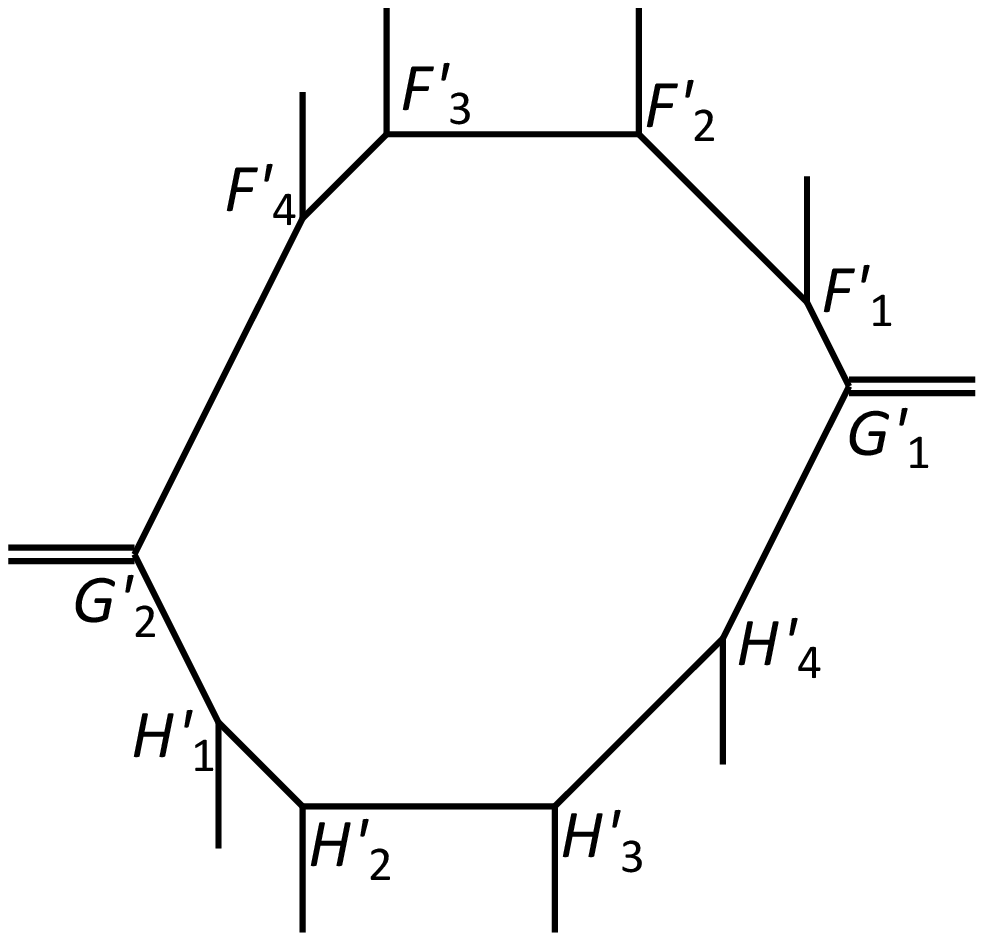}
\caption{Dual diagrams of the toric diagrams for the $E_7$ curve.
The left one is that in the triangular realization while the right one is that in the rectangular realization.}
\label{e7dual}
\end{figure}

Similarly, the dual diagrams of the toric diagrams for the $E_7$ curve in the triangular realization (figure \ref{e7tri}) and the rectangular realization (figure \ref{e7rect}) are depicted in figure \ref{e7dual}.
As previously, we formally assume the inequalities $f_1\ge f_2\ge f_3\ge f_4$, $g_1\ge g_2$ and $h_1\ge h_2\ge h_3\ge h_4$.
In the triangular realization, the coordinates of the vertices are given by
\begin{align}
&F_1:\biggl(f_1,\frac{z'}{f_1^2}\biggr),\quad
F_2:\biggl(f_2,\frac{z'}{f_1f_2}\biggr),\quad
F_3:\biggl(f_3,\frac{z'}{f_1f_2}\biggr),\quad
F_4:\biggl(f_4,\frac{f_4z'}{f_1f_2f_3}\biggr),\nonumber\\
&G_1:\biggl(\sqrt{\frac{f_1f_2f_3f_4g_1}{z'}},g_1\biggr),\quad
G_2:\biggl(\sqrt{\frac{f_1f_2f_3f_4g_1^2}{g_2z'}},g_2\biggr),\nonumber\\
&H_1:\biggl(\frac{h_2h_3h_4z'}{h_1^2},\frac{h_1}{h_2h_3h_4z'}\biggr),\;
H_2:\biggl(\frac{h_3h_4z'}{h_2},\frac{1}{h_3h_4z'}\biggr),\;
H_3:\biggl(h_4z',\frac{1}{h_3h_4z'}\biggr),\;
H_4:\biggl(h_4z',\frac{1}{h_4^2z'}\biggr),
\end{align}
while, in the rectangular realization, they are given by
\begin{align}
&F'_1:\biggl(f_1,\frac{z'}{f_1^2}\biggr),\quad
F'_2:\biggl(f_2,\frac{z'}{f_1f_2}\biggr),\quad
F'_3:\biggl(f_3,\frac{z'}{f_1f_2}\biggr),\quad
F'_4:\biggl(f_4,\frac{f_4z'}{f_1f_2f_3}\biggr),\nonumber\\
&G'_1:\biggl(\sqrt{\frac{z'}{g_1}},g_1\biggr),\quad
G'_2:\biggl(\sqrt{\frac{f_1f_2f_3f_4g_2}{z'}},g_2\biggr),\nonumber\\
&H'_1:\biggl(\frac{1}{g_1h_1},\frac{h_1}{h_2h_3h_4z'}\biggr),\quad
H'_2:\biggl(\frac{1}{g_1h_2},\frac{1}{h_3h_4z'}\biggr),\quad
H'_3:\biggl(\frac{1}{g_1h_3},\frac{1}{h_3h_4z'}\biggr),\quad
H'_4:\biggl(\frac{1}{g_1h_4},\frac{1}{h_4^2z'}\biggr).
\end{align}

Then the area is given by
\begin{align}
&\text{Area}(E_7)=\mu'^2-\frac{i}{2}\biggl(\sum_{i=1}^4\widetilde f_i-\sum_{i=1}^4\widetilde h_i\biggr)\mu'\nonumber\\
&\quad+\frac{\sum_i\widetilde f_i^2-2\sum_{i<j}\widetilde f_i\widetilde f_j}{4}
+\frac{\widetilde g_1^2+\widetilde g_2^2}{2}
+\frac{\sum_i\widetilde h_i^2-2\sum_{i<j}\widetilde h_i\widetilde h_j}{4},
\label{e7area}
\end{align}
where as in the $E_6$ case we have introduced the phase functions $f_i=e^{i\widetilde f_i}$, $g_i=e^{i\widetilde g_i}$ and $h_i=e^{i\widetilde h_i}$ and improved the area using the constraint \eqref{e7constraint}.
As previously, if we redefine the chemical potential by
\begin{align}
\mu=\mu'-\frac{i}{4}\biggl(\sum_{i=1}^4f_i-\sum_{i=1}^4h_i\biggr),
\label{e7muredef}
\end{align}
to remove the imaginary linear terms in $\mu'$ in \eqref{e7area}, the area is given by
\begin{align}
\text{Area}(E_7)=\mu^2+\frac{\Omega(E_7)}{2},
\end{align}
where the quadratic Casimir element of $E_7$ is given by
\begin{align}
\Omega(E_7)=2(2\widetilde t_1^2+2\widetilde t_2^2+2\widetilde t_3^2+2\widetilde t_4^2+2\widetilde t_5^2+6\widetilde t_6^2+3\widetilde t_7^2),
\end{align}
with the variables ${\bm t}$ defined in \eqref{e7t} and $t_i=e^{i\widetilde t_i}$.

\subsection{$E_8$ curve}

\begin{figure}[!t]
\centering\includegraphics[scale=0.45,angle=-90]{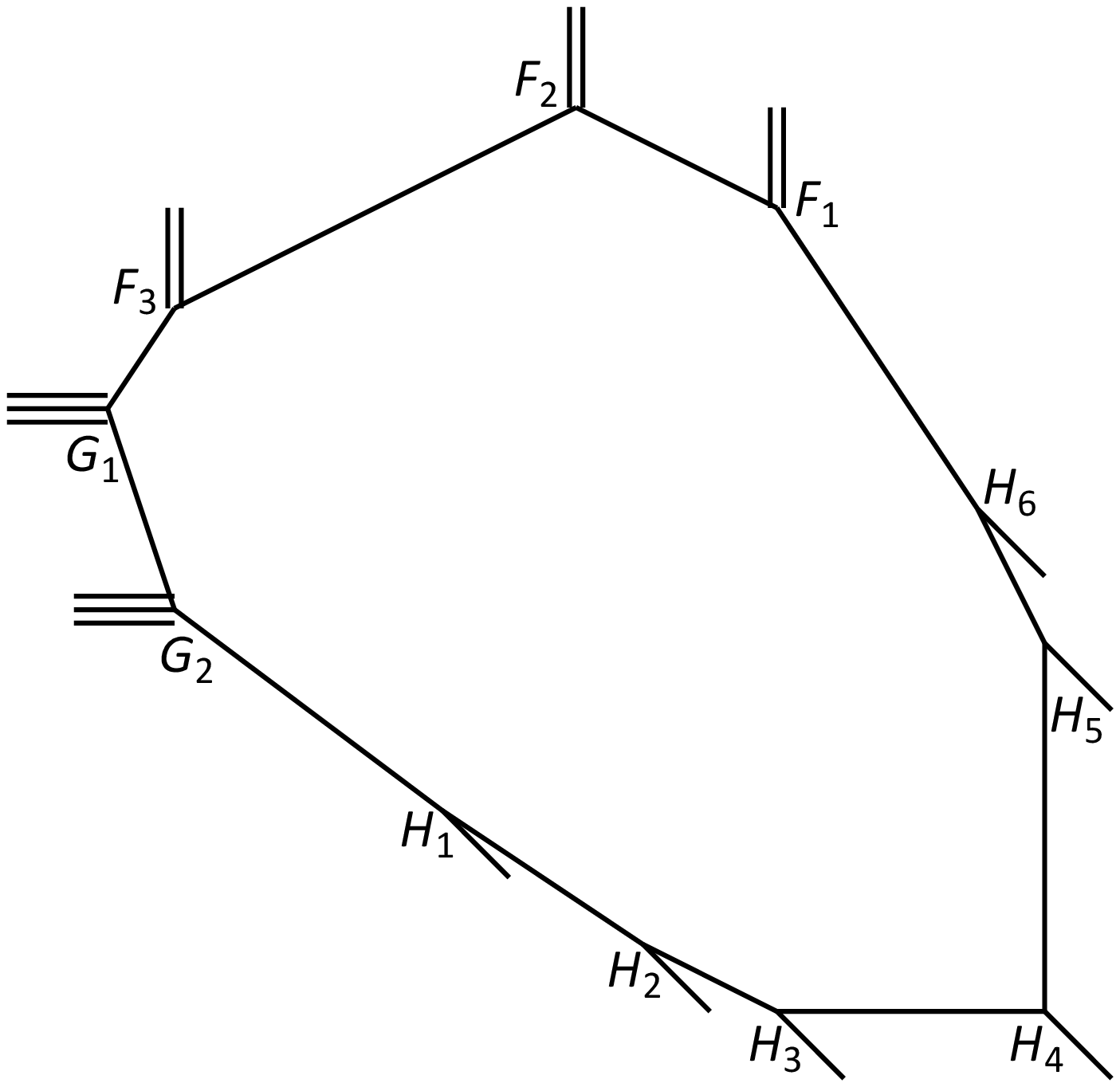}\qquad\qquad\qquad\includegraphics[scale=0.45,angle=-90]{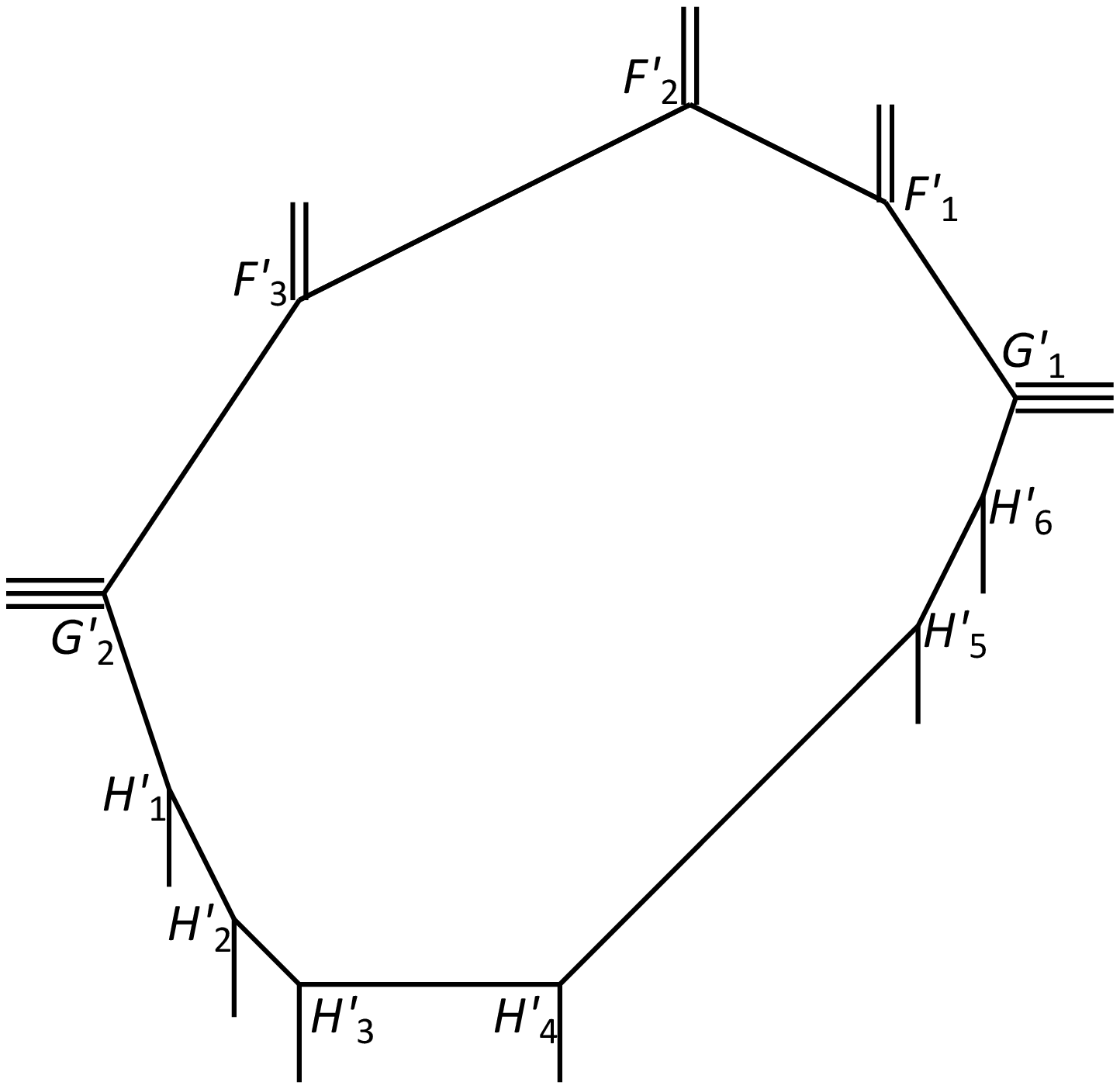}
\caption{Dual diagrams of the toric diagrams for the $E_8$ curve.
The left one is that in the triangular realization while the right one is that in the rectangular realization.}
\label{e8dual}
\end{figure}

Again we can repeat the computation of the area for the dual diagrams of the $E_8$ toric diagrams.
The dual diagrams of the toric diagrams for the $E_8$ curve in the triangular realization (figure \ref{e8tri}) and the rectangular realization (figure \ref{e8rect}) are depicted in figure \ref{e8dual}, where we assume the inequalities $f_1\ge f_2\ge f_3$, $g_1\ge g_2$ and $h_1\ge h_2\ge\cdots\ge h_6$.
In the triangular realization the coordinates of the vertices are given by
\begin{align}
&F_1:\biggl(f_1,\sqrt{\frac{z'}{f_1^3}}\biggr),\quad
F_2:\biggl(f_2,\sqrt{\frac{z'}{f_1^2f_2}}\biggr),\quad
F_3:\biggl(f_3,\sqrt{\frac{f_3z'}{f_1^2f_2^2}}\biggr),\nonumber\\
&G_1:\biggl(\sqrt[\leftroot{-1}\uproot{2}\scriptstyle 3]{\frac{f_1^2f_2^2f_3^2g_1^2}{z'}},g_1\biggr),\quad
G_2:\biggl(\sqrt[\leftroot{-1}\uproot{2}\scriptstyle 3]{\frac{f_1^2f_2^2f_3^2g_1^3}{g_2z'}},g_2\biggr),\nonumber\\
&H_1:\biggl(\frac{z'\prod_{i=2}^6h_i}{h_1^3},\frac{h_1^2}{z'\prod_{i=2}^6h_i}\biggr),\quad\!\!
H_2:\biggl(\frac{z'\prod_{i=3}^6h_i}{h_2^2},\frac{h_2}{z'\prod_{i=3}^6h_i}\biggr),\quad\!\!
H_3:\biggl(\frac{z'\prod_{i=4}^6h_i}{h_3},\frac{1}{z'\prod_{i=4}^6h_i}\biggr),\nonumber\\
&\quad H_4:\biggl(h_5h_6z',\frac{1}{h_4h_5h_6z'}\biggr),\quad\!\!
H_5:\biggl(h_5h_6z',\frac{1}{h_5^2h_6z'}\biggr),\quad\!\!
H_6:\biggl(h_6^2z',\frac{1}{h_6^3z'}\biggr),
\end{align}
while in the rectangular realization the coordinates are
\begin{align}
&F'_1:\biggl(f_1,\sqrt{\frac{z'}{f_1^3}}\biggr),\quad
F'_2:\biggl(f_2,\sqrt{\frac{z'}{f_1^2f_2}}\biggr),\quad
F'_3:\biggl(f_3,\sqrt{\frac{f_3z'}{f_1^2f_2^2}}\biggr),\nonumber\\
&G'_1:\biggl(\sqrt[\leftroot{-1}\uproot{2}\scriptstyle 3]{\frac{z'}{g_1^2}},g_1\biggr),\quad
G'_2:\biggl(\sqrt[\leftroot{-1}\uproot{2}\scriptstyle 3]{\frac{f_1^2f_2^2f_3^2g_2^2}{z'}},g_2\biggr),\nonumber\\
&H'_1:\biggl(\frac{1}{g_1h_1},\frac{h_1^2}{h_2h_3h_4h_5h_6z'}\biggr),\quad
H'_2:\biggl(\frac{1}{g_1h_2},\frac{h_2}{h_3h_4h_5h_6z'}\biggr),\quad
H'_3:\biggl(\frac{1}{g_1h_3},\frac{1}{h_4h_5h_6z'}\biggr),\nonumber\\
&\quad H'_4:\biggl(\frac{1}{g_1h_4},\frac{1}{h_4h_5h_6z'}\biggr),\quad
H'_5:\biggl(\frac{1}{g_1h_5},\frac{1}{h_5^2h_6z'}\biggr),\quad
H'_6:\biggl(\frac{1}{g_1h_6},\frac{1}{h_6^3z'}\biggr).
\end{align}

Then we find that the area is given by
\begin{align}
&\text{Area}(E_8)=\frac{1}{2}\mu'^2-\frac{i}{3}\biggl(\sum_{i=1}^3\widetilde f_i-\sum_{i=1}^6\widetilde h_i\biggr)\mu'\nonumber\\
&+\frac{\sum_i\widetilde f_i^2-4\sum_{i<j}\widetilde f_i\widetilde f_j}{6}
+\frac{\widetilde g_1^2+\widetilde g_2^2}{2}
+\frac{\sum_i\widetilde h_i^2-\sum_{i<j}\widetilde h_i\widetilde h_j}{3}.
\label{e8area}
\end{align}
As perviously, if we redefine the chemical potential by
\begin{align}
\mu=\mu'-\frac{i}{3}\biggl(\sum_{i=1}^3\widetilde f_i-\sum_{i=1}^6\widetilde h_i\biggr),
\label{e8muredef}
\end{align}
to remove the imaginary linear terms in $\mu'$, the area is given by
\begin{align}
\text{Area}(E_8)=\frac{1}{2}\mu^2+\frac{\Omega(E_8)}{2},
\end{align}
where the quadratic Casimir element of $E_8$ is given by
\begin{align}
\Omega(E_8)=2(2\widetilde t_1^2+2\widetilde t_2^2+2\widetilde t_3^2+2\widetilde t_4^2+2\widetilde t_5^2+6\widetilde t_6^2+3\widetilde t_7^2+\widetilde t_8^2),
\end{align}
with the variables ${\bm t}$ defined in \eqref{e8t} and $t_i=e^{i\widetilde t_i}$.

\section{BPS indices}\label{bpslist}

\begin{table}[!ht]
\begin{center}
\begin{tabular}{|c|c|c|}
\hline
$d$&$(j_\text{L},j_\text{R})$&representations\\
\hline\hline
$1$&$(0,0)$&$\overline{\bf 27}$\\\hline
$2$&$(0,\frac{1}{2})$&${\bf 27}$\\\hline
$3$&$(0,1)$&${\bf 78}$\\\cline{2-3}
&$(0,0),(\frac{1}{2},\frac{3}{2})$&${\bf 1}$\\\hline
$4$&$(0,\frac{3}{2})$&$\overline{\bf 351}$\\\cline{2-3}
&$(0,\frac{1}{2}),(\frac{1}{2},2)$&$\overline{\bf 27}$\\\hline
$5$&$(0,2)$&${\bf 1728}+{\bf 27}$\\\cline{2-3}
&$(0,1),(\frac{1}{2},\frac{5}{2})$&${\bf 351}+{\bf 27}$\\\cline{2-3}
&$(0,0),(\frac{1}{2},\frac{3}{2}),(1,3)$&${\bf 27}$\\\hline
\end{tabular}
\begin{tabular}{|c|c|c|}
\hline
$d$&$(j_\text{L},j_\text{R})$&representations\\
\hline\hline
$1$&$(0,0)$&${\bf 248}$\\\cline{2-3}
&$(\frac{1}{2},\frac{1}{2})$&${\bf 1}$\\\hline
$2$&$(0,\frac{1}{2})$&${\bf 3875}+{\bf 1}$\\\cline{2-3}
&$(\frac{1}{2},1)$&${\bf 248}$\\\cline{2-3}
&$(1,\frac{3}{2})$&${\bf 1}$\\\hline
\end{tabular}
\end{center}
\caption{The first few BPS indices for the $E_6$ curve (left) and the $E_8$ curve (right).}
\label{e6e8BPS}
\end{table}

In this appendix we collect a few BPS indices of lower degrees for completeness.
The main reference is \cite{HKP}, though the BPS indices are only given in non-negative integers.
Using the results of the instanton effects in the super Chern-Simons matrix models, it was possible to classify these integers by representations and reduce large integers into small multiplicities.
For the $D_5$ curve and the $E_7$ curve the expression of the BPS indices in terms of multiplicities are given in \cite{MNY}.
Here by applying the decoupling limit for the BPS indices, we are able to identify the BPS indices for the $E_6$ curve and the $E_8$ curve as multiplicities as well.
We list the first few multiplicities in table \ref{e6e8BPS}.

By combining the BPS indices listed in \cite{MNY} and table \ref{e6e8BPS} with the tables of the mirror maps listed in the main text, we are able to describe the non-perturbative terms of the ST/TS correspondence for the del Pezzo geometries completely in the group-theoretical language as explained in section \ref{summary}.
Note also that the BPS indices of the $E_6$ curve matches with \cite{HN} if we drop the spins.

\end{document}